\documentclass[aps,prd,notitlepage,nofootinbib,showpacs,10pt,tightenlines]{revtex4-1}

\usepackage{amssymb,amsmath,bm}
\usepackage{hyperref} 

\usepackage{graphics, graphicx}
\usepackage{color}
\usepackage{mathbbol}

\newcommand{\be}{\begin{equation}}
\newcommand{\ee}{\end{equation}}
\newcommand{\bea}{\begin{eqnarray}}
\newcommand{\eea}{\end{eqnarray}}
\newcommand{\bem}{\begin{multline}}
\newcommand{\eem}{\end{multline}}
\newcommand{\beg}{\begin{gather}}
\newcommand{\eeg}{\end{gather}}

\newcommand{\stackeven}[2]{{{}_{\displaystyle{#1}}\atop\displaystyle{#2}}}

\newcommand{\gsim}{\stackeven{>}{\sim}}
\newcommand{\as}{\alpha_s}

\def\eq#1{{Eq.~(\ref{#1})}}
\def\fig#1{{Fig.~\ref{#1}}}
\newcommand{\ben}{\begin{eqnarray*}}
\newcommand{\een}{\end{eqnarray*}}

\newcommand{\ord}[1]{\mathcal{O}\left(#1\right)}

\begin{document}

\title{Two-Gluon Correlations in Heavy-Light Ion Collisions: \\[2mm]
  Energy and Geometry Dependence, IR Divergences, and
  $k_T$-Factorization}

\author{Yuri~V.~Kovchegov,\footnote{kovchegov.1@asc.ohio-state.edu}
  Douglas~E.~Wertepny\footnote{wertepny.1@osu.edu}}

\affiliation{Department of Physics, The Ohio State University,
  Columbus, OH 43210, USA}

\begin{abstract}
  We study the properties of the cross section for two-gluon
  production in heavy-light ion collisions derived in our previous
  paper \cite{Kovchegov:2012nd} in the saturation/Color Glass
  Condensate framework. Concentrating on the energy and geometry
  dependence of the corresponding correlation functions we find that
  the two-gluon correlator is a much slower function of the
  center-of-mass energy than the one- and two-gluon production cross
  sections. The geometry dependence of the correlation function leads
  to stronger azimuthal near- and away-side correlations in the
  tip-on-tip U+U collisions than in the side-on-side U+U collisions,
  an exactly opposite behavior from the correlations generated by the
  elliptic flow of the quark-gluon plasma: a study of azimuthal
  correlations in the U+U collisions may thus help to disentangle the
  two sources of correlations.

  We demonstrate that the cross section for two-gluon production in
  heavy-light ion collisions contains a power-law infrared (IR)
  divergence even for fixed produced gluon momenta: while saturation
  effects in the target regulate some of the power-law IR divergent
  terms in the lowest-order expression for the two-gluon correlator,
  other power-law IR divergent terms remain, possibly due to absence
  of saturation effects in the dilute projectile. Finally we rewrite
  our result for the two-gluon production cross-section in a
  $k_T$-factorized form, obtaining a new factorized expression
  involving a convolution of one- and two-gluon Wigner distributions
  over both the transverse momenta and impact parameters. We show that
  the two-gluon production cross-section depends on two different
  types of unintegrated two-gluon Wigner distribution functions.
\end{abstract}

\maketitle

%%%%%%%%%%%%%%%%%%%%%%%%%%%%%%%%%%%%%%%%%%%%%%%%%%%%%%%%%%%%%%%%%%%%%%%%%%%%%%%

\section{Introduction}
\label{sec-Intro} 

The discovery of the long-range rapidity correlation known as the
'ridge' in heavy ion collisions at the Relativistic Heavy Ion Collider
(RHIC) \cite{Adams:2005ph,Adare:2008cqb,Alver:2009id,Abelev:2009af}
spurred, among other things, a flurry of activity
\cite{Armesto:2006bv,Armesto:2007ia,Dumitru:2008wn,Gavin:2008ev,Gelis:2008sz,Dusling:2009ni,Dumitru:2010iy,Dumitru:2010mv,Kovner:2010xk,Kovner:2011pe,Levin:2011fb,Kovner:2012jm,Dusling:2012cg,Dusling:2012wy,Dusling:2012iga,Kovchegov:1999ep}
aimed at better understanding two-particle correlations in the parton
saturation/Color Glass Condensate (CGC) physics framework (see
\cite{Gribov:1984tu,Iancu:2003xm,Jalilian-Marian:2005jf,Weigert:2005us,Gelis:2010nm,KovchegovLevin}
for reviews of the saturation/CGC field). Apart from quantifying how
much of the 'ridge' dynamics, which in the meantime was also observed
by experiments at the Large Hadron Collider (LHC) in proton--proton
($pp$) and proton--nucleus ($pA$) collisions
\cite{Khachatryan:2010gv,CMS:2012qk,Abelev:2012aa,Chatrchyan:2013nka},
is due to the initial-state saturation effects, the problem of
two-gluon production in nucleus--nucleus ($AA$) collisions is an
important theoretical problem in its own right, allowing us to gain a
new insight in the nonlinear dynamics of strong gluon fields in the
initial stages of heavy ion collisions.

The strategy for tackling the two-gluon production problem closely
parallels that of the single-gluon production. Any observable in the
saturation/CGC framework is calculated in three stages: (i) First the
observable is evaluated in the quasi-classical Glauber--Mueller (GM)
\cite{Mueller:1989st} / McLerran-Venugopalan (MV)
\cite{McLerran:1994vd,McLerran:1993ka,McLerran:1993ni} approximation;
(ii) then the small-$x$ Balitsky--Kovchegov (BK)
\cite{Balitsky:1996ub,Balitsky:1998ya,Kovchegov:1999yj,Kovchegov:1999ua}
and Jalilian-Marian--Iancu--McLerran--Weigert--Leonidov--Kovner
(JIMWLK)
\cite{Jalilian-Marian:1997dw,Jalilian-Marian:1997gr,Iancu:2001ad,Iancu:2000hn}
nonlinear evolution corrections are included into the expression;
(iii) finally, phenomenological applications
\cite{Albacete:2007sm,ALbacete:2010ad,Albacete:2010sy} demand that the
scale of the running coupling is fixed
\cite{Balitsky:2006wa,Gardi:2006rp,Kovchegov:2006vj,Horowitz:2010yg}
using, e.g., Brodsky--Lepage--Mackenzie (BLM) prescription
\cite{BLM}. Indeed higher-order perturbative corrections (beyond the
running coupling corrections) need to be included as well
\cite{Balitsky:2008zz}, though at this point they have not been
implemented in the existing phenomenology.

While the steps (i), (ii) and (iii) have been fully implemented for
the total cross section in deep inelastic scattering (DIS) and for the
corresponding structure functions, the situation is more complicated
for particle production. The classical gluon production cross section
(step (i) in the above classification) is known analytically in $pp$
collisions (dilute--dilute scattering)
\cite{Kovner:1995ts,Kovner:1995ja,Kovchegov:1997ke} and in $pA$
collisions (dilute--dense scattering)
\cite{Kovchegov:1998bi,Dumitru:2001ux}, while only numerical solution
\cite{Krasnitz:2003jw,Lappi:2003bi} exists for the classical gluon
field in $AA$ collisions (dense--dense scattering). Leading-$\ln 1/x$
evolution corrections (step (ii)) have been included into the $pA$
gluon production cross section in
\cite{Kovchegov:2001sc,Kovner:2006wr}. (The analogous $pp$ result can
be obtained from $pA$ by expansion to the lowest order in saturation
effects in the nucleus.) Running coupling corrections for the gluon
production (step (i)) have only been resummed for the $pp$ case
\cite{Horowitz:2010yg}.

Our understanding of the two-gluon production in the saturation/CGC
framework is somewhat less developed, chiefly due to the complexity of
the problem. The problem of two-gluon production in DIS and $pA$
(dilute--dense) collisions was solved in
\cite{JalilianMarian:2004da,Braun:2000bh,Baier:2005dv,Kovner:2006wr,Iancu:2013uva}
for the gluons produced with the large separation in rapidity, $\Delta
y \gtrsim 1/\as$ with $\as$ the strong coupling constant. To describe
two-gluon production in $AA$ collisions one needs to include
higher-order density effects in the projectile. The lowest-order
quasi-classical two-gluon production cross section in $AA$ was derived
in the original paper \cite{Dumitru:2008wn}, with the gluons close to
each other in rapidity, $\Delta y \lesssim 1/\as$. The corresponding
two-gluon production cross section for the semi-dilute--dense
scattering case (which we will refer to as heavy-light ion collisions)
was found only recently in
\cite{Kovner:2012jm,Kovchegov:2012nd}. (Note that the $pA$ scattering
is not a good model for two-gluon production in dense--dense
collisions in the CGC framework, since in $AA$ collisions the two
gluons are produced mainly from different nucleons: replacing one of
the nuclei by a proton would force two gluons to be emitted off of the
same proton, which is subleading in the $AA$ case. Here we model $AA$
collisions by considering the case when one nucleus is much smaller
than another one, while still having a significant number of nucleons
\cite{Kovchegov:2012nd}: these are the heavy-light ion collisions.)
Since the cross section in \cite{Kovner:2012jm,Kovchegov:2012nd}
expresses the interaction with the dense target nucleus via Wilson
lines, small-$x$ evolution between the produced gluons and the target
can be automatically included into it: however, unlike the $pA$ case
of
\cite{JalilianMarian:2004da,Braun:2000bh,Kovner:2006wr,Iancu:2013uva},
no one has yet included the evolution between the projectile and the
produced gluons in heavy-light ion collisions, allowing the gluons to
be produced at mid-rapidity in RHIC or LHC kinematics. Two-gluon
production in the dense--dense ($AA$) case has so far been tackled
only numerically \cite{Lappi:2009xa} in the quasi-classical limit.

All the above-mentioned results for two-gluon production
\cite{Dumitru:2008wn,Kovner:2012jm,Kovchegov:2012nd,Lappi:2009xa}
indicate that the corresponding cross section exhibits long-range
rapidity correlations, along with enhancement of correlations at the
azimuthal opening angles $\Delta \phi = 0$ and $\Delta \phi = \pi$
between the gluons. A saturation-inspired generalization of the
lowest-order result \cite{Dumitru:2008wn} allowed for successful $pp$
and $pA$ phenomenology
\cite{Dumitru:2010iy,Dusling:2012cg,Dusling:2012wy,Dusling:2012iga}. However,
further theoretical advances are needed to improve the precision of
CGC predictions for di-hadron correlations.

The goal of this paper is to study the properties of the two-gluon
production cross section in heavy-light ion collisions found in
\cite{Kovchegov:2012nd}. While we begin the paper by reviewing the
main results of \cite{Kovchegov:2012nd} in Sec.~\ref{sec-recap}, our
paper is best read in conjunction with \cite{Kovchegov:2012nd}. 

The scope of our paper covers a wide range of issues in our
understanding of two-gluon correlations. Recently there was a
significant amount of discussion in the community of whether the
'ridge' correlations seen in $pp$ and $pA$ collisions at the LHC
\cite{Khachatryan:2010gv,CMS:2012qk,Abelev:2012aa,Chatrchyan:2013nka}
have hydrodynamic or CGC origin \cite{Bzdak:2013zma}. It appears
important to find an experiment being able to disentangle the two
mechanism generating azimuthal correlations. In Sec.~\ref{sec:Geo} we
show that such an experiment could be the study of long-range rapidity
correlations in the $U+U$ collisions. We show that the CGC
correlations are stronger in the tip-on-tip $U+U$ collisions than in
the side-on-side ones. Such behavior is the exact opposite of the
ellipticity-driven hydrodynamic correlations
\cite{Heinz:2004ir,Kuhlman:2005ts,Kuhlman:2006qp}. Perhaps the two
effects could be disentangled by studying the $U+U$ collisions data.

It has been known since the original calculation of
\cite{Dumitru:2008wn} that the two-gluon production cross section in
CGC contains power-law infrared (IR) divergences, even for fixed
momenta of the produced gluons (see \eq{eq:corr_LO} below). This is in
stark contrast to the single-gluon production cross section, which
only contains weak logarithmic IR divergences. The authors of
\cite{Dumitru:2008wn} conjectured that saturation effects in both
colliding nuclei, when included, would regulate this divergence: this
assumption appears to be confirmed by numerical simulations of
\cite{Lappi:2009xa}. The analytical expression for two-gluon
production in heavy-light ion collisions obtained in
\cite{Kovchegov:2012nd} contains all-order saturation effects in one
of the nuclei (the target nucleus). In Sec.~\ref{sec:IRsafe} we study
the result of \cite{Kovchegov:2012nd} to explore whether the
saturation effects in one of the nuclei are sufficient to regulate the
power-law IR divergence of \cite{Dumitru:2008wn}. We find that, while
some of the power-law IR divergent terms are regulated by saturation
effects in the target, other terms with the same divergence remain,
rendering the whole cross-section power-law IR-divergent (see
\eq{crossed_xsect_IR3} below). Such result is indeed worrisome, since
it questions our ability to make controlled phenomenological
predictions for di-hadron correlations: however we find that the
IR-divergent piece does not contain any azimuthally-nontrivial
correlations. Hence the $\Delta \phi$-dependent part of the
correlations is not affected and is safe from this IR divergence.

The cross section for single-gluon production in $pA$ collisions
calculated in approximations (i) and (ii) described above can be cast
in a $k_T$-factorized form
\cite{Braun:2000bh,Kovchegov:2001sc}. Recently, $k_T$-factorization at
the JIMWLK functional level has been proven (in the leading-$\ln 1/x$
approximation) for local in rapidity particle production in
dense--dense collisions in
\cite{Gelis:2008rw,Gelis:2008ad,Gelis:2008sz}. Motivated by these
results we try obtaining a $k_T$-factorized form for two-gluon
production cross section in Sec.~\ref{sec:fact}. The final result,
given in \eq{eq:factorized_final}, is a $k_T$-factorized expression,
whose form is more complicated than that for the single-gluon
production \cite{Braun:2000bh,Kovchegov:2001sc}. The expression
\eqref{eq:factorized_final} contains not only a convolution over
transverse momenta, but also over impact parameters. The objects being
convoluted are not the unintegrated gluon distributions, but rather
one- and two-gluon Wigner distributions, containing the information
about both the gluons momenta and impact parameters. Note that the
production cross section employs two types of two-gluon Wigner
distributions: double-trace \eqref{eq:doubletrace_dist} and quadrupole
\eqref{eq:quad_dist} ones, introduced for the first time in this
paper. (This is to be compared with the dipole single-gluon
distribution entering the single-gluon production
\cite{Braun:2000bh,Kovchegov:2001sc}.)

Another important question concerning the energy-dependence of the
two-gluon production cross section is studied in
Sec.~\ref{sec:energy}. We find that, in the leading power-of-energy
approximation, the two-gluon correlation function in heavy-light ion
collisions is independent of the center-of-mass energy of the
collision. This is a prediction which can be verified
experimentally. Note that, as we have mentioned above, the two-gluon
production cross section of \cite{Kovchegov:2012nd} was derived
without any evolution corrections between the projectile and the
produced gluons. Hence, in this approximation, the net rapidity
interval is equal to the rapidity interval between the produced gluons
and the target. Further studies of energy and rapidity dependence of
two-gluon production should be carried out improving upon our
approximation.

We conclude in Sec.~\ref{sec:conc} by summarizing our main results and
discussing future avenues of research on correlations in the CGC
framework.

%%%%%%%%%%%%%%%%%%%%%%%%%%%%%%%%%%%%%%%%%%%%%%%%%%%%%%%%%%%%%%%%%%%%%%%%%%%%%%%

\section{Brief Summary of the Results for the Two-Gluon Production
  Cross Section}
\label{sec-recap} 

In \cite{Kovchegov:2012nd} we considered two-gluon production in a
collision of a large heavy ion (target) with a lighter nucleus
(projectile). The projectile was considered to be smaller than the
target, but still large enough a nucleus for the two gluons to be
produced in collisions of different nucleons in the projectile with
the target nucleus. Formally we assume that $A_2 \gg A_1 \gg 1$, where
$A_1$ and $A_2$ are respectively the atomic numbers of the projectile
and the target. The target was large enough for the saturation effects
to be important, such that $\as^2 \, A_2^{1/3} \sim 1$. The saturation
effects in the projectile were kept at the lowest order with $\as^2 \,
A_1^{1/3} \ll 1$. In terms of momentum scales the regime of interest
is $k_T \gsim Q_{s1}$, with $k_T$ the transverse momentum of either
one of the produced gluons ($k_1$ or $k_2$) and $Q_{s1}$ the
saturation scale of the projectile which is much smaller than the
saturation scale of the target, $Q_{s2} \gg Q_{s1} \gg
\Lambda_{QCD}$. The approximation used in \cite{Kovchegov:2012nd}
corresponds to calculating the two-gluon production in nucleus-nucleus
($A+A$) collisions in the McLerran-Venugopalan (MV) model
\cite{McLerran:1994vd,McLerran:1993ka,McLerran:1993ni} while keeping
interactions with the projectile nucleus to the lowest non-trivial
order of two interacting nucleons. This setup may not directly
describe the bulk of proton-nucleus ($p+A$) collisions (also known as
the dense-dilute collisions); note however that if one triggers on the
high-multiplicity $p+A$ events one is then probing rare
high-parton-number fluctuations in the proton wave function, which
make the parton density in the proton appear more like that in a small
nucleus. Hence our setup may be relevant for high-multiplicity $p+A$
collisions as well.

The resulting two-gluon production cross section was written in
\cite{Kovchegov:2012nd} as a sum of two terms corresponding to two
different classes of diagrams (labeled 'square' and 'crossed'),
\begin{align}\label{eq_all}
  \frac{d \sigma}{d^2 k_1 dy_1 d^2 k_2 dy_2} = \frac{d
    \sigma_{square}}{d^2 k_1 dy_1 d^2 k_2 dy_2} + \frac{d
    \sigma_{crossed}}{d^2 k_1 dy_1 d^2 k_2 dy_2},
\end{align}
with 
\begin{align} 
\label{eq:2glue_prod_main} 
\frac{d \sigma_{square}}{d^2 k_1 dy_1 d^2 k_2 dy_2} & = \frac{\as^2 \,
  C_F^2}{16 \, \pi^8} \int d^2 B \, d^2 b_1 \, d^2 b_2 \, T_1 ({\bm B}
- {\bm b}_1) \, T_1 ({\bm B} - {\bm b}_2) \, d^2 x_1 \, d^2 y_1 \, d^2
x_2 \, d^2 y_2 \, e^{- i \; {\bm k}_1 \cdot ({\bm x}_1-{\bm y}_1) - i
  \; {\bm k}_2 \cdot ({\bm x}_2-{\bm y}_2)} \notag \\ & \times \,
\frac{ {\bm x}_1 - {\bm b}_1}{ |{\bm x}_1 - {\bm b}_1 |^2 } \cdot
\frac{ {\bm y}_1 - {\bm b}_1}{ |{\bm y}_1 - {\bm b}_1 |^2 } \ \frac{
  {\bm x}_2 - {\bm b}_2}{ |{\bm x}_2 - {\bm b}_2 |^2 } \cdot \frac{
  {\bm y}_2 - {\bm b}_2}{ |{\bm y}_2 - {\bm b}_2 |^2 } \notag \\
& \times \, \left\langle \left( \frac{1}{N_c^2-1} \; \mbox{Tr}[ U_{{\bm x}_1}
    U_{{\bm y}_1}^\dagger ] \; - \; \frac{1}{N_c^2-1} \; \mbox{Tr}[ U_{{\bm
        x}_1} U_{{\bm b}_1}^\dagger ] \; - \; \frac{1}{N_c^2-1} \;
    \mbox{Tr}[ U_{{\bm b}_1} U_{{\bm y}_1}^\dagger ] \; + \; 1 \right) \right. \notag \\
& \times \left. \left( \frac{1}{N_c^2-1} \; \mbox{Tr}[ U_{{\bm x}_2} U_{{\bm
        y}_2}^\dagger ] \; - \; \frac{1}{N_c^2-1} \; \mbox{Tr}[ U_{{\bm x}_2}
    U_{{\bm b}_2}^\dagger ] \; - \; \frac{1}{N_c^2-1} \; \mbox{Tr}[ U_{{\bm
        b}_2} U_{{\bm y}_2}^\dagger ] \; + \; 1 \right) \right\rangle
\end{align}
and
\begin{align}\label{crossed_xsect}
  & \frac{d \sigma_{crossed}}{d^2 k_1 dy_1 d^2 k_2 dy_2} =
  \frac{1}{[2(2 \pi)^3]^2} \, \int d^2 B \, d^2 b_1 \, d^2 b_2 \, T_1
  ({\bm B} - {\bm b}_1) \, T_1 ({\bm B} - {\bm b}_2) \, d^2 x_1 \, d^2
  y_1 \, d^2 x_2 \, d^2 y_2 \notag \\ & \times \, \left[ e^{- i \;
      {\bm k}_1 \cdot ({\bm x}_1-{\bm y}_2) - i \; {\bm k}_2 \cdot
      ({\bm x}_2-{\bm y}_1)} + e^{- i \; {\bm k}_1 \cdot ({\bm
        x}_1-{\bm y}_2) + i \; {\bm k}_2 \cdot ({\bm x}_2-{\bm y}_1)}
  \right] \, \frac{16 \; {\alpha}_s^2}{\pi^2} \, \frac{C_F}{2 N_c} \;
  \frac{ {\bm x}_1 - {\bm b}_1}{|{\bm x}_1 - {\bm b}_1|^2} \cdot
  \frac{{\bm y}_2 - {\bm b}_2 }{ |{\bm y}_2 - {\bm b}_2|^2 } \,
  \frac{{\bm x}_2 - {\bm b}_2}{|{\bm x}_2 - {\bm b}_2|^2} \cdot
  \frac{{\bm y}_1 - {\bm b}_1 }{ |{\bm y}_1 - {\bm b}_1|^2} \notag \\
  & \times \, \bigg[ Q( {\bm x}_1, {\bm y}_1 , {\bm x}_2 , {\bm y}_2 )
  - Q( {\bm x}_1, {\bm y}_1 , {\bm x}_2 , {\bm b}_2 ) - Q( {\bm x}_1,
  {\bm y}_1 , {\bm b}_2 , {\bm y}_2 ) + S_G( {\bm x}_1, {\bm y}_1) -
  Q( {\bm x}_1, {\bm b}_1 , {\bm x}_2 , {\bm y}_2 ) \; + \; Q( {\bm
    x}_1, {\bm b}_1 , {\bm x}_2 , {\bm b}_2 ) \notag \\ & + Q( {\bm
    x}_1, {\bm b}_1 , {\bm b}_2 , {\bm y}_2 ) - S_G( {\bm x}_1, {\bm
    b}_1) - Q( {\bm b}_1, {\bm y}_1 , {\bm x}_2 , {\bm y}_2 ) + Q(
  {\bm b}_1, {\bm y}_1 , {\bm x}_2 , {\bm b}_2 ) + Q( {\bm b}_1, {\bm
    y}_1 , {\bm b}_2 , {\bm y}_2 ) - S_G( {\bm b}_1, {\bm y}_1) + S_G(
  {\bm x}_2, {\bm y}_2)
  \notag \\
  & - S_G( {\bm x}_2, {\bm b}_2) - S_G( {\bm b}_2, {\bm y}_2) +
  1\bigg].
\end{align}

We denote two-dimensional vectors in the transverse plane by ${\bm v}
= (v^x, v^y)$ with their length $v_T \equiv v_\perp \equiv |{\bm
  v}|$. As usual, $\as$ is the strong coupling constant, $N_c$ is the
number of quark colors, and $C_F = (N_c^2 -1)/2 N_c$ is the Casimir
operator of SU($N_c$) in the fundamental representation. The center of
the projectile nucleus is located at impact parameter $\bm B$ with
respect to the center of the target nucleus, with ${\bm b}_1$ and
${\bm b}_2$ the impact parameters of the two interacting nucleons in
the projectile, also measured with respect to the center of the
target. The nuclear profile function $T_1 ({\bm b})$ describes the
distribution of nucleons in the projectile. Angle brackets $\langle
\ldots \rangle$ denote averaging in the target wave function squared.

The interactions with the target are described using
\begin{align}
  \label{eq:Wline}
  U_{\bm x} = \mbox{P} \exp \left\{ i \, g \,
    \int\limits_{-\infty}^\infty \, d x^+ \, {\cal A}^- (x^+, x^-=0,
    {\bm x}) \right\},
\end{align}
which is the Wilson line taken along the $x^+$ light cone with ${\cal
  A}^-$ the gluon field of the target nucleus in the adjoint
representation. The contribution \eqref{crossed_xsect} depends on the
$S$-matrices for the adjoint color-dipole
\begin{align}
  S_G ( {\bm x}_1 , {\bm x}_2 , Y) \equiv \frac{1}{N_c^2-1} \;
  \left\langle \mbox{Tr} [ U_{{\bm x}_1} U_{{\bm x}_2}^\dagger ]
  \right\rangle (Y)
\end{align}
and adjoint color-quadrupole
\begin{align}\label{quad_def}
  Q ( {\bm x}_1 , {\bm x}_2 , {\bm x}_3 , {\bm x}_4, Y ) \equiv
  \frac{1}{N_c^2-1} \; \left\langle \mbox{Tr} [ U_{{\bm x}_1} U_{{\bm
        x}_2}^\dagger U_{{\bm x}_3} U_{{\bm x}_4}^\dagger ]
  \right\rangle (Y),
\end{align}
where, for the future purposes, we now explicitly show the rapidity
($Y$) dependence of the matrix elements.

We impose no ordering on the rapidities $y_1$ and $y_2$ of the two
produced gluons. (We do assume that $|y_1 - y_2| \ll 1/\as$ and $0 < Y
- y_{1,2} \ll 1/\as$, with $Y$ the rapidity of the projectile, such
that no small-$x$ evolution corrections need to be included in the
rapidity intervals between the gluons and between the projectile and
the gluons.) In the case of rapidity-ordered two-gluon production
(say, $y_2 \gg y_1$) the corresponding cross section was found
previously in \cite{Kovner:2010xk} (see also
\cite{Kovner:2011pe,Baier:2005dv}), in apparent agreement with our
\eq{eq:2glue_prod_main} (modulo the nuclear profile functions $T_1
({\bm b})$ we included in the projectile nucleus). An expression for
two-gluon production without rapidity ordering containing both the
double-trace and quadrupole structures of
Eqs.~\eqref{eq:2glue_prod_main} and \eqref{crossed_xsect} was
obtained in \cite{Kovner:2012jm} shortly before our work
\cite{Kovchegov:2012nd}.

The matrix elements of the double-trace, dipole and quadrupole
operators entering Eqs.~\eqref{eq:2glue_prod_main} and
\eqref{crossed_xsect} were evaluated in \cite{Kovchegov:2012nd} using
the Gaussian approximation (the MV model). When using this
approximation one treats both the projectile and the target in the
same consistent way, including only multiple interactions with target
and projectile nucleons in the cross section. The drawback is, of
course, that the resulting two-gluon production cross section is
energy- and rapidity-independent, just like all other observables in
the quasi-classical approximation. Inclusion of the full energy and
rapidity dependence goes beyond the scope of the present
work. However, evolution corrections can be readily included in the
rapidity interval between the produced gluons and the target by
evolving the double-trace, dipole and quadrupole operators using the
BK
\cite{Balitsky:1996ub,Balitsky:1998ya,Kovchegov:1999yj,Kovchegov:1999ua}
and JIMWLK
\cite{Jalilian-Marian:1997dw,Jalilian-Marian:1997gr,Iancu:2001ad,Iancu:2000hn}
evolution equations. This would make the two-gluon production cross
section energy-dependent: the effect of such evolution corrections
will be explored below in Sec.~\ref{sec:energy}.

In the MV model and in the large-$N_c$ approximation\footnote{Note
  that the large-$N_c$ approximation here implies a regular 't Hooft
  large-$N_c$ limit taken while keeping the saturation scale $Q_{s}$
  fixed: this $Q_{s}$ fixing can be achieved within the standard
  large-$N_c$ limit by assuming that nucleons are made out of $\sim
  N_c^2$ valence quarks.} the double-trace and quadrupole operators
entering Eqs.~\eqref{eq:2glue_prod_main} and \eqref{crossed_xsect}
were found in \cite{Kovchegov:2012nd}
(cf. \cite{JalilianMarian:2004da}). The results are as follows. For
the double-trace operator we write
\begin{align}
\label{Ddef}
  \frac{1}{(N_c^2 - 1)^2} \, \langle \mbox{Tr}[ U_{{\bm x}_1} U_{{\bm
      x}_2}^\dagger ] \, \mbox{Tr}[ U_{{\bm x}_3} U_{{\bm x}_4}^\dagger ]
  \rangle = \frac{1}{(N_c^2 - 1)^2} \, \langle \mbox{Tr}[ U_{{\bm x}_1}
  U_{{\bm x}_2}^\dagger ] \rangle \langle \mbox{Tr}[ U_{{\bm x}_3} U_{{\bm
      x}_4}^\dagger ] \rangle + \Delta ( {\bm x}_1 , {\bm x}_2 , {\bm
    x}_3 , {\bm x}_4 ),
\end{align}
where $\Delta$ represents the subleading in $\frac{1}{N_c^2}$ terms in
the matrix element. To leading order in $\frac{1}{N_c^2}$ in the MV
model it is given by
\begin{align}
  \label{Delta_exp}
  \Delta ( {\bm x}_1 , {\bm x}_2 , {\bm x}_3 , {\bm x}_4 ) =
  \frac{(D_3-D_2)^2}{N_c^2} \, & \left[ \frac{e^{D_1}}{D_1-D_2} -
    \frac{2 \, e^{D_1}}{(D_1-D_2)^2} + \frac{e^{D_1}}{D_1-D_3} -
    \frac{2 \,
      e^{D_1}}{(D_1-D_3)^2} \right. \notag \\
  & \left. + \frac{2 \, e^{\frac{1}{2}(D_1+D_2)}}{(D_1-D_2)^2} +
    \frac{2 \, e^{\frac{1}{2}(D_1+D_3)}}{(D_1-D_3)^2} \right] + O
  \left( \frac{1}{N_c^4} \right),
\end{align}
where we have defined
\begin{subequations}
\label{Ds}
\begin{align}
  D_1 & = - \Gamma_G \left( \bm x_1 , \bm x_2 , Y = 0 \right) - \Gamma_G \left( \bm x_3 , \bm x_4 , Y = 0 \right) \\
  D_2 & = - \Gamma_G \left( \bm x_1 , \bm x_3 , Y = 0 \right) - \Gamma_G \left( \bm x_2 , \bm x_4 , Y = 0 \right) \\
  D_3 & = - \Gamma_G \left( \bm x_1 , \bm x_4 , Y = 0 \right) -
  \Gamma_G \left( \bm x_2 , \bm x_3 , Y = 0 \right) .
\end{align}
\end{subequations}
In the MV model used in \cite{Kovchegov:2012nd}
\begin{equation}
\label{gammadef}
\Gamma_G \left( \bm x_1 , \bm x_2 , Y = 0 \right) = 
\frac{Q_{s2}^2}{4} | \bm x_1 - \bm x_2 |^2 \, 
\ln \left( \frac{1}{| \bm x_1 - \bm x_2 | \Lambda } \right)
\end{equation}
with $Q_{s2}$ the target saturation scale for gluons in the
quasi-classical (MV) limit \cite{Mueller:1989st} and $\Lambda$ an
infrared (IR) cutoff. (Note that our notation here is slightly
different from \cite{Kovchegov:2012nd}: we use $Q_{s2}$ instead of
$Q_{s0}$ used in \cite{Kovchegov:2012nd} to denote the same MV
saturation scale of the target nucleus.)

The $S$-matrix $S_G$ for the gluon color dipole interaction with the
target and in the MV model is \cite{Mueller:1989st}
\begin{align}
  \label{eq:SG_GM}
  S_G ({\bm x}_1, {\bm x}_2, Y=0) = e^{- \Gamma_G \left( \bm x_1 , \bm
      x_2 , Y = 0 \right)}.
\end{align}
The gluon color-quadrupole $S$-matrix in the MV model and in the
large-$N_c$ approximation is
\begin{align}
\label{eq:quad_MV}
Q ({\bm x}_1, {\bm x}_2, {\bm x}_3, {\bm x}_4, Y=0) = \left[e^{D_1/2}
  + \frac{D_3 - D_2}{D_1 - D_3} \, \left( e^{D_1/2} - e^{D_3/2}
  \right) \right]^2.
\end{align}

The correlation function is defined by
\begin{align}\label{corr_def}
  C ( {\bm k}_1, {y}_1, {\bm k}_2, {y}_2 ) = {\cal N} \, \frac{\frac{d
      \sigma}{d^2 k_1 dy_1 \, d^2 k_2 dy_2}}{\frac{d \sigma}{d^2 k_1 d y_1}
    \, \frac{d \sigma}{d^2 k_2 d y_2}} - 1
\end{align}
where the normalization factor ${\cal N}$ is usually fixed by
requiring that the numerator of $C$ (after reducing both terms in
\eqref{corr_def} to the common denominator) integrates out to zero
when integrating over the whole sample defined by the cuts. (Our
${\cal N}$ here is defined differently from that in Eq.~(1) of
\cite{Kovchegov:2012nd}.) For instance, the correlator as a function
of gluon rapidities $y_1, y_2$ and azimuthal angles $\phi_1, \phi_2$
for gluons with fixed magnitudes of their transverse momenta $k_1,
k_2$ is given by
\begin{align}\label{corr_def2}
  C ( {\bm k}_1, {y}_1, {\bm k}_2, {y}_2 ) = \frac{\left[ \int d
      \phi_1 \, d y_1 \, \frac{d \sigma}{d^2 k_1 d y_1} \, \int d
      \phi_2 \, d y_2 \, \frac{d \sigma}{d^2 k_2 d y_2} \right] }{
    \left[ \int d \phi_1 \, dy_1 \, d \phi_2 \, dy_2 \, \frac{d
        \sigma}{d^2 k_1 dy_1 \, d^2 k_2 dy_2} \right]} \ \frac{\frac{d
      \sigma}{d^2 k_1 dy_1 \, d^2 k_2 dy_2} }{\frac{d \sigma}{d^2 k_1
      d y_1} \, \frac{d \sigma}{d^2 k_2 d y_2}} - 1.
\end{align}
In \cite{Kovchegov:2012nd} it was shown that the correlations
contained in the cross sections \eqref{eq_all},
\eqref{eq:2glue_prod_main} and \eqref{crossed_xsect} are symmetric
under ${\bm k}_1 \leftrightarrow {\bm k}_2$ and ${\bm k}_2 \rightarrow
- {\bm k}_2$. This implies that the correlation function contains only
even Fourier harmonics in its Fourier decomposition over the azimuthal
opening angle $\Delta \phi = \phi_1 - \phi_2$: therefore the near- and
away-side correlations (that is, correlations around $\Delta \phi =0$
and $\Delta \phi =\pi$ respectively) resulting from the calculation
\cite{Kovchegov:2012nd} are identical. The correlations are also flat
in rapidity up to $|y_1 - y_2| \lesssim 1/\as$, making them a
plausible contributor to the 'ridge' correlation observed in $A+A$,
$p+A$ and $p+p$ collisions at RHIC and LHC.

Substituting Eqs. \eqref{eq_all}, \eqref{eq:2glue_prod_main} and
\eqref{crossed_xsect} along with the lowest-order single-gluon
production cross section
\begin{align}
  \label{eq:1GLO}
  \frac{d \sigma^{pA_2}}{d^2 k \, dy \, d^2 b} = \frac{\as \,
    C_F}{\pi^2} \, \frac{Q_{s2}^2 ({\bm b})}{k_T^4} \, \ln
  \frac{k_T^2}{\Lambda^2}
\end{align}
into \eq{corr_def2} and expanding the resulting correlator to the
lowest non-trivial order in multiple rescattering in the target
(lowest non-trivial order in $Q_{s2}$) one obtains
\cite{Kovchegov:2012nd}
\begin{align}
  \label{eq:corr_LO}
  & C ({\bm k}_1, y_1, {\bm k}_2, y_2)\big|_{LO} = \frac{1}{N_c^2} \,
  \frac{\int d^2 B \, d^2 b \, [T_1 ({\bm B} - {\bm b})]^2 \, Q_{s2}^4
    ({\bm b})}{\int d^2 B \, d^2 b_1 \, d^2 b_2 \, T_1 ({\bm B} - {\bm
      b}_1) \, T_1 ({\bm B} - {\bm b}_2) \, Q_{s2}^2 ({\bm b}_1) \,
    Q_{s2}^2 ({\bm b}_2) } \notag \\ & \times \, \frac{{\bm k}_1^2 \,
    {\bm k}_2^2}{\ln \frac{k_1^2}{\Lambda^2} \, \ln
    \frac{k_2^2}{\Lambda^2}} \, \bigg\{ 2 \, \int\limits_\Lambda
  \frac{d^2 l}{({\bm l}^2)^2} \, \left[ \frac{1}{({\bm k}_1 - {\bm
        l})^2 \, ({\bm k}_2 + {\bm l})^2} + \frac{1}{({\bm k}_1 - {\bm
        l})^2 \, ({\bm k}_2 - {\bm l})^2} \right] \notag \\ & +
  \frac{1}{8} \, \bigg[ \int\limits_\Lambda \frac{d^2 l}{({\bm l}^2)^2
    \, (({\bm l} - {\bm k}_1 + {\bm k}_2)^2)^2 \, (({\bm k}_1 - {\bm
      l})^2)^2 \, (({\bm k}_2 + {\bm l})^2)^2} \, \left[ {\bm l}^2 \,
    ({\bm k}_2 + {\bm l})^2 + ({\bm k}_1 - {\bm l})^2 \, ({\bm l} -
    {\bm k}_1 + {\bm k}_2)^2 - {\bm k}_1^2 \, ({\bm k}_2 - {\bm k}_1 +
    2 \, {\bm l})^2 \right] \notag \\ & \times \, \left[ {\bm l}^2 \,
    ({\bm k}_1 - {\bm l})^2 + ({\bm k}_2 + {\bm l})^2 \, ({\bm l} -
    {\bm k}_1 + {\bm k}_2)^2 - {\bm k}_2^2 \, ({\bm k}_2 - {\bm k}_1 +
    2 \, {\bm l})^2 \right] + ({\bm k}_2 \rightarrow - {\bm k}_2)
  \bigg] \bigg\}.
\end{align}
The momentum-space part of the expression \eqref{eq:corr_LO}
reproduces that derived in the original analyses of the ridge
correlations in the saturation picture
\cite{Dumitru:2008wn,Dusling:2009ni,Dumitru:2010iy}. The prefactor of
\eq{eq:corr_LO} containing impact parameter integrations brings in a
non-trivial dependence of the resulting correlations on geometry,
which will be investigated below in Sec.~\ref{sec:Geo}. The correlator
\eqref{eq:corr_LO} also contains a power-law IR divergence at ${\bm l}
=0$ with the integrand scaling as $\sim 1/l^4$ in that region. This is
a stronger divergence than $\sim 1/l^2$ divergences usually
encountered in single-gluon production cross section
calculations. Elucidating how a part of this divergence is removed by
saturation effects in the target nucleus will be one of the topics
presented below in Sec.~\ref{sec:IRsafe}.

%%%%%%%%%%%%%%%%%%%%%%%%%%%%%%%%%%%%%%%%%%%%%%%%%%%%%%%%%%%%%%%%%%%%%%%%%%%%%

%%%%%%%%%%%%%%%%%%%%%%%%%%%%%%%%%%%%%%%%%%%%%%%%%%%%%%%%%%%%%%%%%%%%%%%%%%%%%

\section{Geometry-Dependent Correlations}
\label{sec:Geo}

In \cite{Kovchegov:2012nd} we pointed out that the geometry of the
collision can have an effect on the correlation function, both through
the so-called geometric correlations introduced in
\cite{Kovchegov:2012nd} (see also
\cite{Frankfurt:2003td,Frankfurt:2010ea} for a discussion of the role
of geometry in di-jet production in $p+p$ collisions) and through a
collision geometry-dependent prefactor of the correlator, like that in
\eq{eq:corr_LO}. Note also that in the approximation considered, the
two-gluon production cross section contains only the even Fourier
harmonics in the azimuthal opening angle $\Delta \phi$: it would be
important to better understand the effect of geometry on the Fourier
expansion coefficients. We know that even Fourier harmonics in the
di-hadron correlators are also generated by the event-averaged
hydrodynamics, describing the flow of the quark-gluon plasma. (The odd
harmonics are generated by the event-by-event hydrodynamic
simulations, including geometry fluctuations \cite{Alver:2010gr}.) It
would be interesting to understand the differences and similarities of
the two types of correlations.

Let us concentrate specifically on the elliptic flow observable $v_2$,
resulting from the 2nd Fourier harmonic of the correlation
function. The value of $v_2$ in the event-averaged hydrodynamics is
driven by the ellipticity of the overlap region of the colliding
nuclei: the larger the ellipticity, the larger is $v_2$. In contrast
to this behavior, the non-flow correlations in
Eqs.~\eqref{eq:2glue_prod_main} and \eqref{crossed_xsect} do not seem
to require any ellipticity at all to produce a second harmonic (and
other even harmonics) in the correlator, resulting in the
geometry-dependent non-flow contribution to $v_2$ which is not
ellipticity-driven. This can be seen from the lowest-order correlator
in \eq{eq:corr_LO}: there the geometry-dependent factor factorizes
from the momentum-dependent term which contains the azimuthal angle
dependence of the correlations. The strength of the correlations in
\eqref{eq:corr_LO} is indeed dependent on the geometry-dependent
prefactor: however, it is not {\sl a priori} clear whether this factor
depends on the ellipticity of the overlap region.

\begin{figure}[h]
  \includegraphics[width=10cm]{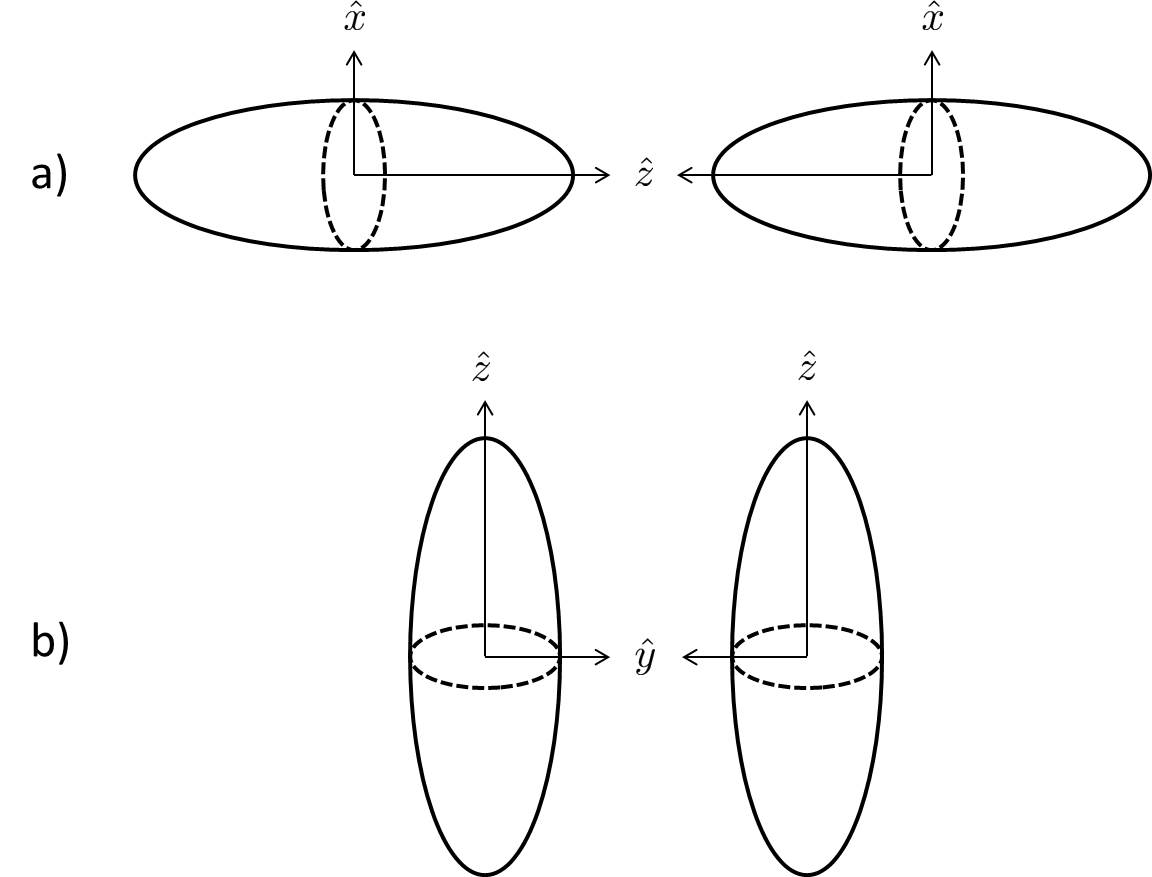}
  \caption{The layout of two possible geometries for the $U+U$
    collisions. The top panel is the tip-on-tip collision, which has
    the z-axis of the two nuclei anti-parallel to each other and
    (anti-)parallel to the collision axis. The bottom diagram is the
    side-on collision, which has the z-axis of the two nuclei parallel
    to each other and perpendicular to the collision axis. }
\label{collision_geometry} 
\end{figure}

To elucidate this issue let us consider uranium-uranium ($U+U$)
collisions. Data from such collisions have been collected at RHIC, in
order to study the properties of hydrodynamic evolution, which
predicts stronger elliptic flow (larger $v_2$) in the side-on-side
collisions (bottom panel in \fig{collision_geometry}) than in the
tip-on-tip collisions (top panel in \fig{collision_geometry}), since
the ellipticity in the former case is much larger than that in the
latter case \cite{Heinz:2004ir,Kuhlman:2005ts,Kuhlman:2006qp}.

To compare this with the behavior of the correlations in the CGC
dynamics we will employ the lowest-order correlator
\eqref{eq:corr_LO}. Note that the higher-order corrections to this
correlator, which are contained in Eqs.~\eqref{eq:2glue_prod_main} and
\eqref{crossed_xsect}, are likely to regulate some of the IR
singularities present in \eqref{eq:corr_LO}, introducing new factors
of the saturation scale $Q_{s2} ({\bm b})$, which may modify the
geometry-dependence of the lowest-order correlator
\eqref{eq:corr_LO}. However, as we will see below, the power-law IR
divergences in \eqref{eq:corr_LO} do not affect the azimuthal
angle-dependent correlations; hence our estimate of the magnitude of
the Fourier harmonics with index $n \ge 2$ should not be affected
qualitatively by higher-order corrections.

To see how the geometry of the collision affects the correlation we
take the ratio of two correlation functions which have different
geometries associated with the $U+U$ collision illustrated in
\fig{collision_geometry}: tip-on-tip (top panel) and side-on-side
(bottom panel). This requires fixing the impact parameter between the
two nuclei, ${\bm B}$, which, in this case, is fixed to ${\bm 0}$ for
both correlations. In the MV model which we have used here $Q_{s2}^2 =
4 \pi \alpha_s^2 T_2 (\bm b)$. In our case the two nuclei involved in
a collision are identical and, hence, have the same nuclear profile
functions, $T_1 (\bm b) = T_2 (\bm b)$. (Note that, while the gluon
production cross section in Eqs.~\eqref{eq:2glue_prod_main} and
\eqref{crossed_xsect} was derived in the $A_2 \gg A_1 \gg 1$ limit
with $k_1, k_2 \gtrsim Q_{s1}$, the lowest-order correlator
\eqref{eq:corr_LO} is valid for $k_1, k_2 \gg Q_{s1}, Q_{s2}$ with the
ordering condition relaxed on $A_1, A_2 \gg 1$.) The difference
between the two geometries in \fig{collision_geometry} is governed by
the nuclear profile function. The ratio between the tip-on-tip and
side-on-side correlation functions \eqref{eq:corr_LO} can be written
as
\begin{align}
\label{georatio}
\frac{C_{tip-on-tip} ({\bm k}_1, y_1, {\bm k}_2,
  y_2)\big|_{LO}}{C_{side-on-side} ({\bm k}_1, y_1, {\bm k}_2,
  y_2)\big|_{LO}} = \frac{\int d^2 b \, [T_{tip-on-tip} ({\bm
    b})]^4}{\left[ \int d^2 b \, [T_{tip-on-tip} ({\bm b})]^2
  \right]^2} \frac{\left[ \int d^2 b \, [T_{side-on-side} ({\bm b})]^2
  \right]^2}{\int d^2 b \, [T_{side-on-side} ({\bm b})]^4}. 
\end{align}
Note that the momentum dependence cancels out in the ratio of two
lowest-order correlators.

%%%%%%%%%%%%%%%%%%%%%%%%%%%%%%%%%%%%%%%%%%%%%%%%%%%%%%%%%%%%%%%%%%%%%%%%%%%%
%%%%%%%%%%%%%%%%%%%%%%%%%%%%%%%%%%%%%%%%%%%%%%%%%%%%%%%%%%%%%%%%%%%%%%%%%%%

For the analytical estimate we are about to perform here we employ a
toy model of a uranium nucleus as a prolate spheroid with the Gaussian
distribution of the nucleon number density
\begin{align}
\label{density}
  \rho (\vec{\bm r}) = \rho_0 \; e^{-\frac{x^2}{R^2}-\frac{y^2}{R^2}-\frac{\lambda^2}{R^2}z^2}
\end{align}
where $\lambda \approx 0.79$ is related to the ellipticity $\epsilon$
of the spheroid by $\lambda = \sqrt{1 - \epsilon^2}$. To translate
this into a nuclear profile function we integrate over one of the
spatial coordinates: $z$ for the tip-on-tip collisions and $y$ for the
side-on-side collisions (see \fig{collision_geometry}). Thus we have
\begin{align}
\label{nuc_thick}
& T_{tip-on-tip}({\bm b} = (x,y)) = \int\limits_{-\infty}^\infty d z
\, \rho (\vec{\bm r}) = \sqrt{\pi} \; \frac{R}{\lambda} \; \rho_0 \;
e^{-\frac{b^2}{R^2}} \notag \\ & T_{side-on-side}({\bm b} = (z,x)) =
\int\limits_{-\infty}^\infty d y \, \rho (\vec{\bm r}) = \sqrt{\pi} \;
R \; \rho_0 \; e^{-\frac{x^2}{R^2}-\frac{\lambda^2}{R^2}z^2}.
\end{align}

Plugging these results into \eq{georatio} and integrating we arrive at
\begin{align}
  \frac{C_{tip-on-tip} ({\bm k}_1, y_1, {\bm k}_2,
    y_2)\big|_{LO}}{C_{side-on-side} ({\bm k}_1, y_1, {\bm k}_2,
    y_2)\big|_{LO}} = \frac{1}{\lambda} \approx 1.26 \ \ \ (\mbox{for}
  \ U + U).
\end{align}
Thus a tip-on-tip collision enhances the initial-state (CGC)
correlation between two gluons as compared to the side-on-side
collision. We have checked this conclusion numerically by using more
realistic nuclear density profiles in \eq{georatio}, invariably
getting stronger correlations in the tip-on-tip versus side-on-side
collisions.

We conclude that, at least at the lowest order, the two-gluon
correlations behave in an exactly opposite way from hydrodynamics:
while hydrodynamic contribution to $v_2$ is ellipticity-driven, and is
hence larger in the side-on-side $U+U$ collisions, the CGC
correlations considered here give stronger correlations for the
tip-on-tip $U+U$ collisions. This difference in geometry dependence
should allow these two effects to be experimentally
distinguishable. Further work is needed to understand the geometry
dependence of the full correlator resulting from the two-gluon
production cross section in Eqs.~\eqref{eq:2glue_prod_main} and
\eqref{crossed_xsect}.

%%%%%%%%%%%%%%%%%%%%%%%%%%%%%%%%%%%%%%%%%%%%%%%%%%%%%%%%%%%%%%%%%%%%%%%%%%%%%

\section{IR Divergences}
\label{sec:IRsafe}

Saturation effects are known to regulate IR divergences in total and
production cross sections, along with related observables. For
instance, the unintegrated gluon distribution function at the lowest
order has a power-law IR divergence, $\phi (k_T) \sim 1/k_T^2$ for
$k_T \to 0$; saturation effects in the MV model reduce this IR
divergence to a logarithmic integrable singularity
\cite{Jalilian-Marian:1997xn}, $\phi (k_T) \sim \ln (Q_s^2/k_T^2)$. It
is likely that similar IR screening takes place in the two-gluon
production cross section at hand.

An analysis of the pole structure in the lowest-order correlator of
\eq{eq:corr_LO} reveals poles at
\begin{align}
  {\bm l} = \; {\bm 0}, \; {\bm k}_1, \; {\bm k}_2, \; -{\bm k}_2, \;
  {\bm k}_1 - {\bm k}_2.
\end{align}
Taking a closer look at these poles we see that the majority of them
are proportional to $\frac{1}{p^2}$ as $p \rightarrow 0$ which, after
integration over momentum $p$, gives rise to logarithmic divergences,
likely to be absorbed into gluon distributions of the nucleons
\cite{Mueller:1989st}. However, the pole at ${\bm l}={\bm 0}$ in the
first term in the curly brackets scales proportional to
$\frac{1}{l^4}$ which, after integration, gives rise to a power-law IR
divergence. Such power-law divergence is quite rare in the
quasi-classical MV limit, and it appears important to us to verify
that it is indeed regulated by the saturation effects in the full
cross section given by Eqs.~\eqref{eq:2glue_prod_main} and
\eqref{crossed_xsect}, such that the corresponding correlator, which
would include all-order saturation effects in the target nucleus,
would not depend on the IR cutoff in the power-law way. We will show
in this Section that saturation effects in the target do indeed
regulate the IR power-law divergence present in
\eq{eq:2glue_prod_main}. However, the IR power-law divergence from
\eq{crossed_xsect} is not regularized by the saturation effects in the
target, and is probably regularized by the projectile saturation
effects not included in our analysis \cite{Kovchegov:2012nd}.

\begin{figure}[h]
  \includegraphics[width=10cm]{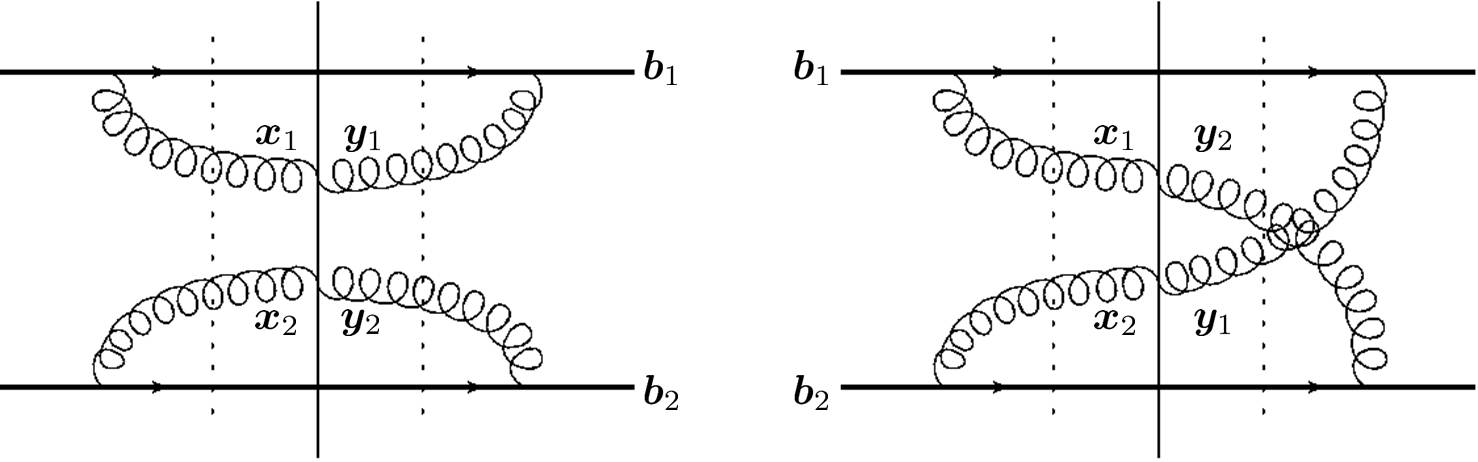}
  \caption{Examples of a 'square' diagram (left panel) and a 'crossed'
    diagram (right panel) with corresponding two-vectors ${\bm x}_1,
    {\bm x}_2, {\bm y}_1,$ and ${\bm y}_2$ labeling the transverse
    positions of the gluons and with ${\bm b}_1$, ${\bm b}_2$ denoting
    the transverse locations of the projectile valence quarks. The
    interaction with the target is shown by the vertical dotted
    lines.}
\label{squareandcrosseddiagrams} 
\end{figure}

There are two different classes of diagrams contributing to the cross
section associated with the correlation \eqref{eq:corr_LO}, the
'square' (separated) diagrams and the 'crossed' diagrams, with
examples of both shown in \fig{squareandcrosseddiagrams} in $A^+ =0$
gauge (with the projectile moving in the light-cone ``+'' direction)
and contributing the expressions \eqref{eq:2glue_prod_main} and
\eqref{crossed_xsect} correspondingly to the two-gluon production
cross section. To analyze the IR divergences it is necessary to look
at each diagram class individually. First we start with the 'square'
cross section.

%%%%%%%%%%%%%%%%%%%%%%%%%%%%%%%%%%%%%%%%%%%%%%%%%%%%%%%%%%%%%%%%%%%%%%%%%%%%
%%%%%%%%%%%%%%%%%%%%%%%%%%%%%%%%%%%%%%%%%%%%%%%%%%%%%%%%%%%%%%%%%%%%%%%%%%%

The cross section for the 'square' diagrams is given by
\eq{eq:2glue_prod_main}. The connected part of this cross section that
contributes the (non-geometric) correlations is obtained by keeping
only the $\Delta$-labeled parts of the double-trace correlators using
\eq{Ddef}. This gives \cite{Kovchegov:2012nd}
\begin{align} 
\label{eq:square_cross_fact} 
\frac{d \sigma_{square}^{(corr)}}{d^2 k_1 dy_1 d^2 k_2 dy_2} & =
\frac{\as^2 \, C_F^2}{16 \, \pi^8} \int d^2 B \, d^2 b_1 \, d^2 b_2 \,
T_1 ({\bm B} - {\bm b}_1) \, T_1 ({\bm B} - {\bm b}_2) \, d^2 x_1 \,
d^2 y_1 \, d^2 x_2 \, d^2 y_2 \, e^{- i \; {\bm k}_1 \cdot ({\bm
    x}_1-{\bm y}_1) - i \; {\bm k}_2 \cdot ({\bm x}_2-{\bm y}_2)}
\notag \\ & \times \, \frac{ {\bm x}_1 - {\bm b}_1}{ |{\bm x}_1 - {\bm
    b}_1 |^2 } \cdot \frac{ {\bm y}_1 - {\bm b}_1}{ |{\bm y}_1 - {\bm
    b}_1 |^2 } \ \frac{ {\bm x}_2 - {\bm b}_2}{ |{\bm x}_2 - {\bm b}_2
  |^2 } \cdot \frac{
  {\bm y}_2 - {\bm b}_2}{ |{\bm y}_2 - {\bm b}_2 |^2 } \notag \\
& \times \, \left[ \Delta( {\bm x}_1 , {\bm y}_1 , {\bm x}_2 , {\bm
    y}_2 ) - \Delta ( {\bm x}_1 , {\bm y}_1 , {\bm x}_2 , {\bm b}_2 )
  - \Delta ( {\bm x}_1 , {\bm y}_1 , {\bm b}_2 , {\bm y}_2 ) - \Delta
  ( {\bm x}_1 , {\bm b}_1 , {\bm x}_2 , {\bm y}_2 )
  - \Delta ( {\bm b}_1 , {\bm y}_1 , {\bm x}_2 , {\bm y}_2 ) \right. \notag \\
& \left. + \Delta ( {\bm x}_1 , {\bm b}_1 , {\bm x}_2 , {\bm b}_2 ) +
  \Delta ( {\bm x}_1 , {\bm b}_1 , {\bm b}_2 , {\bm y}_2 ) + \Delta (
  {\bm b}_1 , {\bm y}_1 , {\bm x}_2 , {\bm b}_2 ) + \Delta ( {\bm b}_1
  , {\bm y}_1 , {\bm b}_2 , {\bm y}_2 ) \right].
\end{align}
Let us introduce the variable ${\bm b}=\frac{1}{2}({\bm b}_1 + {\bm
  b}_2)$, which is the transverse position of the center of mass of
the two quarks, and ${\bm \Delta b}={\bm b}_1 - {\bm b}_2$, which is
the transverse separation between the two quarks. We also shift the
coordinates of the gluons such that
\begin{align}
  {\tilde {\bm x}}_1 = {\bm x}_1 - {\bm b}_1, \; {\tilde {\bm y}}_1 =
  {\bm y}_1 - {\bm b}_1, \; {\tilde {\bm x}}_2 = {\bm x}_2 - {\bm b}_2,
  \; {\tilde {\bm y}}_2 = {\bm y}_2 - {\bm b}_2.
\end{align}

For connected diagrams like those that gave rise to
\eq{eq:square_cross_fact} the distance ${\bm \Delta b}$ has to be
perturbatively small: we assume that $\Delta b \ll 1/\Lambda$ with
$\Lambda$ some IR cutoff of the order of the QCD confinement scale
$\Lambda_{QCD}$. Since the corresponding distance $1/\Lambda$ is of
the order of a nucleon size, it is much smaller than the radius of a
large projectile nucleus, $1/\Lambda \ll R_1 \approx
A_1^{1/3}/\Lambda$, such that the nuclear profile function does not
vary much over the distances of the order of $\Delta b$. This allows
for the approximation
\begin{align}
  \label{Tapp}
  T_1({\bm B} - {\bm b}_1)T_1({\bm B} - {\bm b}_2) \; = \; T_1 \left(
    {\bm B} - {\bm b} - \frac{{\bm \Delta b}}{2} \right) \, T_1 \left(
    {\bm B} - {\bm b} + \frac{{\bm \Delta b}}{2} \right) \; \approx \;
  \left[ T_1({\bm B} - {\bm b}) \right]^2.
\end{align}

With this approximation and employing coordinate redefinitions
outlined above, the cross-section \eqref{eq:square_cross_fact} can be
written as
\begin{align} 
\label{eq:square_cross_approx} 
\frac{d \sigma_{square}^{(corr)}}{d^2 k_1 dy_1 d^2 k_2 dy_2} =
\frac{\as^2 \, C_F^2}{16 \, \pi^8} \int d^2 B \ d^2 b \ & d^2 {\Delta
  b} \ [T_1 ({\bm B} - {\bm b})]^2 \, d^2 {\tilde x}_1 \, d^2 {\tilde
  y}_1 \, d^2 {\tilde x}_2 \, d^2 {\tilde y}_2 \, e^{- i \; {\bm k}_1
  \cdot ({\tilde {\bm x}}_1-{{\tilde {\bm y}}}_1) - i \; {\bm k}_2
  \cdot ({\tilde {\bm x}}_2- {\tilde {\bm y}_2)}} \notag \\ \times \,
\frac{ {\tilde {\bm x}}_1}{ |{\tilde {\bm x}}_1|^2 } \cdot \frac{
  {\tilde {\bm y}}_1}{ |{\tilde {\bm y}}_1|^2 } \, \frac{ {\tilde {\bm
      x}}_2}{ |{\tilde {\bm x}}_2|^2 } \cdot \frac{ {\tilde {\bm
      y}}_2}{ |{\tilde {\bm y}}_2|^2 } \, \bigg[ & \Delta \left(
  {\tilde {\bm x}}_1 + {\bm b} + \frac{1}{2} \, {\bm \Delta b} ,
  {\tilde {\bm y}}_1 + {\bm b} + \frac{1}{2} \, {\bm \Delta b} ,
  {\tilde {\bm x}}_2 + {\bm b} - \frac{1}{2} \, {\bm \Delta b} ,
  {\tilde {\bm y}}_2 + {\bm b} - \frac{1}{2} \, {\bm \Delta b} \right) \notag \\
- & \Delta \left( {\tilde {\bm x}}_1 + {\bm b} + \frac{1}{2} \, {\bm
    \Delta b} , {\tilde {\bm y}}_1 + {\bm b} + \frac{1}{2} \, {\bm
    \Delta b} , {\tilde {\bm x}}_2 + {\bm b} - \frac{1}{2} \, {\bm
    \Delta b} , {\bm b} - \frac{1}{2} \, {\bm \Delta b} \right) \notag \\
- & \Delta \left( {\tilde {\bm x}}_1 + {\bm b} + \frac{1}{2} \, {\bm
    \Delta b} , {\tilde {\bm y}}_1 + {\bm b} + \frac{1}{2} \, {\bm
    \Delta b} , {\bm b} - \frac{1}{2} \, {\bm \Delta b} ,
  {\tilde {\bm y}}_2 + {\bm b} - \frac{1}{2} \, {\bm \Delta b} \right) \notag \\
- & \Delta \left( {\tilde {\bm x}}_1 + {\bm b} + \frac{1}{2} \, {\bm
    \Delta b} , {\bm b} + \frac{1}{2} \, {\bm \Delta b} , {\tilde {\bm
      x}}_2 + {\bm b} - \frac{1}{2} \, {\bm \Delta b} ,
  {\tilde {\bm y}}_2 + {\bm b} - \frac{1}{2} \, {\bm \Delta b} \right) \notag \\
- & \Delta \left( {\bm b} + \frac{1}{2} \, {\bm \Delta b} , {\tilde
    {\bm y}}_1 + {\bm b} + \frac{1}{2} \, {\bm \Delta b} , {\tilde
    {\bm x}}_2 + {\bm b} - \frac{1}{2} \, {\bm \Delta b} ,
  {\tilde {\bm y}}_2 + {\bm b} - \frac{1}{2} \, {\bm \Delta b} \right) \notag \\
+ & \Delta \left( {\tilde {\bm x}}_1 + {\bm b} + \frac{1}{2} \, {\bm
    \Delta b} , {\bm b} + \frac{1}{2} \, {\bm \Delta b} , {\tilde {\bm
      x}}_2 + {\bm b} - \frac{1}{2} \, {\bm \Delta b} ,
  {\bm b} - \frac{1}{2} \, {\bm \Delta b} \right) \notag \\
+ & \Delta \left( {\tilde {\bm x}}_1 + {\bm b} + \frac{1}{2} \, {\bm
    \Delta b} , {\bm b} + \frac{1}{2} \, {\bm \Delta b} , {\bm b} -
  \frac{1}{2} \, {\bm \Delta b} ,
  {\tilde {\bm y}}_2 + {\bm b} - \frac{1}{2} \, {\bm \Delta b} \right) \notag \\
+ & \Delta \left( {\bm b} + \frac{1}{2} \, {\bm \Delta b} , {\tilde
    {\bm y}}_1 + {\bm b} + \frac{1}{2} \, {\bm \Delta b} , {\tilde
    {\bm x}}_2 + {\bm b} - \frac{1}{2} \, {\bm \Delta b} ,
  {\bm b} - \frac{1}{2} \, {\bm \Delta b} \right) \notag \\
+ & \Delta \left( {\bm b} + \frac{1}{2} \, {\bm \Delta b} , {\tilde
    {\bm y}}_1 + {\bm b} + \frac{1}{2} \, {\bm \Delta b} , {\bm b} -
  \frac{1}{2} \, {\bm \Delta b} , {\tilde {\bm y}}_2 + {\bm b} -
  \frac{1}{2} \, {\bm \Delta b} \right) \bigg].
\end{align}
The advantage of this form of the cross section is that all the
$\Delta b$-dependence is now in the $\Delta$-terms. 

Our next step is to identify the coordinate-space IR divergence
corresponding to the ${\bm l} =0$ singularity in \eq{eq:corr_LO}. The
cross sections in Eqs.~\eqref{eq:2glue_prod_main} and
\eqref{crossed_xsect}, and, therefore, the cross section in
\eq{eq:square_cross_approx} are all written as convolutions in the
transverse coordinate space. We need to identify which transverse
coordinate integral in \eq{eq:square_cross_approx} corresponds to the
$1/l^4$ divergence in \eq{eq:corr_LO}.

%%%%%%%%%%%%%%%%%%%%%%%%%%%%%%%%%%%%%%%%%%%%%%%%%%%%%%%%%%%%%%%%%%%%%%%%%%%%
\begin{figure}[h]
  \includegraphics[width=7cm]{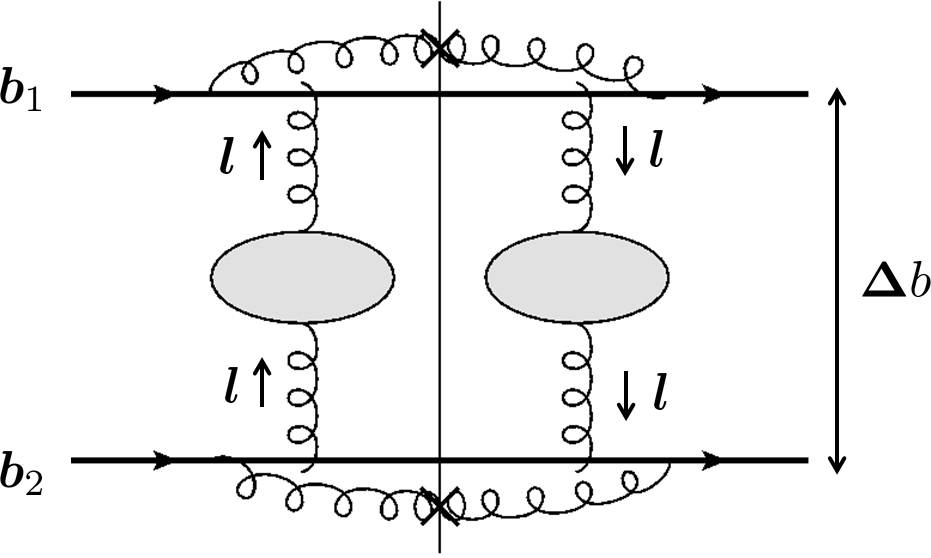}
  \caption{A pictorial representation of the quadratic IR divergence
    in \eq{eq:corr_LO}: here ${\bm l}$ is the transverse momentum
    transferred between the two quarks in the amplitude. It is
    conjugate to ${\bm \Delta b}$, the separation between the two
    quarks. Gray ovals denote two nucleons in the target
    nucleus. Disconnected $t$-channel gluon lines indicate that the
    $t$-channel gluons can couple to either the $s$-channel gluons or
    to valence quarks.}
\label{momentumflowl} 
\end{figure}
%%%%%%%%%%%%%%%%%%%%%%%%%%%%%%%%%%%%%%%%%%%%%%%%%%%%%%%%%%%%%%%%%%%%%%%%%%%

Such identification can be done by analyzing the lowest-order diagrams
giving rise to the correlation \eqref{eq:corr_LO}: the diagrams are
shown in \cite{Kovchegov:2012nd} (see Fig. 8 there). Studying those
diagrams we see that the transverse momentum ${\bm l}$ corresponds to
the momentum transferred between the two 'systems' or 'clusters'
consisting of the valence quarks along with the produced gluons, as
seen in \fig{momentumflowl}. Already in the left panel of
\fig{squareandcrosseddiagrams} we see that the two-gluon production in
this channel consists of two independent quark-gluon 'clusters', with
the correlation (cross-talk) between them generated through the
interaction with the target. As illustrated in \fig{momentumflowl},
${\bm l}$ is conjugate to the separation between the two quarks, ${\bm
  \Delta b}$: this momentum flows through a nucleon from one
quark-gluon 'system' to another in the amplitude, and in the opposite
direction in the complex conjugate amplitude. (Note that the diagram
in \fig{momentumflowl} is only one example of a set of diagrams
generating the two-gluon correlations at hand.) Thus the large
${\Delta b}$ limit corresponds to the IR divergence at ${\bm l}
\approx 0$ in \eqref{eq:corr_LO}. (The four $t$-channel gluon
propagators in \fig{momentumflowl} give us a $\sim 1/({\bm l}^2)^4$
contribution, with the gauge-invariance of the coupling of these
$t$-channel gluons to the color-neutral quark-gluon system giving a
factor of $({\bm l}^2)^2$ at small-$l$, resulting in the net $1/({\bm
  l}^2)^2$ infrared divergence, as seen in \eq{eq:corr_LO}.) To study
the effects of saturation corrections on this divergence we need to
study the large-$\Delta b$ behavior of \eq{eq:square_cross_approx}
(and, in the next step, of \eq{crossed_xsect}).

As we have already noted, the ${\bm \Delta b}$-dependence of the
integrand in \eq{eq:square_cross_approx} is now purely in the
$\Delta$-terms. To cross-check our conclusion identifying the
power-law IR divergence with the large-${\bm \Delta b}$ behavior of
that integrand, we need to make sure that to lowest order in
saturation effects this cross-section has a quadratic IR
divergence. Our goal after that would be to verify that if we include
saturation effects in the target to all orders the divergence would
become at most logarithmic.

At lowest order in multiple rescatterings in the MV model we have   
\begin{align}
  \label{eq:Delta_LO}
  \Delta_{LO} ( {\bm x}_1 , {\bm x}_2 , {\bm x}_3 , {\bm x}_4 ) =
  \frac{(D_3 - D_2)^2}{2 \, N_c^2}.
\end{align}
which is obtained by expanding \eq{Delta_exp} to the lowest order in
$D_i$'s defined in \eq{Ds} \cite{Kovchegov:2012nd}. Our next step is
to substitute this into \eq{eq:square_cross_approx} and integrate over
${\bm \Delta b}$ concentrating on the large-${\bm \Delta b}$
behavior. To do this we use the large-${\bm \Delta b}$ expansion
\begin{align}
  \label{eq:expansion}
  ({\bm x} + {\bm \Delta b})^2 \, \ln \frac{1}{|{\bm x} + {\bm \Delta
      b}| \, \Lambda} = \Delta b^2 \, \ln \frac{1}{\Delta b \,
    \Lambda} + 2 \, {\bm x} \cdot {\bm \Delta b} \, \left( \ln
    \frac{1}{\Delta b \, \Lambda} - \frac{1}{2} \right) + {\bm x}^2 \,
  \left( \ln \frac{1}{\Delta b \, \Lambda} - \frac{1}{2} \right) -
  \frac{[{\bm x} \cdot {\bm \Delta b}]^2}{\Delta b^2} + {\cal O}
  \left(\frac{1}{\Delta b}\right)
\end{align}
to write the $D_i$'s entering the expression for, say, $ \Delta( {\bm
  x}_1 , {\bm y}_1 , {\bm x}_2 , {\bm y}_2 )$ as
\begin{subequations}\label{eq:Dis}
  \begin{align}
    D_1 = & - \frac{Q_{s2}^2}{4} \, \left[ ({\tilde {\bm x}}_1 -
      {\tilde {\bm y}}_1)^2 \, \ln \frac{1}{|{\tilde {\bm x}}_1 -
        {\tilde {\bm y}}_1| \, \Lambda} + ({\tilde {\bm x}}_2 -
      {\tilde {\bm y}}_2)^2 \, \ln \frac{1}{|{\tilde {\bm x}}_2 -
        {\tilde {\bm
            y}}_2| \, \Lambda} \right] \\
    D_2 = & - \frac{Q_{s2}^2}{4} \, \bigg[ 2 \, \Delta b^2 \, \ln
    \frac{1}{\Delta b \, \Lambda} + 2 \, ({\tilde {\bm x}}_1 - {\tilde
      {\bm x}}_2 + {\tilde {\bm y}}_1 - {\tilde {\bm y}}_2) \cdot {\bm
      \Delta b} \, \left( \ln \frac{1}{\Delta b \, \Lambda} -
      \frac{1}{2} \right) \notag \\ + & \left[ ({\tilde {\bm x}}_1 -
      {\tilde {\bm x}}_2)^2 + ({\tilde {\bm y}}_1 - {\tilde {\bm
          y}}_2)^2 \right] \, \left( \ln \frac{1}{\Delta b \, \Lambda}
      - \frac{1}{2} \right) - \frac{[({\tilde {\bm x}}_1 - {\tilde
        {\bm x}}_2) \cdot {\bm \Delta b}]^2}{\Delta b^2} -
    \frac{[({\tilde {\bm y}}_1 - {\tilde {\bm y}}_2) \cdot {\bm \Delta
        b}]^2}{\Delta b^2} + {\cal O}
    \left(\frac{1}{\Delta b}\right)  \bigg] \\
    D_3 = & - \frac{Q_{s2}^2}{4} \, \bigg[ 2 \, \Delta b^2 \, \ln
    \frac{1}{\Delta b \, \Lambda} + 2 \, ({\tilde {\bm x}}_1 - {\tilde
      {\bm y}}_2 + {\tilde {\bm y}}_1 - {\tilde {\bm x}}_2) \cdot {\bm
      \Delta b} \, \left( \ln \frac{1}{\Delta b \, \Lambda} -
      \frac{1}{2} \right) \notag \\ + & \left[ ({\tilde {\bm x}}_1 -
      {\tilde {\bm y}}_2)^2 + ({\tilde {\bm y}}_1 - {\tilde {\bm
          x}}_2)^2 \right] \, \left( \ln \frac{1}{\Delta b \, \Lambda}
      - \frac{1}{2} \right) - \frac{[({\tilde {\bm x}}_1 - {\tilde
        {\bm y}}_2) \cdot {\bm \Delta b}]^2}{\Delta b^2} -
    \frac{[({\tilde {\bm y}}_1 - {\tilde {\bm x}}_2) \cdot {\bm \Delta
        b}]^2}{\Delta b^2} + {\cal O} \left(\frac{1}{\Delta b}\right)
    \bigg].
  \end{align}
\end{subequations}
Using these in \eq{eq:Delta_LO} we see that
\begin{align}
  \label{eq:delta_expansion}
  \Delta_{LO} ( {\bm x}_1 , {\bm y}_1 , {\bm x}_2 , {\bm y}_2 ) =
  \frac{Q_{s2}^4}{8 \, N_c^2} \left[ ({\tilde {\bm x}}_1 - {\tilde
      {\bm y}}_1) \cdot ({\tilde {\bm x}}_2 - {\tilde {\bm y}}_2)
    \left( \ln \frac{1}{\Delta b \, \Lambda} - \frac{1}{2} \right) -
    \frac{({\tilde {\bm x}}_1 - {\tilde {\bm y}}_1) \cdot {\bm \Delta
        b} \ ({\tilde {\bm x}}_2 - {\tilde {\bm y}}_2) \cdot {\bm
        \Delta b}}{\Delta b^2} \right]^2 \! \! + {\cal O} \!
  \left(\frac{1}{\Delta b}\right)
\end{align}
such that when we integrate $\Delta_{LO}$ over ${\bm \Delta b}$ up to
some IR cutoff $1/\Lambda_{\text{IR}}$ we arrive at an IR-divergent
expression
\begin{align}
  \label{eq:Delta_LO_int}
  \int\limits^{1/\Lambda_{\text{IR}}^2} d^2 & \Delta b \; \Delta_{LO}
  ( {\bm x}_1 , {\bm y}_1 , {\bm x}_2 , {\bm y}_2 ) \notag \\ & =
  \frac{Q_{s2}^4}{64 \, N_c^2} \, \frac{\pi}{\Lambda_{\text{IR}}^2} \,
  \left[ ({\tilde {\bm x}}_1 - {\tilde {\bm y}}_1)^2 \, ({\tilde {\bm
        x}}_2 - {\tilde {\bm y}}_2)^2 + 4 \, \left(2 \ln^2 \left(
        \frac{\Lambda_{\text{IR}}}{\Lambda} \right) - 2 \ln \left(
        \frac{\Lambda_{\text{IR}}}{\Lambda} \right) + 1 \right) \,
    \left[ ({\tilde {\bm x}}_1 - {\tilde {\bm y}}_1) \cdot ({\tilde
        {\bm x}}_2 - {\tilde {\bm y}}_2) \right]^2 \right].
\end{align}
Note that since expressions like \eq{gammadef} are valid for distances
much smaller than $1/\Lambda$ we assume that $1/\Lambda_{\text{IR}} <
1/\Lambda$. It is easy to generalize the expression \eqref{gammadef}
to the case of larger distances (as long as perturbation theory
applies), making the $1/\Lambda_{\text{IR}} < 1/\Lambda$ condition not
necessary: however, such generalization would complicate the algebra
and would not bring any new physics insight. Therefore we will proceed
here with the unmodified expression along with the
$\Lambda_{\text{IR}} > \Lambda$ assumption.

Inserting \eq{eq:Delta_LO_int} along with the similar ${\bm \Delta
  b}$-integrals for other $\Delta$'s into \eq{eq:square_cross_approx}
gives
\begin{align} 
\label{eq:square_cross_LO} 
& \frac{d \sigma_{square,LO}^{(corr)}}{d^2 k_1 dy_1 d^2 k_2 dy_2}
\approx \frac{\as^2 \, C_F^2}{16 \, \pi^8} \int d^2 B \, d^2 b \, [T_1
({\bm B} - {\bm b})]^2 \, d^2 {\tilde x}_1 \, d^2 {\tilde y}_1 \, d^2
{\tilde x}_2 \, d^2 {\tilde y}_2 \, e^{- i \; {\bm k}_1 \cdot ({\tilde
    {\bm x}}_1-{{\tilde {\bm y}}}_1) - i \; {\bm k}_2 \cdot ({\tilde
    {\bm x}}_2- {\tilde {\bm y}_2)}} \; \frac{ {\tilde {\bm x}}_1}{
  |{\tilde {\bm x}}_1|^2 } \cdot \frac{ {\tilde {\bm y}}_1}{ |{\tilde
    {\bm y}}_1|^2 } \notag \\ & \times \, \frac{ {\tilde {\bm x}}_2}{
  |{\tilde {\bm x}}_2|^2 } \cdot \frac{ {\tilde {\bm y}}_2}{ |{\tilde
    {\bm y}}_2|^2 } \, \frac{Q_{s2}^4}{16 \, N_c^2}
\frac{\pi}{\Lambda_\text{IR}^2} \left[ {\tilde {\bm x}}_1 \cdot
  {\tilde {\bm y}}_1 \; {\tilde {\bm x}}_2 \cdot {\tilde {\bm y}}_2 +
  2 \left(2 \ln^2 \left( \frac{\Lambda_{\text{IR}}}{\Lambda} \right) -
    2 \ln \left( \frac{\Lambda_{\text{IR}}}{\Lambda} \right) + 1
  \right) \, ({\tilde {\bm x}}_1 \cdot {\tilde {\bm y}}_2 \; {\tilde
    {\bm x}}_2 \cdot {\tilde {\bm y}}_1 +{\tilde {\bm x}}_1 \cdot
  {\tilde {\bm x}}_2 \; {\tilde {\bm y}}_1 \cdot {\tilde {\bm y}}_2)
\right].
\end{align}
This expression diverges as $\sim 1/\Lambda_\text{IR}^2$ in the IR, as
we expected from the lowest-order contribution.

Now let us see whether this power-law IR divergence is cured by
saturation effects. From Eqs.~\eqref{eq:Dis} and
\eqref{eq:delta_expansion} we conclude that at large $\Delta b$ the
quantities $D_2$ and $D_3$ are large and negative, while $D_1$ is
constant and $D_2 - D_3$ is approximately constant (up to a
logarithm). Even though this was shown for the $D_i$'s contributing to
$\Delta_{LO} ( {\bm x}_1 , {\bm y}_1 , {\bm x}_2 , {\bm y}_2 )$, these
conclusions are also true for all other $\Delta$'s in
\eq{eq:square_cross_approx}. Employing \eq{Delta_exp} we can
approximate any of these $\Delta$'s as
\begin{align}
  \label{eq:Dapprox}
  \Delta \approx - \frac{(D_3 - D_2)^2}{N_c^2} \, e^{D_1} \, \left(
    \frac{1}{D_2} + \frac{1}{D_3} \right) \approx \frac{(D_3 -
    D_2)^2}{N_c^2} \, e^{D_1} \, \frac{4}{Q_{s2}^2} \, \frac{1}{\Delta
    b^2 \, \ln \left( \frac{1}{\Delta b \, \Lambda} \right)}.
\end{align} 
We see right away that, neglecting logarithms, $\Delta \sim 1/\Delta
b^2$, such that the ${\bm \Delta b}$-integral of $\Delta$ is only
logarithmically divergent in the IR and the power-law divergence is
regulated!

A more detailed calculation yields the same conclusion:
\begin{align}
  \label{eq:Delta_all_int}
  \int\limits_{1/\mu^2}^{1/\Lambda_{\text{IR}}^2} d^2 \Delta b \;
  \Delta ( {\bm x}_1 , {\bm y}_1 , {\bm x}_2 , {\bm y}_2 ) \approx &
  \frac{Q_{s2}^2}{4 \, N_c^2} \, e^{D_1} \, \pi \, \bigg[ ({\tilde
    {\bm x}}_1 - {\tilde {\bm y}}_1)^2 \, ({\tilde {\bm x}}_2 -
  {\tilde {\bm y}}_2)^2 \, \ln \frac{\ln \mu/\Lambda}{\ln
    \Lambda_\text{IR}/\Lambda} \notag \\ & + \left[ ({\tilde {\bm
        x}}_1 - {\tilde {\bm y}}_1) \cdot ({\tilde {\bm x}}_2 -
    {\tilde {\bm y}}_2) \right]^2 \, \left( \ln
    \frac{\mu^2}{\Lambda_\text{IR}^2} \, \ln \frac{\mu^2 \,
      \Lambda_\text{IR}^2}{\Lambda^4} - 8 \, \ln
    \frac{\mu^2}{\Lambda_\text{IR}^2} + 8 \, \ln \frac{\ln
      \mu/\Lambda}{\ln \Lambda_\text{IR}/\Lambda} \right) \bigg],
\end{align}
where $\mu$ is an ultraviolet (UV) cutoff, $\mu \gg \Lambda_\text{IR}
> \Lambda$. Typically the role of $\mu$ will be played by the
perturbatively short inverse transverse distances $1/{\tilde x}_i$ or
$1/{\tilde y}_i$, while after the transverse coordinate integrations
are carried out $\mu$ would be a combination of $k_1$, $k_2$ and
$Q_{s2}$. The exact value of $\mu$, while important for the exact
evaluation of the integrals in \eq{eq:square_cross_approx}, is not
important for our goal of determining the degree of the IR divergence
in the expression. 

Substituting \eq{eq:Delta_all_int} along with the similar ${\bm \Delta
  b}$-integrals for other $\Delta$'s into \eq{eq:square_cross_approx}
we would obtain a cross section containing at most $\ln
\Lambda_\text{IR}$ divergences. Since the corresponding expression is
rather cumbersome, we do not show it here explicitly: instead, to
demonstrate that these $\ln \Lambda_\text{IR}$ divergences do not
cancel out, we present the $k_1, k_2 \gg Q_{s2}$ limit of the 'square'
diagrams cross section:
\begin{align}
  \label{eq:square_cross_all}
  \frac{d \sigma_{square}^{(corr)}}{d^2 k_1 dy_1 d^2 k_2 dy_2}
  \Bigg|_{k_1, k_2 \gg Q_{s2}} \! \! \approx \frac{\as^2 \, C_F^2}{16
    \, \pi^8} \int & d^2 B \, d^2 b \, [T_1 ({\bm B} - {\bm b})]^2 \,
  d^2 {\tilde x}_1 \, d^2 {\tilde y}_1 \, d^2 {\tilde x}_2 \, d^2
  {\tilde y}_2 \, e^{- i \; {\bm k}_1 \cdot ({\tilde {\bm
        x}}_1-{{\tilde {\bm y}}}_1) - i \; {\bm k}_2 \cdot ({\tilde
      {\bm x}}_2- {\tilde {\bm y}_2)}} \; \frac{ {\tilde {\bm x}}_1}{
    |{\tilde {\bm x}}_1|^2 } \cdot \frac{ {\tilde {\bm y}}_1}{
    |{\tilde {\bm y}}_1|^2 } \notag \\ \times \, \frac{ {\tilde {\bm
        x}}_2}{ |{\tilde {\bm x}}_2|^2 } \cdot \frac{ {\tilde {\bm
        y}}_2}{ |{\tilde {\bm y}}_2|^2 } \, \frac{\pi \,
    Q_{s2}^4}{N_c^2} \, \bigg[ & {\tilde {\bm x}}_1 \cdot {\tilde {\bm
      y}}_1 \; {\tilde {\bm x}}_2 \cdot {\tilde {\bm y}}_2 \, \ln
  \frac{\ln \mu/\Lambda}{\ln \Lambda_\text{IR}/\Lambda} + \frac{1}{2}
  \, ({\tilde {\bm x}}_1 \cdot {\tilde {\bm y}}_2 \; {\tilde {\bm
      x}}_2 \cdot {\tilde {\bm y}}_1 +{\tilde {\bm x}}_1 \cdot {\tilde
    {\bm
      x}}_2 \; {\tilde {\bm y}}_1 \cdot {\tilde {\bm y}}_2) \notag \\
  & \times \, \left( \ln \frac{\mu^2}{\Lambda_\text{IR}^2} \, \ln
    \frac{\mu^2 \, \Lambda_\text{IR}^2}{\Lambda^4} - 8 \, \ln
    \frac{\mu^2}{\Lambda_\text{IR}^2} + 8 \, \ln \frac{\ln
      \mu/\Lambda}{\ln \Lambda_\text{IR}/\Lambda} \right) \bigg].
\end{align}
Noticing that the IR divergence in \eq{eq:square_cross_all} is at most
logarithmic in $\Lambda_\text{IR}$ (and in $\Lambda$) we conclude that
the all-order multiple rescatterings in the target nucleus regulate
the power-law IR divergence of Eqs.~\eqref{eq:square_cross_LO} and
\eq{eq:corr_LO}. (One should not worry about the potential singularity
of \eq{eq:square_cross_all} in the $\Lambda_\text{IR} \to \Lambda$
limit: as we mentioned above, since \eq{eq:square_cross_all} was
derived in the $\Lambda_\text{IR} \gg \Lambda$ approximation, the
$\Lambda_\text{IR} \to \Lambda$ divergence is regularized if one
includes a more careful treatment for the scattering on a single
nucleon than in \eq{gammadef}.)

%%%%%%%%%%%%%%%%%%%%%%%%%%%%%%%%%%%%%%%%%%%%%%%%%%%%%%%%%%%%%%%%%%%%%%%%%%%%
\begin{figure}[ht]
  \includegraphics[width=7cm]{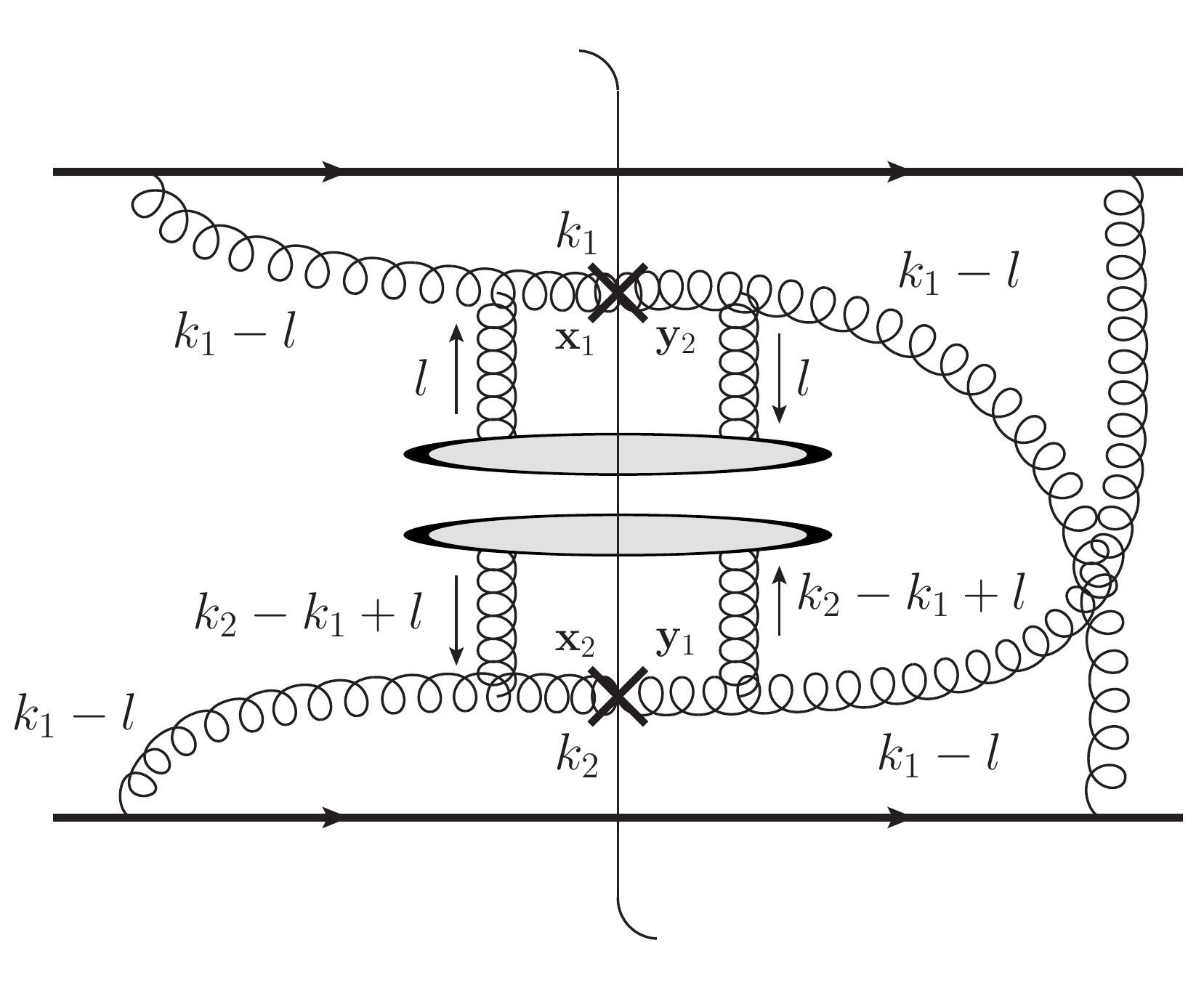}
  \caption{An examples of a 'crossed' diagram containing a power-law
    IR divergence, as described in the text.}
\label{crossedmomentum}
\end{figure}
%%%%%%%%%%%%%%%%%%%%%%%%%%%%%%%%%%%%%%%%%%%%%%%%%%%%%%%%%%%%%%%%%%%%%%%%%%%

Unfortunately similar screening of the IR power-law divergence does
not take place in the second part of the two-gluon production cross
section \eqref{crossed_xsect} corresponding to the sum of the
'crossed' diagrams. To show this we will use a diagrammatic
argument. Start by noticing that the origin of the divergence in the
'square' diagrams at the lowest order, as shown in
\fig{momentumflowl}, is in the four gluon propagators connected to two
nucleons in the target carrying the same momentum $l$. From this we
surmise that the IR divergence in the 'crossed' case originates from
four gluon propagators with the same momentum connecting to the two
nucleons in the projectile nucleus. (At the lowest order the 'square'
and 'crossed' diagrams are related to each other by interchanging the
target and the projectile.) An example of the 'crossed' diagram with
the IR divergence is shown in \fig{crossedmomentum}, where the
momentum labeling clearly demonstrates that the four gluon propagators
attached to the projectile quark lines carry the same momentum $k_1
-l$. Each of those propagators gives a factor of $({\bm k}_1 - {\bm
  l})/({\bm k}_1 - {\bm l})^2$ in the projectile ($A^+ =0$) light-cone
gauge. These factors are dotted pairwise with each other, giving
\begin{align}
  \label{eq:div_orig}
  \left[ \frac{{\bm k}_1 - {\bm l}}{({\bm k}_1 - {\bm l})^2} \cdot
    \frac{{\bm k}_1 - {\bm l}}{({\bm k}_1 - {\bm l})^2} \right]^2 =
  \frac{1}{[({\bm k}_1 - {\bm l})^2]^2},
\end{align}
containing the power-law IR divergence in question (at ${\bm l} = {\bm
  k}_1$ instead of ${\bm l}=0$ due to a different choice of momentum
labeling from that in \eq{eq:corr_LO}).

For the divergence to appear it is essential that the target nucleons
interact only with the $s$-channel gluons: one can easily see that if
either nucleon interacts with the valence quarks the power-law IR
divergence disappears, since we would not have four gluon propagators
with identical momenta in such a case. This implies that of all the
terms in the square brackets of \eq{crossed_xsect} containing
quadrupole and dipole interactions with the target, the IR divergence
may only come from the first quadrupole term, $Q ( {\bm x}_1, {\bm
  y}_1 , {\bm x}_2 , {\bm y}_2 )$. The Fourier exponentials in
\eq{crossed_xsect} make sure that $|{\bm x}_1 - {\bm y}_2| < 1/k_{1}$
and $|{\bm x}_2 - {\bm y}_1| < 1/k_{2}$: therefore, the IR divergence
may only arise from keeping the distances $|{\bm x}_1 - {\bm y}_2|$
and $|{\bm x}_2 - {\bm y}_1|$ fixed, while sending the pairs ${\bm
  x}_1, {\bm y}_2$ and ${\bm x}_2, {\bm y}_1$ far apart from each
other (cf. the analysis of the IR divergences in
\eq{eq:2glue_prod_main}). This limit corresponds to keeping $D_3$
fixed while taking $D_1, \, D_2$ to be large, such that (see
\eq{eq:quad_MV})
\begin{align}
  \label{eq:Qapprox}
  Q ( {\bm x}_1, {\bm y}_1 , {\bm x}_2 , {\bm y}_2 ) \approx e^{D_3} =
  \exp \left[ -\frac{1}{4} \, ({\bm x}_1 - {\bm y}_2)^2 \, Q_{s2}^2 \,
    \ln \frac{1}{|{\bm x}_1 - {\bm y}_2| \, \Lambda} - \frac{1}{4} \,
    ({\bm x}_2 - {\bm y}_1)^2 \, Q_{s2}^2 \, \ln \frac{1}{|{\bm x}_2 -
      {\bm y}_1| \, \Lambda} \right].
\end{align}
Defining
\begin{align}
  \label{eq:rdef}
  {\bm r}_1 = {\bm x}_1 - {\bm y}_2, \ \ \ {\bm r}_2 = {\bm x}_2 - {\bm y}_1
\end{align}
and using the approximation \eqref{Tapp} we rewrite the potentially
IR-divergent part of \eq{crossed_xsect} as
\begin{align}\label{crossed_xsect_IR}
  & \frac{d \sigma_{crossed}}{d^2 k_1 dy_1 d^2 k_2 dy_2}
  \Bigg|_{IR-div} = \int \frac{d^2 B \, d^2 b}{[2(2 \pi)^3]^2} \, d^2
  \Delta b \, \left[ T_1 ({\bm B} - {\bm b}) \right]^2 \, d^2 r_1 \,
  d^2 r_2 \, d^2 y_1 \, d^2 y_2 \left[ e^{- i \; {\bm k}_1 \cdot {\bm
        r}_1 - i \; {\bm k}_2 \cdot {\bm r}_2} + e^{- i \; {\bm k}_1
      \cdot {\bm r}_1 + i \; {\bm k}_2 \cdot {\bm r}_2} \right] \notag
  \\ & \times \, \frac{16 \; {\alpha}_s^2}{\pi^2} \, \frac{C_F}{2 N_c}
  \; \frac{ {\bm r}_1 + {\bm y}_2 - {\bm b} - \tfrac{1}{2} \, {\bm
      \Delta b}}{|{\bm r}_1 + {\bm y}_2 - {\bm b} - \tfrac{1}{2} \,
    {\bm \Delta b}|^2} \cdot \frac{{\bm y}_2 - {\bm b} + \tfrac{1}{2}
    \, {\bm \Delta b} }{ |{\bm y}_2 - {\bm b} + \tfrac{1}{2} \, {\bm
      \Delta b}|^2 } \, \frac{{\bm r}_2 + {\bm y}_1 - {\bm b} +
    \tfrac{1}{2} \, {\bm \Delta b}}{|{\bm r}_2 + {\bm y}_1 - {\bm b} +
    \tfrac{1}{2} \, {\bm \Delta b}|^2} \cdot \frac{{\bm y}_1 - {\bm b}
    - \tfrac{1}{2} \, {\bm \Delta b} }{ |{\bm y}_1 - {\bm b} -
    \tfrac{1}{2} \, {\bm \Delta b}|^2} \notag \\ & \times \,
  e^{-\frac{1}{4} \, {\bm r}_1^2 \, Q_{s2}^2 \, \ln \frac{1}{|{\bm
        r}_1| \, \Lambda} - \frac{1}{4} \, {\bm r}_2^2 \, Q_{s2}^2 \,
    \ln \frac{1}{|{\bm r}_2| \, \Lambda}}.
\end{align}
Integrating \eqref{crossed_xsect_IR} over ${\bm y}_1$ and ${\bm y}_2$
yields
\begin{align}
  \label{crossed_xsect_IR2}
  \frac{d \sigma_{crossed}}{d^2 k_1 dy_1 d^2 k_2 dy_2} \Bigg|_{IR-div}
  = & \int \frac{d^2 B \, d^2 b}{[2(2 \pi)^3]^2} \, d^2 \Delta b \,
  \left[ T_1 ({\bm B} - {\bm b}) \right]^2 \, d^2 r_1 \, d^2 r_2
  \left[ e^{- i \; {\bm k}_1 \cdot {\bm r}_1 - i \; {\bm k}_2 \cdot
      {\bm r}_2} + e^{- i \; {\bm k}_1 \cdot {\bm r}_1 + i \; {\bm
        k}_2 \cdot {\bm r}_2} \right] \notag \\ & \times \, 32 \;
  {\alpha}_s^2 \, \frac{C_F}{N_c} \; \ln (|{\bm r}_1 - {\bm \Delta b}|
  \, \Lambda) \, \ln (|{\bm r}_2 + {\bm \Delta b}| \, \Lambda) \,
  e^{-\frac{1}{4} \, {\bm r}_1^2 \, Q_{s2}^2 \, \ln \frac{1}{|{\bm
        r}_1| \, \Lambda} - \frac{1}{4} \, {\bm r}_2^2 \, Q_{s2}^2 \,
    \ln \frac{1}{|{\bm r}_2| \, \Lambda}}.
\end{align}
We see that the $d^2 \Delta b$-integral in \eq{crossed_xsect_IR2}
diverges as a power of the IR cutoff, such that
\begin{align}
  \label{crossed_xsect_IR3}
  \frac{d \sigma_{crossed}}{d^2 k_1 dy_1 d^2 k_2 dy_2} \Bigg|_{IR-div}
  \sim \frac{1}{\Lambda_\text{IR}^2}.
\end{align}
Since the expression \eqref{crossed_xsect_IR2} contains the only
potentially-divergent term in \eq{crossed_xsect}, we see that the
divergence \eqref{crossed_xsect_IR3} is not canceled by other terms in
\eq{crossed_xsect}. We conclude that the saturation effects in the
target nucleus do not regulate a part of the IR divergence present in
the lowest-order result \eqref{eq:corr_LO} that originates in
\eqref{crossed_xsect}.

%%%%%%%%%%%%%%%%%%%%%%%%%%%%%%%%%%%%%%%%%%%%%%%%%%%%%%%%%%%%%%%%%%%%%%%%%%%%
\begin{figure}[ht]
  \includegraphics[width= 10cm]{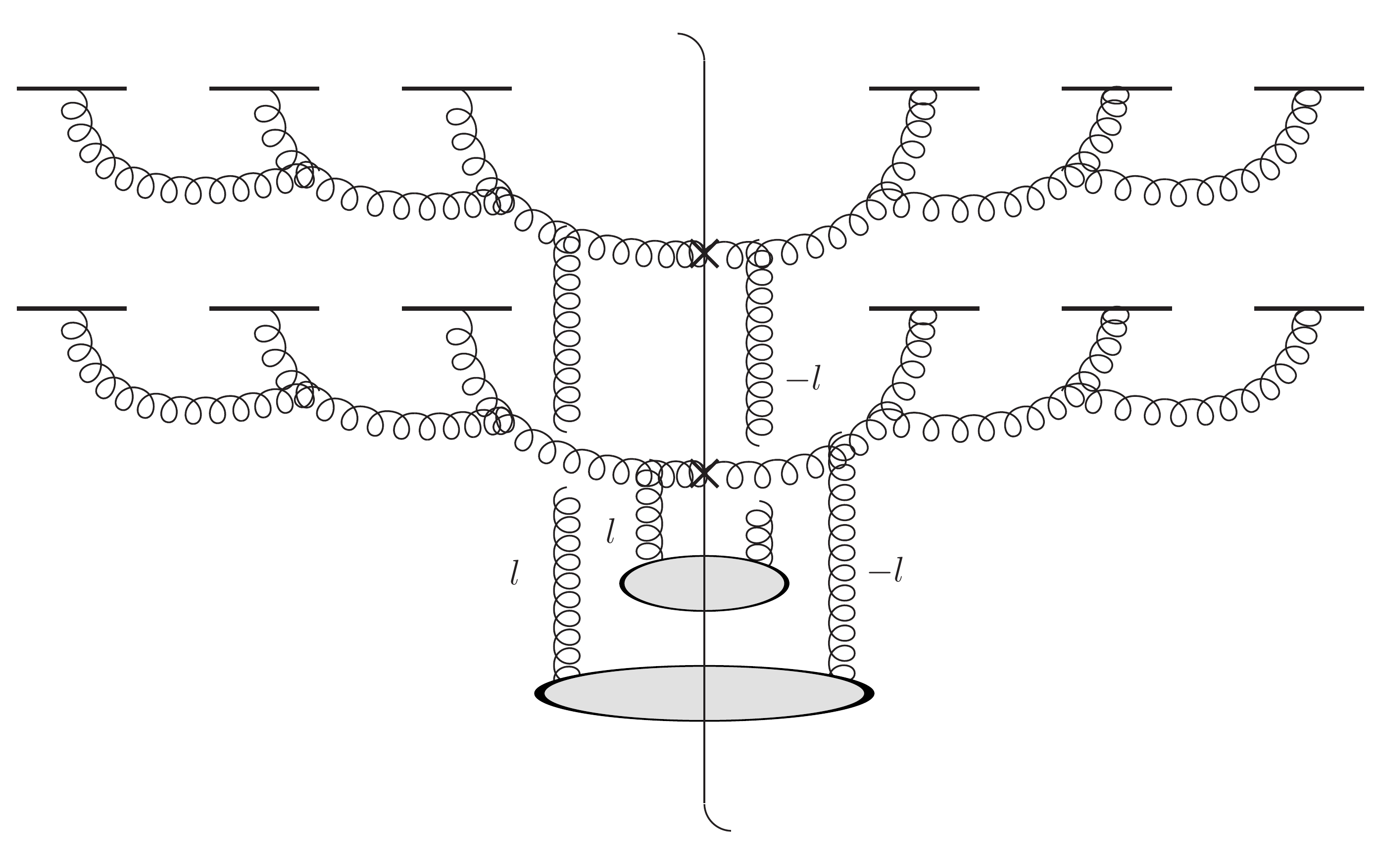}
  \caption{A diagram in the light-cone gauge of the target containing
    the power-law IR divergence of \eq{crossed_xsect_IR3}.}
\label{wwdiagram}
\end{figure}
%%%%%%%%%%%%%%%%%%%%%%%%%%%%%%%%%%%%%%%%%%%%%%%%%%%%%%%%%%%%%%%%%%%%%%%%%%%

We would like to propose a possible physical interpretation of this
divergence illustrated in \fig{wwdiagram}, which shows a diagram
contributing to the two-gluon production cross section in the $A^- =0$
light-cone gauge of the target nucleus. The target nucleus is shown at
the top of \fig{wwdiagram}, while the projectile nucleus is at the
bottom. We assume that the saturation effects come into the two-gluon
production cross section as the non-Abelian Weizs\"{a}cker-Williams
(WW) gluon fields
\cite{Kovchegov:1996ty,Jalilian-Marian:1997xn,Kovchegov:1997pc,Kovchegov:1998bi},
responsible for the gluon mergers in the top part of
\fig{wwdiagram}. If the extra nucleons (and the gluon mergers) are
removed in \fig{wwdiagram}, we would readily recover one of the
original graphs considered in \cite{Dumitru:2008wn} giving long-range
rapidity correlations.

The graph in \fig{wwdiagram} diverges for the same reason as
\fig{momentumflowl}: four $t$-channel gluon propagators give $1/({\bm
  l}^2)^4$, while the color-neutrality of the target gives a factor of
$({\bm l}^2)^2$, altogether resulting in $1/({\bm l}^2)^2$ factor in
the ${\bm l}$-integral at small-$l_T$. At any order in the target
saturation effects we still have this IR divergence, with the degree
of divergence being the same at any order in the powers of $\as^2 \,
A_2^{1/3}$ (the parameter corresponding to resumming the WW saturation
effects in the target \cite{Kovchegov:1997pc}). This is why the IR
divergence \eqref{crossed_xsect_IR3} survives the inclusion of
saturation effects in the target wave function.

The interpretation proposed in \fig{wwdiagram} allows one to hope that
the IR divergence would be removed after inclusion of multiple
rescatterings in the projectile wave function, similar to how the IR
divergence of \fig{momentumflowl} was regulated by the multiple
rescatterings in the target. Numerical simulations of the classical
two-gluon correlations in the nucleus--nucleus collisions
\cite{Lappi:2009xa} appear to support this conjecture.

To conclude this section let us note that, as follows from
\eq{crossed_xsect_IR2}, the ${\bm k}_1$- and ${\bm k}_2$-dependences
factorize in the coefficient of $1/\Lambda_\text{IR}^2$ in
\eqref{crossed_xsect_IR3}. Hence the IR-divergent term from
\eqref{crossed_xsect} does not generate two-gluon correlations with a
non-trivial azimuthal angular dependence. This conclusion is
consistent with the same azimuthal angle-independence of the power-law
IR-divergent part of the lowest-order correlator \eqref{eq:corr_LO},
generalizing the latter to all orders in the multiple rescatterings in
the target. Thus, while the power-law IR divergence of
\eq{crossed_xsect_IR3} does present a theoretical problem for the
quasi-classical two-gluon production cross section in heavy-light ion
collisions, such divergence does not affect the azimuthal
angle-dependent part of the correlation function needed for
phenomenology of the 'ridge' correlations.

%%%%%%%%%%%%%%%%%%%%%%%%%%%%%%%%%%%%%%%%%%%%%%%%%%%%%%%%%%%%%%%%%%%%%%%%%%%%%
%%%%%%%%%%%%%%%%%%%%%%%%%%%%%%%%%%%%%%%%%%%%%%%%%%%%%%%%%%%%%%%%%%%%%%%%%%%%%

\section{$k_T$-Factorization}
\label{sec:fact}

It is a well-known result of saturation physics that the single-gluon
production cross section in the proton-nucleus ($pA$) collisions
calculated either in the quasi-classical or leading-$\ln 1/x$
evolution approximations can be cast in the form consistent with
$k_T$-factorization
\cite{Braun:2000bh,Kovchegov:2001sc,Kharzeev:2003wz} (see
\cite{Jalilian-Marian:2005jf,KovchegovLevin} for pedagogical
presentations of these results). By proton-nucleus collisions we
denote dilute-dense scattering where the projectile wave function
contains no saturation effects. Such scattering is slightly different
from the two-gluon production in heavy-light ion collisions at hand:
in obtaining Eqs.~\eqref{eq:2glue_prod_main} and \eqref{crossed_xsect}
we considered two gluons originating in the projectile wave function,
which could be deemed a ``saturation effect'' compared to the single
gluon needed for quasi-classical gluon production in $pA$
collisions. It appears to be interesting to investigate whether the
two-gluon production cross section \eqref{eq_all} could also be
written in a $k_T$-factorized form.

The cross-section for the production of a single gluon in a $pA$
collision, calculated in the quasi-classical and/or leading-$\ln 1/x$
approximations, can be written as a convolution of two different
unintegrated gluon distributions \cite{Kovchegov:2001sc}\footnote{The
  normalization of this result has recently been questioned in
  \cite{Avsar:2012tz}. The worry of \cite{Avsar:2012tz} is, however,
  unjustified: the factor of $\pi^2$ difference between
  \eq{eq:singlegluon} and the corresponding result of
  \cite{Avsar:2012tz} (given by Eq.~(A21) there) is due to the
  difference in the definitions of gluon distributions. The
  unintegrated gluon distributions $\phi$ used in \eq{eq:singlegluon}
  are normalized to give the number of gluons per $d k_T^2$ element of
  transverse momentum phase space, whereas the gluon transverse
  momentum distributions (TMDs) used in \cite{Avsar:2012tz} are
  defined to give the number of gluons per $d^2 k$ phase space. The
  resulting factor of $\pi$ difference in each distribution function
  leads to an overall factor of $\pi^2$ difference in the
  normalizations of \eq{eq:singlegluon} and Eq.~(A21) from
  \cite{Avsar:2012tz}. Thus the discrepancy is entirely due to a
  different convention.}
\begin{align} 
\label{eq:singlegluon} 
\frac{d \sigma_g}{d^2 k \, d y} = \frac{2 \as}{C_F} \frac{1}{{\bm
    k}^2} \int d^2 q \; \left\langle \phi_{A_1} ({\bm q}, Y-y)
\right\rangle_{A_1} \, \left\langle \phi_{A_2} ({\bm k}-{\bm q}, y)
\right\rangle_{A_2}
\end{align}
where we replaced the proton by the light ion $A_1$, implying that no
saturation effects are included in the light ion wave function, which
makes it equivalent to a proton for the purpose of the single-gluon
production calculation. The angle brackets $\langle \ldots
\rangle_{A_1}$ and $\langle \ldots \rangle_{A_2}$ denote averaging in
the projectile and target wave functions respectively.

The unintegrated gluon distribution for the light ion is
\begin{equation} 
\label{eq:dipole_wave_int} 
\left\langle \phi_{A_1} ({\bm q}, y) \right\rangle_{A_1} =
\frac{C_F}{\as ( 2 \pi)^3} \int d^2 b \; d^2 r \; 
e^{-i {\bm q} \cdot {\bm r}} \; \nabla_{{\bm r}}^2 \; n_G ({\bm b} + {\bm r}, {\bm b}, y),
\end{equation}
where $n_G ({\bm b} + {\bm r}, {\bm b}, y)$ is the gluon dipole
scattering amplitude on the projectile evaluated without saturation
effects (no multiple rescatterings, only linear BFKL evolution). The
two gluons in the dipole are located at transverse positions ${\bm b}
+ {\bm r}$ and ${\bm b}$, and the rapidity interval for the scattering
is $y$. In the quasi-classical limit one has
\begin{align} 
\label{eq:dipole_amp} 
n_G ({\bm b} + {\bm r}, {\bm b}, y=0) = \pi \, \as^2 \, r_\perp^2 \ln
\left( \frac{1}{|{\bm r}| \Lambda} \right) T_1 ({\bm b}).
\end{align}

The unintegrated gluon distribution for the heavy ion is defined as
\begin{align} 
\label{eq:trace_wave} 
\left\langle \phi_{A_2} ({\bm q}, y) \right\rangle_{A_2} =
\frac{C_F}{\as ( 2 \pi)^3} \int d^2 b \; d^2 r \; e^{-i {\bm q} \cdot
  {\bm r}} \; \nabla_{{\bm r}}^2 \; N_{G} ({\bm b} + {\bm r}, {\bm b},
y)
\end{align}
where we use the following convention for the imaginary part of the
forward scattering amplitude for the gluon dipole on the target
nucleus,
\begin{align} 
\label{eq:trace_amp} 
N_{G} ({\bm x}, {\bm y}, Y) = \; \frac{1}{N_c^2-1} \; \left\langle
  \mbox{Tr} \left[ \mathbb{1} - U_{{\bm x}} U_{{\bm y}}^\dagger
  \right] \right\rangle_{A_2} (Y).
\end{align}
The correlator in \eq{eq:trace_amp} is evaluated either in the MV
model (classical limit) or with the full nonlinear BK/JIMWLK
evolution.

In Eqs.~\eqref{eq:dipole_amp} and \eqref{eq:trace_amp} the vector
${\bm r}$ is the transverse size of the gluon dipole and ${\bm b}$ can
be thought of as the impact parameter of the dipole. Normally the
impact parameter is defined as the transverse position of the center
of mass of the dipole: we introduced a slightly different notation
here for the future convenience.

The distribution functions \eqref{eq:dipole_wave_int} and
\eqref{eq:trace_wave} defined above are needed for the $k_T$
factorization expression \eqref{eq:singlegluon} of the single gluon
production cross-section in $pA$ (or heavy-light ion)
collisions. However, when we are dealing with the two-gluon production
cross-section \eqref{eq_all}, these distribution function are likely
not to be adequate. First we notice that the only Wilson line operator
in the single-gluon production case is the gluon dipole
\eqref{eq:trace_amp}. In the expression for the two-gluon production
cross section \eqref{eq_all} we have both the quadruple operator
\eqref{quad_def}, and the double trace operator \eqref{Ddef}, which
would lead to different distribution functions.

Secondly, the two gluon production cross-section has geometric
correlations \cite{Kovchegov:2012nd}, which arise purely from the
integration over the impact parameters $B$, $b_1$ and $b_2$ in
Eqs.~\eqref{eq:2glue_prod_main} and \eqref{crossed_xsect}. This
prevents the integrals over the impact parameters from being contained
within the distribution functions themselves. This will end up
drastically changing the nature of the distribution functions and thus
the final factorized from.

The last major difference comes from the 'crossed' diagrams. These
diagrams contain the interference of the wave functions of the
incoming nucleons, which generates a significant ``cross-talk''
between different parts of the diagram; it is, therefore, {\it a
  priori} unlikely that factorization would take place. As we will see
below, the factorized form of the expression cannot be written purely
as a convolution of distribution functions without additional factors,
like in \eq{eq:singlegluon}. While factorized form can be achieved, it
would also contain an extra factor (a ``coefficient function'') in the
final result for the convolution.

With these considerations in mind we first should take a look at the
nature of the distribution functions needed for the $k_T$-factorized
expression for the two-gluon production.

%%%%%%%%%%%%%%%%%%%%%%%%%%%%%%%%%%%%%%%%%%%%%%%%%%%%%%%%%%%%%%%%%%%%%%%%%%%%%

\subsection{One- and Two-Gluon Distribution Functions}
\label{sec:fact-distribution}

As we mentioned above, the impact parameter convolutions in
Eqs.~\eqref{eq:2glue_prod_main} and \eqref{crossed_xsect} do not
appear to be factorizable into the integral over the distances between
the gluons and the projectile and a separate integral over the
distances between the gluons and the target, in stark contrast to the
single-gluon production case
\cite{Braun:2000bh,Kovchegov:2001sc}. Therefore, any factorization
expression we could obtain for the two-gluon production has to have an
explicit convolution over the impact parameters. Therefore, we first
need to rewrite the single-gluon distribution functions introduced
above for the fixed impact parameter. We can easily recast
Eqs.~\eqref{eq:dipole_wave_int} and \eqref{eq:trace_wave} as
\begin{align} 
\label{eq:dipole_wave} 
\left\langle \frac{d \phi_{A_1} ({\bm q}, y)}{d^2 b} \right\rangle_{A_1} =
\frac{C_F}{\as ( 2 \pi)^3} \int d^2 r \; 
e^{-i {\bm q} \cdot {\bm r}} \; \nabla_{{\bm r}}^2 \; n_G ({\bm b} + {\bm r}, {\bm b}, y)
\end{align}
and
\begin{align} 
\label{eq:dipole_dist} 
\left\langle \frac{d \phi_{A_2} ({\bm q}, y)}{d^2 b}
\right\rangle_{A_2} = \frac{C_F}{\as ( 2 \pi)^3} \int d^2 r \; e^{-i
  {\bm q} \cdot {\bm r}} \; \nabla_{{\bm r}}^2 \; N_G ({\bm b} + {\bm
  r}, {\bm b}, y).
\end{align}
Since now these distribution functions fix both the momentum of the
gluon $\bm q$ and its (approximate) position in the transverse
coordinate space $\bm b$, along with its rapidity $y$ specifying the
value of Bjorken-$x$ variable, we identify the differential
unintegrated gluon distribution functions in
Eqs.~\eqref{eq:dipole_wave} and \eqref{eq:dipole_dist} with the Wigner
distribution \cite{Wigner:1932eb} for gluons (see
\cite{Belitsky:2002sm,Accardi:2012qut} and references therein for
applications of Wigner distributions in perturbative QCD).

\begin{figure}[t]
  \includegraphics[width=10cm]{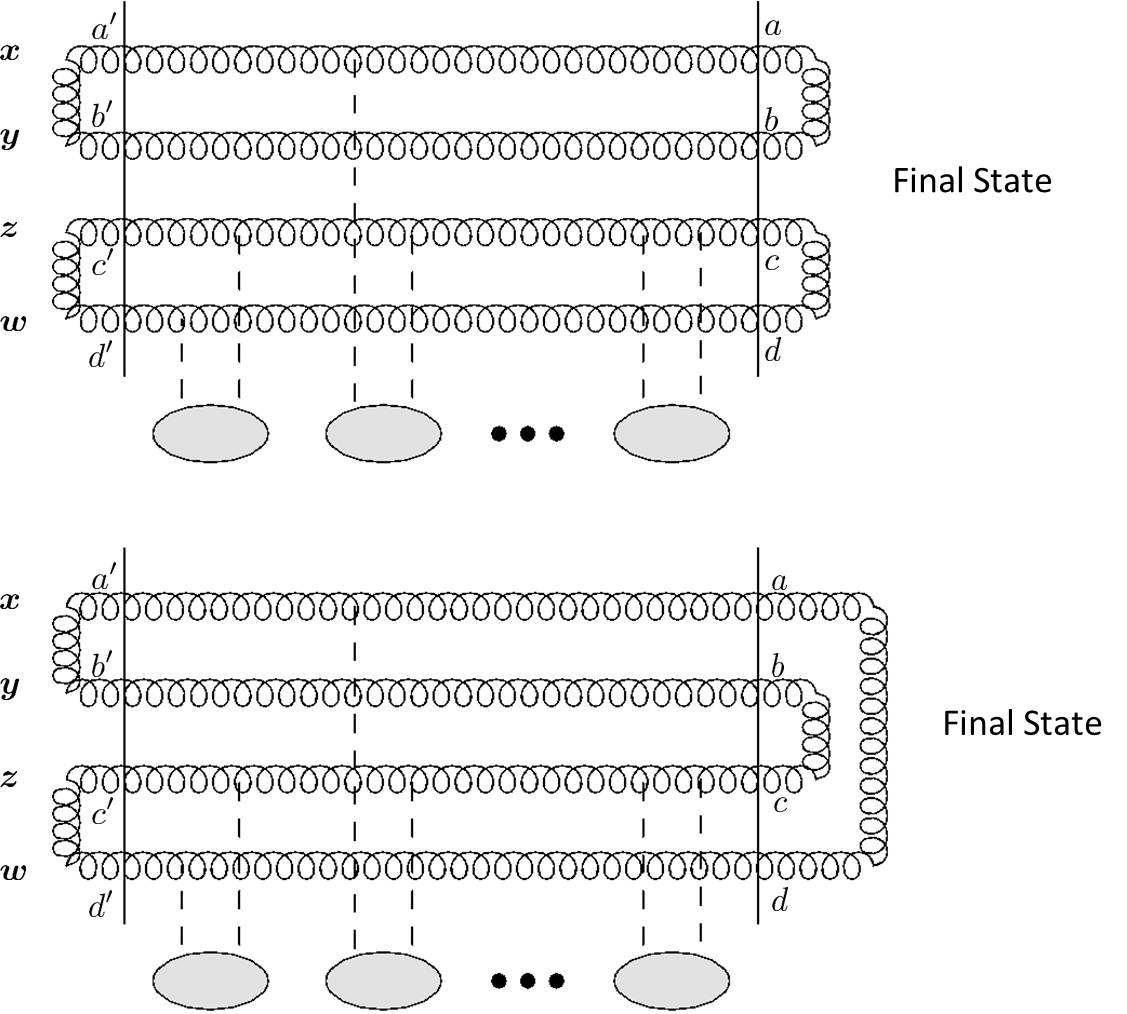}
  \caption{The top panel represents the forward amplitude for the
    scattering of two gluon dipoles on a target nucleus in the
    quasi-classical approximation: this is an essential contribution
    to the definition of the two-gluon distribution in
    \eq{eq:doubletrace_dist}. The bottom panel represents the
    quadrupole scattering on the target, as in the definition of the
    two-gluon distribution in \eq{eq:quad_dist}. The vectors ${\bm
      x}$, ${\bm y}$, ${\bm z}$, and ${\bm w}$ label the positions of
    the gluon Wilson lines. Vertical solid lines denote the initial
    (left) and final (right) states. The final state of the gluons is
    labeled to stress that the difference between the two panels is in
    the color configurations of the final state.}
\label{distributions} 
\end{figure}

Here we introduce two different distribution functions which are
associated with the two Wilson line operators entering the two-gluon
production cross-section \eqref{eq_all}, the gluon quadrupole and the
double-trace operators. The two-gluon distribution function associated
with the gluon double-trace operator is
\begin{align} 
\label{eq:doubletrace_dist} 
\left\langle \frac{d \phi_{A_2}^{D} ({\bm q}_1, {\bm q}_2, y)}{d^2 b_1
    \; d^2 b_2} \right\rangle_{A_2} = \left( \frac{C_F}{\as ( 2
    \pi)^3} \right)^2 \int d^2 r_1 \; d^2 r_2 \, e^{-i {\bm q}_1 \cdot
  {\bm r}_1 -i {\bm q}_2 \cdot {\bm r}_2} \; \nabla_{{\bm r}_1}^2 \;
\nabla_{{\bm r}_2}^2 \; N_D ({\bm b}_1 + {\bm r}_1, {\bm b}_1, {\bm
  b}_2 + {\bm r}_2, {\bm b}_2, y),
\end{align} 
where
\begin{align} 
\label{eq:doubletrace_amp} 
N_D ({\bm x}, {\bm y}, {\bm z}, {\bm w}, Y) = \; \frac{1}{(N_c^2-1)^2}
\; \left\langle \mbox{Tr} \left[ \mathbb{1} - U_{{\bm x}} U_{{\bm
        y}}^\dagger \right] \mbox{Tr} \left[ \mathbb{1} - U_{{\bm z}}
    U_{{\bm w}}^\dagger \right] \right\rangle_{A_2} (Y).
\end{align} 
The correlator $N_D$ is illustrated diagrammatically in the top panel
of \fig{distributions} for the quasi-classical approximation. The
distribution function \eqref{eq:doubletrace_dist} gives us the number
density for pairs of gluons, with the transverse momenta ${\bm q}_1,
{\bm q}_2$ and positions ${\bm b}_1, {\bm b}_2$ of the gluons fixed
and with the rapidity of both gluons being close to $y$ (up to $\ll
1/\as$ variations): we can think of this distribution function as a
two-gluon Wigner distribution.

The distribution function associated with the gluon quadrupole
operator is 
\begin{align} 
\label{eq:quad_dist} 
\left\langle \frac{d \phi_{A_2}^{Q} ({\bm q}_1, {\bm q}_2, y)}{d^2 b_1
    \; d^2 b_2} \right\rangle_{A_2} = \left( \frac{C_F}{\as ( 2
    \pi)^3} \right)^2 \int d^2 r_1 \; d^2 r_2 \, e^{-i {\bm q}_1 \cdot
  {\bm r}_1 -i {\bm q}_2 \cdot {\bm r}_2} \; \nabla_{{\bm r}_1}^2 \;
\nabla_{{\bm r}_2}^2 \; N_{Q} ({\bm b}_1 + {\bm r}_1, {\bm b}_1, {\bm
  b}_2 + {\bm r}_2, {\bm b}_2, y)
\end{align}
with
\begin{align} 
\label{eq:quad_amp} 
N_{Q} ({\bm x}, {\bm y}, {\bm z}, {\bm w}, Y) = \; \frac{1}{N_c^2-1}
\; \left\langle \mbox{Tr} \left[ \left( \mathbb{1} - U_{{\bm x}}
      U_{{\bm y}}^\dagger \right) \left( \mathbb{1} - U_{{\bm z}}
      U_{{\bm w}}^\dagger \right) \right] \right\rangle_{A_2} (Y).
\end{align} 
The definition \eqref{eq:quad_dist} is illustrated diagrammatically in
the lower panel of \fig{distributions} in the quasi-classical
approximation. The object defined in \eqref{eq:quad_dist} can, similar
to \eqref{eq:doubletrace_dist}, be thought of as a (different)
two-gluon Wigner distribution.

Notice how both the two-dipole \eqref{eq:doubletrace_dist} and
quadrupole \eqref{eq:quad_dist} two-gluon distribution are composed of
Wilson line operators. This is natural for distribution functions
entering production cross section, since in high energy scattering all
cross sections are expressed in terms of Wilson lines. This is in
exact parallel to the single-gluon distribution
\eqref{eq:dipole_dist}, which is related to the adjoint dipole
operator. Note that since the single-gluon production cross section
depends only on the adjoint dipole operator, one can express it only
in terms of the single-gluon distribution \eqref{eq:trace_wave}. For
the two-gluon production \eqref{eq_all}, which contains both the
double-trace and quadrupole operators, we end up with two different
two-gluon distributions \eqref{eq:doubletrace_dist} and
\eqref{eq:quad_dist}.

%%%%%%%%%%%%%%%%%%%%%%%%%%%%%%%%%%%%%%%%%%%%%%%%%%%%%%%%%%%%%%%%%%%%%%%%%%%%
%%%%%%%%%%%%%%%%%%%%%%%%%%%%%%%%%%%%%%%%%%%%%%%%%%%%%%%%%%%%%%%%%%%%%%%%%%%

There is also an alternative single-gluon distribution, the so-called
Weizs\"{a}cker-Williams (WW) distribution
\cite{Jalilian-Marian:1997xn,Kovchegov:1998bi,Kovchegov:2001sc,Kharzeev:2003wz},
which was found to be related to the $q\bar q$ back-to-back jet
production in DIS \cite{Dominguez:2011wm}. In the quasi-classical MV
picture the Weizs\"{a}cker-Williams two-gluon distribution, given by
the correlator of four different gluon fields, would simply factorize
into a product of two single-gluon WW distributions. It is possible,
however, that beyond the quasi-classical limit the two-gluon WW
distribution (properly defined in terms of Wilson line operators along
the lines of the single-gluon WW distribution from
\cite{Dominguez:2011wm}) would constitute an independent new object,
related to some observables. Investigating this possibility further
is beyond the scope of this work.

Since there exists more experience in the field with dipole
distribution functions \eqref{eq:dipole_dist}, it would be nice to be
able to write the two-gluon distributions \eqref{eq:doubletrace_dist}
and \eqref{eq:quad_dist} as combinations of dipole
distributions. Unfortunately this is not possible in general; however
each distribution does contain a piece that can be written in terms of
dipole distributions.

The most obvious is the double-trace two-gluon distribution function
\eqref{eq:doubletrace_dist}. Since in the large-$N_c$ limit
\begin{align} 
\label{eq:singlet_amp_dipole}
  N_{D} ({\bm x}, {\bm y}, {\bm z}, {\bm w}) \bigg|_{\mbox{large}-N_c} = \;
  N_{G} ({\bm x}, {\bm y}) \; N_{G} ({\bm z}, {\bm w})
\end{align}
with $N_G = 1 - S_G$, we can see by plugging this result into
\eq{eq:doubletrace_dist} and comparing to \eq{eq:dipole_dist} that
\begin{align} 
\label{eq:singlet_app} 
\left\langle \frac{d \phi_{A_2}^{D} ({\bm q}_1, {\bm q}_2, y)}{d^2 b_1
    \; d^2 b_2} \right\rangle_{\! \! A_2} \Bigg|_{\mbox{large}-N_c} =
\left\langle \frac{d \phi_{A_2} ({\bm q}_1, y)}{d^2 b_1}
\right\rangle_{\! \! A_2} \left\langle \frac{d \phi_{A_2} ({\bm q}_2,
    y)}{d^2 b_2} \right\rangle_{\! \! A_2}.
\end{align}
The double-trace two-gluon distribution function factorizes into two
dipole distribution functions only in the large-$N_c$
limit. Unfortunately, the only correlations left in the two-gluon
production cross section \eqref{eq_all} evaluated in the large-$N_c$
limit are the geometric correlations \cite{Kovchegov:2012nd}. All of
the other correlations contained in \eqref{eq_all} are subleading in
$N_c$; for instance, the correlations \eqref{eq:corr_LO} are
explicitly $\ord{1/N_c^2}$. 

In order to isolate the dipole contribution to the two-gluon
quadrupole distribution \eqref{eq:quad_dist} we cannot just take the
large-$N_c$ limit like we did for the singlet distribution. (In
addition the whole corresponding contribution to the cross section
\eqref{crossed_xsect} is $\ord{1/N_c^2}$ when compared to
\eq{eq:2glue_prod_main}.) Instead we can single out the part of the
two-gluon quadrupole distribution which is expressible in terms of
single-gluon dipole distributions: we will show later that this is
exactly the part that gives rise to the early-time
Hanbury-Brown--Twiss (HBT) correlations \cite{HanburyBrown:1956pf}
discussed in \cite{Kovchegov:2012nd}.

First let us analyze the quadrupole operator (cf. \eq{quad_def})
\begin{align} 
\label{eq:quad_color1} 
Q({\bm x}, {\bm y}, {\bm z}, {\bm w}, Y) = \frac{1}{N_c^2-1} \;
\left\langle \mbox{Tr} \left[ U_{{\bm x}} U_{{\bm y}}^\dagger U_{{\bm
        z}} U_{{\bm w}}^\dagger \right] \right\rangle_{A_2} (Y) =
\frac{1}{N_c^2-1} \; \left\langle \delta^{ad} \delta^{bc} \; U_{{\bm
      x}}^{a a'} U_{{\bm y}}^{b b'} U_{{\bm z}}^{c c'} U_{{\bm w}}^{d
    d'} \; \delta^{a'b'} \delta^{c'd'} \right\rangle_{\! A_2} (Y). 
\end{align}
Here we have written out the color structure implied by the trace
notation in terms of the adjoint color indices $a$, $b$, $c$, $d$ (and
the corresponding primed variables) shown in the lower panel of
\fig{distributions}. The four gluon lines in the final state in
\fig{distributions} carrying indices $a$, $b$, $c$, $d$ are in a net
color-neutral state. This allows us to classify the color states in
the quadrupole operator by the color states of the two gluons with
indices $a$ and $b$. Choosing the color state of gluons $a$ and $b$
sets the color state of gluons $c$ and $d$ due to the color neutrality
of all four gluons in the final state. The same applies to the initial
state gluons with the color indices $a'$, $b'$, $c'$, and $d'$.

A pair of gluons may be found in either of the following irreducible
representations of SU($N_c$)
\begin{align}
\label{8x8}
  & {(N_c^2 - 1)} \otimes {(N_c^2 - 1)} = V_1 \oplus V_2 \oplus V_3
  \oplus V_4 \oplus V_5 \oplus V_6 \oplus V_7 \notag \\ & = {\bm 1}
  \oplus {(N_c^2 -1)} \oplus \frac{N_c^2 (N_c -3) (N_c +1)}{4} \oplus
  \frac{N_c^2 (N_c + 3) (N_c - 1)}{4} \oplus {(N_c^2 -1)} \oplus
  \frac{(N_c^2 -1) (N_c^2 - 4)}{4} \oplus \frac{(N_c^2 -1) (N_c^2 -
    4)}{4}.
\end{align}
In \eq{8x8} we follow the notation for the irreducible representations
introduced in \cite{Cvitanovic:2008zz}, see page 120 there. We will,
however, use a different normalization scheme from the projection
operators $P_i^{abcd}$'s. We normalize the states such that
$P_i^{abcd} \; P_i^{abcd} = 1$ (summation over repeated indices is
implied), which implies, due to the orthonormality of the projection
operators,
\begin{align} 
\label{eq:unit_op}
\mathbb{1}^{abcd, \; a'b'c'd'} = \sum_{i=1}^7 P_i^{abcd} \;
P_i^{a'b'c'd'}.
\end{align}

The only projection operator we need to know explicitly for the
following calculation is the singlet projector,
\begin{align} 
\label{eq:blank1} 
  P_1^{abcd} = \frac{1}{N_c^2-1} \delta^{ab} \delta^{cd}.
\end{align}
Using the singlet projection and the unit operator \eqref{eq:unit_op}
we can rewrite \eq{eq:quad_color1} as (dropping the $A_2$ subscript
and not showing rapidity dependence for brevity)
\begin{align} 
\label{eq:quad_color2} 
\frac{1}{N_c^2-1} \; \left\langle \mbox{Tr} \left[ U_{{\bm x}} U_{{\bm
        y}}^\dagger U_{{\bm z}} U_{{\bm w}}^\dagger \right]
\right\rangle = \sum_{i=1}^7 P_i^{a''b''b''a''} \; \left\langle
  P_i^{abcd} \; U_{{\bm x}}^{a a'} U_{{\bm y}}^{b b'} U_{{\bm z}}^{c
    c'} U_{{\bm w}}^{d d'} \; P_1^{a'b'c'd'} \right\rangle .
\end{align}
We can isolate the part that gives the factorized dipole contribution
in the sum of \eq{eq:quad_color2}. This contribution comes from the
large-$N_c$ part of the double-dipole operator, which, in turn,
originates in the $P_1$-term in the sum in
\eqref{eq:quad_color2}. Isolating the double trace operator from the
rest of the expression in \eq{eq:quad_color2} we arrive at
\begin{align} 
\label{eq:quad_color3} 
\frac{1}{N_c^2-1} \; \left\langle \mbox{Tr} \left[ U_{{\bm x}} U_{{\bm
        y}}^\dagger U_{{\bm z}} U_{{\bm w}}^\dagger \right]
\right\rangle & = \frac{1}{(N_c^2-1)^2} \; \left\langle \mbox{Tr}
  \left[ U_{{\bm x}} U_{{\bm y}}^\dagger \right] \mbox{Tr} \left[
    U_{{\bm z}} U_{{\bm w}}^\dagger \right] \right\rangle \notag \\ &
+ \, \sum_{i=2}^7 P_i^{a''b''b''a''} \; \left\langle P_i^{abcd} \;
  U_{{\bm x}}^{a a'} U_{{\bm y}}^{b b'} U_{{\bm z}}^{c c'} U_{{\bm
      w}}^{d d'} \; P_1^{a'b'c'd'} \right\rangle .
\end{align}
The double trace operator comes with a prefactor of $\frac{1}{(N_c^2 -
  1)^2}$, which means that when we combine \eq{eq:quad_color3} with
\eq{eq:quad_amp} we arrive at
\begin{align} 
\label{eq:quad_amp_dipole}
  N_{Q} ({\bm x}, {\bm y}, {\bm z}, {\bm w}) = \;
  N_{G} ({\bm x}, {\bm y}) \; N_{G} ({\bm z}, {\bm w})
  + \cdots .
\end{align}
The ellipses in \eqref{eq:quad_amp_dipole} represent the remaining
contributions which are not contained in the factorized gluon dipoles,
the $P_2$ through $P_7$ terms and the sub-leading in $N_c$ terms from
the double trace operator in \eq{eq:quad_color3}. Plugging
\eq{eq:quad_amp_dipole} into \eq{eq:quad_dist} we arrive at
\begin{align} 
\label{eq:quad_approx} 
\left\langle \frac{d \phi_{A_2}^{Q} ({\bm q}_1, {\bm q}_2, y)}{d^2 b_1
    \; d^2 b_2} \right\rangle_{\! \! A_2} = \left\langle \frac{d
    \phi_{A_2} ({\bm q}_1, y)}{d^2 b_1} \right\rangle_{\! \! A_2}
\left\langle \frac{d \phi_{A_2} ({\bm q}_2, y)}{d^2 b_2}
\right\rangle_{\! \! A_2} + \cdots,
\end{align}
where we have isolated the factorized dipole distributions from the
rest of the expression. Let us stress that the terms represented by
ellipsis in \eq{eq:quad_approx} are not suppressed by any parameter
involved in the problem: these corrections are comparable to the term
shown explicitly on the right of \eqref{eq:quad_approx}. Hence even if
we took the leading-$N_c$ limit of the two-gluon quadrupole
distribution there would still be terms that would not be contained
inside the two factorized gluon distributions of
\eqref{eq:quad_approx}.

%%%%%%%%%%%%%%%%%%%%%%%%%%%%%%%%%%%%%%%%%%%%%%%%%%%%%%%%%%%%%%%%%%%%%%%%%%%%%

\subsection{Derivation of the Factorized Forms}
\label{sec:fact-main}

Now that we have defined the necessary distribution functions we can
start constructing the factorized form of the two-gluon production
cross-section. Each of the parts of the cross section \eqref{eq_all}
given by Eqs.~\eqref{eq:2glue_prod_main} and \eqref{crossed_xsect}
factorizes differently.

The easiest case to factorize, and thus the first one we will cover,
is the 'square' diagram component
\eqref{eq:2glue_prod_main}. Separating the transverse vectors
associated with either one of the valence quarks and emitted gluons,
we can write \eq{eq:2glue_prod_main} in the following form,
\begin{align} 
\label{eq:square_cross_form2} 
\frac{d \sigma_{square}}{d^2 k_1 dy_1 d^2 k_2 dy_2} & = \frac{\as^2 \,
  C_F^2}{16 \, \pi^8} \int d^2 B \, \left\langle \int \, d^2 x_1 \,
  d^2 y_1 \, d^2 b_1 \, T_1 ({\bm B} - {\bm b}_1) e^{- i \; {\bm k}_1
    \cdot ({\bm x}_1-{\bm y}_1)} \frac{ {\bm x}_1 - {\bm b}_1}{ |{\bm
      x}_1 - {\bm b}_1 |^2 } \cdot \frac{ {\bm y}_1 - {\bm b}_1}{
    |{\bm y}_1 - {\bm b}_1 |^2 } \right.  \notag \\ & \times \, \left(
  \frac{1}{N_c^2-1} \; \mbox{Tr}[ U_{{\bm x}_1} U_{{\bm y}_1}^\dagger
  ] \; - \; \frac{1}{N_c^2-1} \; \mbox{Tr}[ U_{{\bm x}_1} U_{{\bm
      b}_1}^\dagger ] \; - \; \frac{1}{N_c^2-1} \; \mbox{Tr}[ U_{{\bm
      b}_1} U_{{\bm y}_1}^\dagger ] \; + \; 1 \right) \notag \\ &
\times \, \int \, d^2 x_2 \, d^2 y_2 \, d^2 b_2 \, T_1 ({\bm B} - {\bm
  b}_2) e^{- i \; {\bm k}_2 \cdot ({\bm x}_2-{\bm y}_2)} \frac{ {\bm
    x}_2 - {\bm b}_2}{ |{\bm x}_2 - {\bm b}_2 |^2 } \cdot \frac{ {\bm
    y}_2 - {\bm b}_2}{ |{\bm y}_2 - {\bm b}_2 |^2 } \notag \\ & \times
\, \left. \left( \frac{1}{N_c^2-1} \; \mbox{Tr}[ U_{{\bm x}_2} U_{{\bm
        y}_2}^\dagger ] \; - \; \frac{1}{N_c^2-1} \; \mbox{Tr}[
    U_{{\bm x}_2} U_{{\bm b}_2}^\dagger ] \; - \; \frac{1}{N_c^2-1} \;
    \mbox{Tr}[ U_{{\bm b}_2} U_{{\bm y}_2}^\dagger ] \; + \; 1 \right)
\right\rangle_{A_2}.
\end{align}
Notice that the first two lines in \eq{eq:square_cross_form2} are the
only two lines that contain the variables ${\bm x}_1, \; {\bm y}_1, \;
{\bm b}_1$, while the next two lines are the only ones that contain
the variables ${\bm x}_2, \; {\bm y}_2, \; {\bm b}_2$. In the limit we
are dealing with ${\bm x}_1, \; {\bm y}_1, \; {\bm b}_1$ are
perturbatively close to each other. Since $T_1({\bm b})$ is slowly
varying it is approximately constant over perturbatively short
scales. Thus we can make the approximation
\begin{align}
 \label{Tapprox}
  T_1({\bm B} - {\bm b}_1) \; \approx \; T_1({\bm B} - {\bm x}_1)
  \; \approx \; T_1({\bm B} - {\bm y}_1).
\end{align}
This same approximation also applies to ${\bm x}_2, \; {\bm y}_2, \;
{\bm b}_2$. Notice that the second line of
\eqref{eq:square_cross_form2} has four different terms in the
parentheses, each of which is at most a function of two of the three
variables ${\bm x}_1, \; {\bm y}_1, \; {\bm b}_1$. Combining this fact
with the approximation \eqref{Tapprox} we can perform one of the ${\bm
  x}_1, \; {\bm y}_1, \; {\bm b}_1$ integrals over a different
variable for each term in the second line depending on which variable
is not in the trace. A similar thing is done with the ${\bm x}_2, \;
{\bm y}_2, \; {\bm b}_2$ integral. After doing this and integrating by
parts we arrive at
\begin{align} 
\label{eq:square_cross_form3} 
\frac{d \sigma_{square}}{d^2 k_1 dy_1 d^2 k_2 dy_2} & = \frac{ \as^2
  \, C_F^2 }{4 \, \pi^6} \frac{1}{{\bm k}_1^2 \; {\bm k}_2^2} \int d^2
B \, d^2 b_1 \, d^2 b_2 \, d^2 x_1 \, d^2 x_2 \, T_1 ({\bm B} - {\bm
  b}_1) \; T_1 ({\bm B} - {\bm b}_2) \ln \left( \frac{1}{|{\bm x}_1 -
    {\bm b}_1| \Lambda} \right) \ln \left( \frac{1}{|{\bm x}_2 - {\bm
      b}_2| \Lambda} \right) \notag \\ & \times \, e^{- i \; {\bm k}_1
  \cdot ({\bm x}_1-{\bm b}_1)- i \; {\bm k}_2 \cdot ({\bm x}_2-{\bm
    b}_2)} \; \nabla_{{\bm x}_1}^2 \; \nabla_{{\bm x}_2}^2 \;
\frac{1}{(N_c^2-1)^2} \left\langle \mbox{Tr} \left[ \mathbb{1} -
    U_{{\bm x}_1} U_{{\bm b}_1}^\dagger \right] \mbox{Tr} \left[
    \mathbb{1} - U_{{\bm x}_2} U_{{\bm b}_2}^\dagger \right]
\right\rangle_{\! A_2}.
\end{align}

From here we can manipulate this expression into a form reminiscent of
\eq{eq:singlegluon} but not quite the same. As mentioned in the
discussion at the beginning of Sec.~\ref{sec:fact}, the integrals over
the impact parameters cannot be absorbed into the distribution
functions. This is now manifest in \eq{eq:square_cross_form3}: we have
three integrals (over ${\bm B}$, ${\bm b}_1$ and ${\bm b}_2$) and four
impact parameter-related distances (${\bm B} - {\bm b}_1$, ${\bm B} -
{\bm b}_2$, ${\bm b}_1$ and ${\bm b}_2$). We conclude that we must use
the new distribution functions defined in Eqs.~\eqref{eq:dipole_wave}
and \eqref{eq:doubletrace_dist} while convoluting them over the impact
parameters ${\bm B}$, ${\bm b}_1$ and ${\bm b}_2$. Employing
Eqs.~\eqref{eq:dipole_wave} and \eqref{eq:doubletrace_dist} we can
rewrite the 'square' diagrams contribution to the two-gluon production
cross section \eqref{eq:2glue_prod_main} in the factorized form
\begin{align} 
\label{eq:factorized_square} 
\frac{d \sigma_{square}}{d^2 k_1 dy_1 d^2 k_2 dy_2} & = \left( \frac{2
    \; \as}{C_F} \right)^2 \frac{1}{k_1^2 \; k_2^2} \int d^2 B \, d^2
b_1 \, d^2 b_2 \int d^2 q_1 \, d^2 q_2 \, \notag \\ & \times \;
\left\langle \frac{d \phi_{A_1} ({\bm q}_1, y=0)}{d^2 ({\bm B}-{\bm
      b}_1)} \right\rangle_{\! \! A_1} \left\langle \frac{d \phi_{A_1}
    ({\bm q}_2, y=0)}{d^2 ({\bm B}-{\bm b}_2)} \right\rangle_{\! \!
  A_1} \left\langle \frac{d \phi_{A_2}^{D} ({\bm q}_1 - {\bm k}_1,
    {\bm q}_2 - {\bm k}_2, y \approx y_1 \approx y_2)}{d^2 b_1 \; d^2
    b_2} \right\rangle_{\! \! A_2}.
\end{align}
The asymmetry in rapidity arguments of the distribution entering
\eq{eq:factorized_square} is due to the fact that the projectile in
the original \eq{eq:2glue_prod_main} was treated in the lowest-order
quasi-classical approximation, while the whole non-linear evolution
\cite{Balitsky:1996ub,Balitsky:1998ya,Kovchegov:1999yj,Kovchegov:1999ua,Jalilian-Marian:1997dw,Jalilian-Marian:1997gr,Iancu:2001ad,Iancu:2000hn}
is included in the rapidity interval between the produced gluons and
the target by the use of the Wilson lines.  As mentioned previously,
\eq{eq:factorized_square} is similar to \eq{eq:singlegluon} but has a
few key differences. \eq{eq:singlegluon} employs unintegrated gluon
distributions (gluon transverse momentum distributions (TMDs)), while
\eq{eq:factorized_square} uses one- and two-gluon Wigner
distributions. Related to that, in \eq{eq:singlegluon} the convolution
happens only over transverse momentum, while \eq{eq:factorized_square}
also contains integrals over impact parameters ${\bm B}, \; {\bm
  b}_1$, and ${\bm b}_2$.

One may also note that \eq{eq:factorized_square} is not
target-projectile symmetric: the target is described by a single
two-gluon distribution, while the projectile is represented by two
single-gluon distributions. In contrast, \eq{eq:singlegluon} is
completely target-projectile symmetric. In fact, \eq{eq:singlegluon}
is often generalized to the case of nucleus--nucleus ($AA$) collisions
by using \eq{eq:trace_wave} for both unintegrated gluon distributions
in it. While such generalization allows for successful phenomenology
(see e.g. \cite{ALbacete:2010ad}), it is theoretically not justified
below the saturation scales of both nuclei. Moreover, there is
numerical evidence \cite{Blaizot:2010kh} demonstrating that the
$k_T$-factorization formula \eqref{eq:singlegluon} is not valid in
$AA$ collisions. Therefore, it appears that the apparent
target-projectile symmetry of \eq{eq:singlegluon} is, in fact,
somewhat misleading: the equation was derived in the limit where the
projectile is dilute, while the target may or may not be dense,
leading to the difference in the definitions of the unintegrated gluon
distributions of the target and the projectile in
Eqs.~\eqref{eq:dipole_wave_int} and \eqref{eq:trace_wave}. It is
likely that \eq{eq:singlegluon} is not valid for dense-dense
scattering \cite{Blaizot:2010kh}, and is thus not truly
target-projectile symmetric due to the underlying assumptions.

With the 'square' diagrams contribution to the cross section cast in a
factorized form we now turn our attention to the 'crossed' diagrams
contribution \eqref{crossed_xsect}. It is helpful to write out the
crossed diagrams part of the cross section, \eq{crossed_xsect}, in the
following form,
\begin{align} 
\label{crossed_xsect_fac} 
\frac{d \sigma_{crossed}}{d^2 k_1 dy_1 d^2 k_2 dy_2} & = \frac{\as^2
  \, C_F^2}{16 \, \pi^8} \int d^2 B \, \left\langle \int \; d^2 x_1 \;
  d^2 y_1 \; d^2 b_1 \; T_1 ({\bm B} - {\bm b}_1) \; e^{- i \; {\bm
      k}_1 \cdot {\bm x}_1 + i \; {\bm k}_2 \cdot {\bm y}_1} \left[
    \frac{ {\bm x}_1 - {\bm b}_1}{ |{\bm x}_1 - {\bm b}_1 |^2 }
  \right]_i \left[ \frac{ {\bm y}_1 - {\bm b}_1}{ |{\bm y}_1 - {\bm
        b}_1 |^2 } \right]_j \right.  \notag \\ & \times \;
\frac{1}{N_c^2-1} \left[ U_{{\bm x}_1} U_{{\bm y}_1}^\dagger \; - \;
  U_{{\bm x}_1} U_{{\bm b}_1}^\dagger \; - \; U_{{\bm b}_1} U_{{\bm
      y}_1}^\dagger \; + \; \mathbb{1} \right]^{ab} \notag \\ & \times
\; \int \; d^2 x_2 \; d^2 y_2 \; d^2 b_2 \; T_1 ({\bm B} - {\bm b}_2)
\; e^{- i \; {\bm k}_2 \cdot {\bm x}_2 + i \; {\bm k}_1 \cdot {\bm
    y}_2} \left[ \frac{ {\bm x}_2 - {\bm b}_2}{ |{\bm x}_2 - {\bm b}_2
    |^2 } \right]_j \left[ \frac{ {\bm y}_2 - {\bm b}_2}{ |{\bm y}_2 -
    {\bm b}_2 |^2 } \right]_i \notag \\ & \times \; \frac{1}{N_c^2-1}
\left. \left[ U_{{\bm x}_2} U_{{\bm y}_2}^\dagger \; - \; U_{{\bm
        x}_2} U_{{\bm b}_2}^\dagger \; - \; U_{{\bm b}_2} U_{{\bm
        y}_2}^\dagger \; + \; \mathbb{1} \right]^{ba}
\right\rangle_{A_2} \; + \; ({\bm k}_2 \rightarrow - {\bm k}_2),
\end{align}
where $i,j = 1,2$ are transverse vector indices and $a,b = 1, \ldots ,
N_c^2 -1$ are adjoint color indices, with summation assumed over
repeated indices. Here we have again separated the terms that depend
on ${\bm x}_1, \; {\bm y}_1, \; {\bm b}_1$ from the terms that depend
on ${\bm x}_2, \; {\bm y}_2, \; {\bm b}_2$. Using the same trick we
employed when factorizing the 'square' diagrams contribution, we
evaluate the ${\bm x}_1, \; {\bm y}_1, \; {\bm b}_1$, and ${\bm x}_2,
\; {\bm y}_2, \; {\bm b}_2$ integrals piece by piece arriving at
(after transverse vector relabeling)
\begin{align} 
  \label{crossed_xsect_fac2}
  & \frac{d \sigma_{crossed}}{d^2 k_1 dy_1 d^2 k_2 dy_2} = \frac{
    \as^2 \; C_F^2 }{4 \; \pi^6} \int d^2 B \; d^2 b_1 \; d^2 b_2 \;
  d^2 x_1 \; d^2 x_2 \; T_1 ({\bm B} - {\bm b}_1) \; T_1 ({\bm B} -
  {\bm b}_2) \notag \\ & \times \; \left\{ \frac{1}{2} \delta_{ij} \ln
    \left( \frac{1}{|{\bm x}_1 - {\bm b}_1| \Lambda} \right) \; - \;
    \frac{ \left[ {\bm x}_1 - {\bm b}_1 \right]_i \; \left[ {\bm x}_1
        - {\bm b}_1 \right]_j}{2 \, |{\bm x}_1 - {\bm b}_1|^2} \; - i
    \left[ \frac{{\bm k}_1}{k_1^2} \right]_i \left[ \frac{ {\bm x}_1 -
        {\bm b}_1 }{ |{\bm x}_1 - {\bm b}_1|^2} \right]_j \; - i
    \left[ \frac{ {\bm x}_1 - {\bm b}_1 }{ |{\bm x}_1 - {\bm b}_1|^2}
    \right]_i \left[ \frac{{\bm k}_2}{k_2^2} \right]_j \right\} \notag
  \\ & \times \; \left\{ \frac{1}{2} \delta_{ij} \ln \left(
      \frac{1}{|{\bm x}_2 - {\bm b}_2| \Lambda} \right) \; - \; \frac{
      \left[ {\bm x}_2 - {\bm b}_2 \right]_i \; \left[ {\bm x}_2 -
        {\bm b}_2 \right]_j}{2 \, |{\bm x}_2 - {\bm b}_2|^2} \; - i
    \left[ \frac{{\bm k}_1}{k_1^2} \right]_i \left[ \frac{ {\bm x}_2 -
        {\bm b}_2 }{ |{\bm x}_2 - {\bm b}_2|^2} \right]_j \; - i
    \left[ \frac{ {\bm x}_2 - {\bm b}_2 }{ |{\bm x}_2 - {\bm b}_2|^2}
    \right]_i \left[ \frac{{\bm k}_2}{k_2^2} \right]_j \right\} \notag
  \\ & \times \; e^{- i \; {\bm k}_1 \cdot ({\bm x}_1-{\bm b}_2)- i \;
    {\bm k}_2 \cdot ({\bm x}_2-{\bm b}_1)} \; \frac{1}{(N_c^2-1)^2}
  \left\langle \mbox{Tr} \left[ \left( \mathbb{1} - U_{{\bm x}_1}
        U_{{\bm b}_1}^\dagger \right) \left( \mathbb{1} - U_{{\bm
            x}_2} U_{{\bm b}_2}^\dagger \right) \right]
  \right\rangle_{A_2} \; + \; ({\bm k}_2 \rightarrow - {\bm k}_2),
\end{align}
where we have employed
\begin{align}
  \label{formula1}
  \int d^2 b \, \left[ \frac{ {\bm x} - {\bm b}}{ |{\bm x} - {\bm
        b}|^2 } \right]_i \left[ \frac{ {\bm y} - {\bm b}}{ |{\bm y} -
      {\bm b}|^2 } \right]_j = \pi \left\{ \delta_{ij} \, \ln \left(
      \frac{1}{|{\bm x} - {\bm y}| \, \Lambda} \right) - \frac{\left[
        {\bm x} - {\bm y} \right]_i \; \left[ {\bm x} - {\bm y}
      \right]_j}{|{\bm x} - {\bm y}|^2} \right\}
\end{align}
along with other, more common, two-dimensional integrals (see, e.g.,
Appendix A.2 of \cite{KovchegovLevin} for a list of useful integrals).

To proceed we rewrite \eq{crossed_xsect_fac2} as
\begin{align} 
  \label{crossed_xsect_fac3}
  & \frac{d \sigma_{crossed}}{d^2 k_1 dy_1 d^2 k_2 dy_2} = \frac{
    \as^2 \; C_F^2 }{4^3 \; \pi^6} \frac{1}{k_1^4 \, k_2^4} \int d^2 B
  \; d^2 b_1 \; d^2 b_2 \; d^2 x_1 \; d^2 x_2 \, e^{- i \; {\bm k}_1
    \cdot ({\bm x}_1-{\bm b}_2)- i \; {\bm k}_2 \cdot ({\bm x}_2-{\bm
      b}_1)} \notag \\ & \times \; \left\{ \left[
      \overleftarrow{\nabla}_{x_1}^2 \, \overleftarrow{\nabla}_{b_1}^2
      \, \nabla^i_{x_1} \, \nabla^j_{x_1} +
      \overleftarrow{\nabla}_{b_1}^2 \, \overleftarrow{\nabla}_{x_1}^i
      \, \nabla^j_{x_1} \, \nabla^2_{x_1} -
      \overleftarrow{\nabla}_{x_1}^2 \, \overleftarrow{\nabla}^j_{b_1}
      \, \nabla^i_{x_1} \, \nabla^2_{x_1} \right] ({\bm x}_1 - {\bm
      b}_1)^2 \, \ln \left( \frac{1}{|{\bm x}_1 - {\bm b}_1| \Lambda}
    \right) \, T_1 ({\bm B} - {\bm b}_1) \right\} \notag \\ & \times
  \; \left\{ \left[ \overleftarrow{\nabla}_{x_2}^2 \,
      \overleftarrow{\nabla}_{b_2}^2 \, \nabla^i_{x_2} \,
      \nabla^j_{x_2} - \overleftarrow{\nabla}_{x_2}^2 \,
      \overleftarrow{\nabla}_{b_2}^i \, \nabla^j_{x_2} \,
      \nabla^2_{x_2} + \overleftarrow{\nabla}_{b_2}^2 \,
      \overleftarrow{\nabla}^j_{x_2} \, \nabla^i_{x_2} \,
      \nabla^2_{x_2} \right] ({\bm x}_2 - {\bm b}_2)^2 \, \ln \left(
      \frac{1}{|{\bm x}_2 - {\bm b}_2| \Lambda} \right) \; T_1 ({\bm
      B} - {\bm b}_2) \right\} \notag \\ & \times \;
  \frac{1}{(N_c^2-1)^2} \left\langle \mbox{Tr} \left[ \left(
        \mathbb{1} - U_{{\bm x}_1} U_{{\bm b}_1}^\dagger \right)
      \left( \mathbb{1} - U_{{\bm x}_2} U_{{\bm b}_2}^\dagger \right)
    \right] \right\rangle_{\! \! A_2} + ({\bm k}_2 \rightarrow - {\bm
    k}_2),
\end{align}
where $\nabla$'s denote transverse coordinate derivatives and the left
arrow over $\nabla$ indicates that the derivative is acting on the
exponential to the left of the curly brackets.

Notice the non-trivial transverse index structure in
\eq{crossed_xsect_fac3}: this drastically alters the factorized form
of the expression, as compared to, say, \eq{eq:factorized_square}.
Inverting Fourier transforms in Eqs.~\eqref{eq:dipole_wave} and
\eqref{eq:quad_dist}, employing \eq{eq:dipole_amp}, and substituting
the results into \eq{crossed_xsect_fac3} yields, after a fair bit of
algebra,
\begin{align} 
\label{eq:factorized_crossed} 
& \frac{d \sigma_{crossed}}{d^2 k_1 dy_1 d^2 k_2 dy_2} = \left(
  \frac{2 \; \as}{C_F} \right)^2 \frac{1}{k_1^2 \; k_2^2} \int d^2 B
\; d^2 b_1 \; d^2 b_2 \int d^2 q_1 \; d^2 q_2 \; \frac{\mathcal{K} (
  {\bm b}_1, {\bm b}_2, {\bm k}_1, {\bm k}_2, {\bm q}_1, {\bm
    q}_2)}{N_c^2-1} \notag \\ & \times \; \left\langle \frac{d
    \phi_{A_1} ({\bm q}_1, y=0)}{d^2 ({\bm B}-{\bm b}_1)}
\right\rangle_{\! \! A_1} \left\langle \frac{d \phi_{A_1} ({\bm q}_2,
    y=0)}{d^2 ({\bm B}-{\bm b}_2)} \right\rangle_{\! \! A_1}
\left\langle \frac{d \phi_{A_2}^{Q} ({\bm k}_1 - {\bm q}_1, {\bm k}_2
    - {\bm q}_2, y \approx y_1 \approx y_2)}{d^2 b_1 \; d^2 b_2}
\right\rangle_{\! \! A_2} \; + \; ({\bm k}_2 \rightarrow - {\bm k}_2),
\end{align}
where the ``coefficient function'' is defined as
\begin{align} 
  \label{eq:kernel_crossed}
  \mathcal{K} ( {\bm b}_1, {\bm b}_2, {\bm k}_1, {\bm k}_2, {\bm q}_1,
  {\bm q}_2) & = \frac{1}{q_1^2 \; q_2^2 \; ({\bm k}_1-{\bm q}_1)^2
    ({\bm k}_2-{\bm q}_2)^2} \; e^{-i \, ( {\bm k}_1 - {\bm k}_2 )
    \cdot ( {\bm b}_1 - {\bm b}_2 )} \; \left\{ k_1^2 \; k_2^2 ({\bm
      q}_1 \cdot {\bm q}_2)^2 \right.  \notag \\ & - \; k_1^2 \; ({\bm
    q}_1 \cdot {\bm q}_2) \left[ ({\bm k}_2 \cdot {\bm q}_1) \; q_2^2
    \; + \; ({\bm k}_2 \cdot {\bm q}_2) \; q_1^2 \; - \; q_1^2 \;
    q_2^2 \right] \notag \\ & - \; k_2^2 \; ({\bm q}_1 \cdot {\bm
    q}_2) \left[ ({\bm k}_1 \cdot {\bm q}_1) \; q_2^2 \; + \; ({\bm
      k}_1 \cdot {\bm q}_2) \; q_1^2 \; - \; q_1^2 \; q_2^2 \right]
  \notag \\ & \left.  + \; q_1^2 \; q_2^2 \; \left[ ({\bm k}_1 \cdot
      {\bm q}_1) ({\bm k}_2 \cdot {\bm q}_2) \; + \; ({\bm k}_1 \cdot
      {\bm q}_2) ({\bm k}_2 \cdot {\bm q}_1) \right] \right\}
\end{align}
with $q_i = |{\bm q}_i|$, $k_i = |{\bm k}_i|$.

Inserting \eq{eq:factorized_square} and \eq{eq:factorized_crossed}
into \eq{eq_all} we arrive at the $k_T$-factorized form for the two
gluon production cross section in heavy-light ion collisions
\begin{align} 
\label{eq:factorized_final} 
& \frac{d \sigma}{d^2 k_1 dy_1 d^2 k_2 dy_2} = \left( \frac{2 \;
    \as}{C_F} \right)^2 \frac{1}{k_1^2 \; k_2^2} \int d^2 B \; d^2 b_1
\; d^2 b_2 \int d^2 q_1 \; d^2 q_2 \; \left\langle \frac{d \phi_{A_1}
    ({\bm q}_1, y=0)}{d^2 ({\bm B}-{\bm b}_1)} \right\rangle_{\! \!
  A_1} \left\langle \frac{d \phi_{A_1} ({\bm q}_2, y=0)}{d^2 ({\bm
      B}-{\bm b}_2)} \right\rangle_{\! \! A_1} \notag \\ & \times
\left\{ \left\langle \frac{d \phi_{A_2}^{D} ({\bm q}_1 - {\bm k}_1,
      {\bm q}_2 - {\bm k}_2, y)}{d^2 b_1 \; d^2 b_2} \right\rangle_{\!
    \! A_2} \! \! + \left[ \frac{\mathcal{K} ( {\bm b}_1, {\bm b}_2,
      {\bm k}_1, {\bm k}_2, {\bm q}_1, {\bm q}_2)}{N_c^2-1}
    \left\langle \frac{d \phi_{A_2}^{Q} ({\bm q}_1 - {\bm k}_1, {\bm
          q}_2 - {\bm k}_2, y)}{d^2 b_1 \; d^2 b_2} \right\rangle_{\!
      \! A_2} \! \! + ({\bm k}_2 \rightarrow - {\bm k}_2) \right]
\right\}
\end{align}
with $y \approx y_1 \approx y_2$ in the curly
brackets. \eq{eq:factorized_final} is the main result of this Section.

Notice that \eq{eq:factorized_final} has all of the properties we
expected: it contains the convolution over the impact parameters ${\bm
  B}$, ${\bm b}_1$, and ${\bm b}_2$ along with different two-gluon
distribution functions. As advertised, \eq{eq:factorized_final} also
contains a ``coefficient function'' associated with the factorized
form of 'crossed' diagrams.

The convolution over impact parameters in \eq{eq:factorized_final}
appears to imply that the 2-gluon production cross section is
sensitive to the $b$-dependence of the one- and two-gluon
distributions $\phi$, $\phi^D$, $\phi^Q$. From
Eqs.~\eqref{eq:dipole_dist}, \eqref{eq:doubletrace_dist}, and
\eqref{eq:quad_dist} we see that the $b$-dependence of those gluon
distributions is related to that of the dipole, double-trace and
quadrupole operators. It is known that any perturbative approach, such
as the CGC formalism employed here, cannot describe correctly the
$b$-dependence of scattering amplitudes in peripheral collisions due
to the importance of non-perturbative effects
\cite{Kovner:2001bh}. It, therefore, appears that the two-gluon
production cross-section is also sensitive to the non-perturbative
large-$b$ physics. Note, however, that this conclusion also applies to
the single-gluon production in \eq{eq:singlegluon}, since the impact
parameter integral in \eq{eq:dipole_wave_int} is also sensitive to
large-$b$ physics. Recent studies \cite{Levin:2014bwa} appear to
indicate that this sensitivity to non-perturbative effects at the
periphery is not very strong, and may be negligible at high energies.

Unfortunately the factorization expression \eqref{eq:factorized_final}
is different from that used in
\cite{Dusling:2012cg,Dusling:2012wy,Dusling:2012iga}. The expression
in those references was motivated by extrapolation of the
dilute--dilute scattering case to the dense--dense scattering by
analogy with the single-gluon production \eq{eq:singlegluon}. While
our result is valid only for the dense-dilute scattering, we can
conclude that the extrapolation suggested in
\cite{Dumitru:2010mv,Dumitru:2010iy,Dusling:2012cg,Dusling:2012wy,Dusling:2012iga}
does not work in the dense-dilute case, and is, therefore, unlikely to
be valid in the dense-dense scattering case either.

Just like \eq{eq:factorized_square}, the expression
\eqref{eq:factorized_final} is not projectile-target symmetric. While
this is natural due to the asymmetric treatment of the target and
projectile in our dense-dilute scattering approximation, this
asymmetry also means that a simple generalization to the case of
nucleus--nucleus scattering along the lines of what was done with
\eq{eq:singlegluon} in
\cite{Kharzeev:2001gp,Kharzeev:2001yq,ALbacete:2010ad,Albacete:2007sm}
appears to be impossible for \eq{eq:factorized_final}.

The factorized form of the two-gluon production cross section
\eqref{eq:factorized_final} contains a few interesting properties. If
we look at the large-$N_c$ limit the 'crossed' diagrams contribution
can be neglected and, using \eq{eq:singlet_app}, the two-gluon singlet
distribution function factorizes into two single gluon distribution
functions,
\begin{align} 
\label{eq:factorized_largeNc} 
\frac{d \sigma}{d^2 k_1 dy_1 d^2 k_2 dy_2} & \
\bigg|_{\mbox{large}-N_c} = \left( \frac{2 \; \as}{C_F} \right)^2
\frac{1}{k_1^2 \; k_2^2} \int d^2 B \, d^2 b_1 \, d^2 b_2 \int d^2 q_1
\, d^2 q_2 \notag \\ & \times \; \left\langle \frac{d \phi_{A_1} ({\bm
      q}_1, y=0)}{d^2 ({\bm B}-{\bm b}_1)} \right\rangle_{\! \!  A_1}
\left\langle \frac{d \phi_{A_1} ({\bm q}_2, y=0)}{d^2 ({\bm B}-{\bm
      b}_2)} \right\rangle_{\!  \!  A_1} \left\langle \frac{d
    \phi_{A_2} ({\bm q}_1 - {\bm k}_1, y)}{d^2 b_1} \right\rangle_{\!
  \!  A_2} \left\langle \frac{d \phi_{A_2} ({\bm q}_2 - {\bm k}_2,
    y)}{d^2 b_2} \right\rangle_{\! \!  A_2}.
\end{align}
This equation can only generate correlations between the two gluons
through the convolution over the impact parameters, which are
geometric correlations \cite{Kovchegov:2012nd}. This form does not
contain the information needed to, say, calculate the correlation
function \eqref{eq:corr_LO}, since for that one needs terms that are
subleading in the large-$N_c$ limit.

Another interesting property is that we can isolate the part of the
cross-section that gives rise to HBT correlations
\cite{HanburyBrown:1956pf}. Due to the nature of HBT correlations, the
only way the correlations can be generated is through interference
effects. Thus only the 'crossed' diagrams contribute. In addition, for
the correlation to be pure HBT, the two produced gluons should have
the same colors (to be identical particles): imposing same-color
requirement on the 'crossed' diagrams is equivalent to the projection
employed in arriving at \eq{eq:quad_approx}. We conclude that the only
part of the quadrupole two-gluon distribution function
\eqref{eq:quad_dist} that contributes to HBT correlations is the
portion that can be factorized into two single-gluon distributions
shown in \eqref{eq:quad_approx}. With the help of \eq{eq:quad_approx}
the HBT part of the two-gluon production cross-section can be written
as
\begin{align} 
\label{eq:factorized_HBT} 
\frac{d \sigma_{HBT}}{d^2 k_1 dy_1 d^2 k_2 dy_2} & = \left( \frac{2 \;
    \as}{C_F} \right)^2 \frac{1}{k_1^2 \; k_2^2} \int d^2 B \; d^2 b_1
\; d^2 b_2 \int d^2 q_1 \; d^2 q_2 \; \frac{\mathcal{K} ( {\bm b}_1,
  {\bm b}_2, {\bm k}_1, {\bm k}_2, {\bm q}_1, {\bm q}_2)}{N_c^2-1}
\notag \\ \times \; \left\langle \frac{d \phi_{A_1} ({\bm q}_1,
    y=0)}{d^2 ({\bm B}-{\bm b}_1)} \right\rangle_{\! \!  A_1} &
\left\langle \frac{d \phi_{A_1} ({\bm q}_2, y=0)}{d^2 ({\bm B}-{\bm
      b}_2)} \right\rangle_{\! \!  A_1} \left\langle \frac{d
    \phi_{A_2} ({\bm q}_1 - {\bm k}_1, y)}{d^2 b_1} \right\rangle_{\!
  \!  A_2} \left\langle \frac{d \phi_{A_2} ({\bm q}_2 - {\bm k}_2,
    y)}{d^2 b_2} \right\rangle_{\! \!  A_2} \; + \; ({\bm k}_2
\rightarrow - {\bm k}_2).
\end{align}

To summarize this Section let us stress that we were able to find a
factorized form for the two-gluon production cross section in
heavy-light ion collisions, given by \eq{eq:factorized_final}. It had
to be written in a different form than that of the single gluon
production cross-section \eqref{eq:singlegluon}. In particular, in
\eq{eq:factorized_final} we have a convolution over the impact
parameters which requires that the distribution functions have to be
written as differentials with respect to impact parameters, that is as
gluon Wigner distributions. There was also a ``coefficient function''
factor \eqref{eq:kernel_crossed} that was associated with the
interference effects included in the 'crossed' diagrams. These facts
may have important implications for $k_T$-factorization when it comes
to multi-gluon cross-sections and could possibly give insight into the
nature of $k_T$-factorization in general.

%%%%%%%%%%%%%%%%%%%%%%%%%%%%%%%%%%%%%%%%%%%%%%%%%%%%%%%%%%%%%%%%%%%%%%%%%%%%%%%%%%

\section{Energy Dependence of the Correlations}
\label{sec:energy}

Our goal now is to study the energy and rapidity dependence of the
correlator in \eq{corr_def}. As mentioned before, we are working in
the regime where the rapidities of the produced gluons, $y_1, y_2$,
are sufficiently close to the rapidity of the projectile $Y$ such that
no small-$x$ evolution needs to be included in the $[y_1, y_2]$ and
$[y_2, Y]$ rapidity intervals (for $y_2 > y_1$). This implies that
$|y_1 - y_2| \ll 1/\as$ and $0 < Y - y_{1,2} \ll 1/\as$. Taking the
rapidity of the target to be $0$, we see that the results of
\cite{Kovchegov:2012nd} outlined above apply to the case when $Y$,
$y_1$, and $y_2$ are all large enough to necessitate the inclusion of
small-$x$ evolution in the rapidity interval between the target and
the produced gluons. Since, from the standpoint of the
leading-logarithmic small-$x$ evolution, the rapidities of the gluons
and the projectile are close enough to be considered identical, $y_1
\approx y_2 \approx Y$, including the evolution between the gluons and
the projectile would only generate the dependence of the cross section
\eqref{eq_all} on the net rapidity interval $Y$ (or, equivalently, on
the center-of-mass energy of the collision), without producing any
additional $y_1$ and $y_2$ dependence of the cross section. To
generate the latter one has to include the small-$x$ evolution between
the two gluons and between the gluons and the projectile: this is left
for future work. Therefore, here we will not distinguish between the
energy and rapidity dependence of the cross section \eqref{eq_all},
since the two are identical in the approximation used.

In the leading-$\ln 1/x$ approximation, the effects of the nonlinear
BK/JIMWLK evolution are included in the cross section \eqref{eq_all}
by simply evolving the Wilson-line correlators in
Eqs.~\eqref{eq:2glue_prod_main} and \eqref{crossed_xsect} up to
rapidity $Y$. The corresponding evolution equations can be easily
obtained by applying the JIMWLK evolution to the correlators. The
results are listed in the Appendix, with the evolution equations for
the gluon dipole, gluon quadrupole, and the double-trace operator
given by the Eqs.~\eqref{dip_JIMWLK}, \eqref{quad_JIMWLK}, and
\eqref{double_tr_JIMWLK} respectively. Unfortunately none of these
equations is a closed integro-differential equation: their
right-hand-sides contain higher-order Wilson-line correlators, which
in turn would obey other evolution equations, involving yet
higher-order correlators, etc., forming the whole infinite tower of
the Balitsky hierarchy
\cite{Balitsky:1996ub,Balitsky:1998ya}. Unfortunately the hierarchy by
itself can not be solved neither numerically nor analytically: instead
one can solve the JIMWLK functional evolution equation numerically,
and construct all the correlators by averaging the corresponding
operators over all color field configurations
\cite{Rummukainen:2003ns,Kovchegov:2008mk,Dumitru:2011vk}.

Instead, here we will try to evaluate the energy dependence of the
cross section \eqref{eq_all} by using an approximate analytic method,
based on the Gaussian truncation of the JIMWLK evolution
\cite{Kovner:2001vi,Kovchegov:2008mk}. It is based on the observation
that relations between different Wilson line correlators calculated in
the MV model remain approximately valid for JIMWLK-evolved
correlators. For instance, as was shown in \cite{Kovchegov:2008mk},
the relation
\begin{align}
  \label{eq:Casimir}
  S_G ({\bm x}_1, {\bm x}_2, Y) = \left[ S_{q\bar q} ({\bm x}_1, {\bm
      x}_2, Y) \right]^{N_c/C_F}
\end{align}
between the adjoint and fundamental ($S_{q\bar q}$) dipole
$S$-matrices, valid strictly-speaking only in the MV model (that is,
at $Y=0$), is also preserved (with high accuracy, but not exactly) by
the leading-$\ln 1/x$ JIMWLK evolution. In fact, both dipole
$S$-matrices in \eq{eq:Casimir} can be well approximated if one writes
\begin{align}
  \label{eq:BKconn}
  S_{q\bar q} ({\bm x}_1, {\bm x}_2, Y) = \left[ S_{BK} ({\bm x}_1,
    {\bm x}_2, Y) \right]^{2 \, C_F/N_c},
\end{align}
where $S_{BK}$ is found from the BK evolution equation,
\begin{align}
  \label{eq:BK}
  \partial_Y S_{BK} ({\bm x}, {\bm y}, Y) = \frac{\as \, N_c}{2 \,
    \pi^2} \, \int d^2 z \frac{({\bm x} - {\bm y})^2}{({\bm x} - {\bm
      z})^2 \, ({\bm y} - {\bm z})^2} \, \left[ S_{BK} ({\bm x}, {\bm
      z}, Y) \, S_{BK} ({\bm z}, {\bm y}, Y) - S_{BK} ({\bm x}, {\bm
      y}, Y) \right].
\end{align}
In the large-$N_c$ approximation for which the BK equation is valid,
\eq{eq:BKconn} reduces to $S_{q\bar q} = S_{BK}$, a relation which
also seems to hold well numerically if $S_{q\bar q}$ is found using
JIMWLK evolution \cite{Rummukainen:2003ns}. However, the full
\eq{eq:BKconn} provides an even better numerical agreement between
JIMWLK-evolved $S_{q\bar q}$ and BK-evolved $S_{BK}$
\cite{Kovchegov:2008mk}. 

The validity of the Gaussian truncation for higher-order correlators
of Wilson lines was studied numerically in \cite{Dumitru:2011vk},
while analytic arguments in support of the Gaussian approximation have
been proposed in \cite{Iancu:2011nj,Iancu:2011ns}.

Here we will assume that the Gaussian truncation is valid, and apply it
to determine the energy dependence of the two-gluon production cross
section \eqref{eq_all}. While the exact analytic solution of the BK
equation is not known, a good approximation exists immediately outside
the saturation region \cite{Iancu:2002tr,Mueller:2002zm}
\begin{align}
  \label{eq:Nevol}
  N ({\bm x}_1, {\bm x}_2, Y) \propto \big[ | \bm x_1 - \bm x_2 | \,
  Q_s \left( Y \right) \big]^{1+ 2 i \; \nu_0},
\end{align}
where the (imaginary part of the) dipole forward scattering amplitude
is
\begin{align}
  \label{eq:Ndef}
  N ({\bm x}_1, {\bm x}_2, Y) = 1 - S_{BK} ({\bm x}_1, {\bm x}_2, Y)
\end{align}
and $\nu_0 \approx - 0.1275 \, i$. The approximate solution
\eqref{eq:Nevol} was derived in the limit where the impact-parameter
dependence of the dipole scattering amplitude could be neglected. The
proportionality \eqref{eq:Nevol} becomes an equality if one includes
an energy-dependent prefactor \cite{Iancu:2002tr,Mueller:2002zm},
though its energy dependence is much slower than that of the factor
shown in \eq{eq:Nevol}: as we neglect this prefactor, we have to
remember that all our conclusions here will be valid up to a factor
which may vary slowly with energy. The solution \eqref{eq:Nevol} is
valid in the extended geometric scaling region, $1/k_{geom} \lesssim |
\bm x_1 - \bm x_2 | \lesssim 1/Q_s \left( Y \right)$, with $k_{geom}
\approx Q_s^2 (Y)/Q_s (0) \gg Q_s (Y)$ \cite{Iancu:2002tr}. There the
BK equation \eqref{eq:BK} is linearized in the powers of the dipole
amplitude $N$, and the resulting BFKL equation is solved with an IR
saturation boundary \cite{Mueller:2002zm}.

To find the gluon dipole $S$ matrix we use Eqs.~\eqref{eq:Casimir} and
\eqref{eq:BKconn} to write 
\begin{align}
  \label{eq:SGevol}
  1 - S_{G} ({\bm x}_1, {\bm x}_2, Y) = 1 - \left[ S_{BK} ({\bm x}_1,
    {\bm x}_2, Y) \right]^2 = 2 \, N ({\bm x}_1, {\bm x}_2, Y) -
  \left[ N ({\bm x}_1, {\bm x}_2, Y) \right]^2 \notag \\ \approx 2 \,
  N ({\bm x}_1, {\bm x}_2, Y) \propto \big[ | \bm x_1 - \bm x_2 | \,
  Q_s \left( Y \right) \big]^{1+ 2 i \; \nu_0}.
\end{align}
Comparing this to \eq{eq:SG_GM} and postulating the latter to be valid
at all rapidities (thus defining $\Gamma_G$ for $Y \neq 0$), we
conclude that, in this linearized regime,
\begin{align}\label{Gevol}
  \Gamma_G \left( \bm x_1 , \bm x_2 , Y \right) \approx 1 - S_{G}
  ({\bm x}_1, {\bm x}_2, Y) \propto \left( | \bm x_1 - \bm x_2 | \,
    Q_s \left( Y \right) \right)^{1+ 2 i \; \nu_0}.
\end{align}

We now want to determine the energy-dependence of the two-gluon
production cross section in Eqs.~\eqref{eq:2glue_prod_main} and
\eqref{crossed_xsect} outside the saturation region, that is for $k_1,
k_2 \gg Q_{s2} (Y)$. From the experience with the single inclusive
gluon production
\cite{Kharzeev:2002pc,Albacete:2003iq,Kharzeev:2003wz} it appears that
the behavior of the cross section in the $k_1, k_2 \approx Q_{s2} (Y)$
regime (at the edge of the saturation region) is qualitatively similar
to that for $k_1, k_2 \gg Q_{s2} (Y)$ (outside the saturation
region). We, therefore, hope that by studying the energy-dependence of
two-gluon production at $k_1, k_2 \gg Q_{s2} (Y)$ we would obtain a
good estimate of the energy dependence in other kinematic regions as
well.

Let us begin with the 'square' diagram case. Employing the results of
the previous Section we see that, for \eq{eq:square_cross_form3} the
$k_1, k_2 \gg Q_{s2} (Y)$ limit implies that ${\bm x}_1$ is close to
${\bm b}_1$ while ${\bm x}_2$ is close to ${\bm b}_2$. It is also a
good approximation to assume that the two pairs of points, ${\bm x}_1,
{\bm b}_1$ and ${\bm x}_2, {\bm b}_2$, are close to each other. To
justify this assume that ${\bm x}_1, {\bm b}_1$ and ${\bm x}_2, {\bm
  b}_2$ are far apart, further away than $1/Q_{s2} (Y)$. Using the
Gaussian truncation along with \eq{eq:Dapprox} we can evaluate the
connected part of the double-trace operator as proportional to
\begin{align}
  \label{dtrace_app}
  (D_2 - D_3)^2 \, \left( \frac{1}{D_2} + \frac{1}{D_3} \right),
\end{align}
where
\begin{subequations} \label{Dagain}
\begin{align}
  D_1 & = - \Gamma_G ({\bm x}_1, {\bm b}_1, Y) - \Gamma_G ({\bm x}_2,
  {\bm b}_2, Y) \ll 1 \\
  D_2 & = - \Gamma_G ({\bm x}_1, {\bm x}_2, Y) - \Gamma_G ({\bm b}_1,
  {\bm b}_2, Y) \gg 1 \\
  D_3 & = - \Gamma_G ({\bm x}_1, {\bm b}_2, Y) - \Gamma_G ({\bm b}_1,
  {\bm x}_2, Y) \gg 1.
\end{align}
\end{subequations}
Since for very large dipoles (with sizes much larger than $1/Q_{s2}
(Y)$) the $S$-matrix is given by the Levin-Tuchin formula
\cite{Levin:1999mw}
\begin{align}
  \label{eq:LT}
  S_G ({\bm x}_1, {\bm x}_2, Y) \propto e^{ - \mbox{const} \, \ln^2
    \left( |{\bm x}_1 - {\bm x}_2| \, Q_{s2} (Y) \right)}
\end{align}
we see that 
\begin{align}
  \label{Glarge}
  \Gamma_G ({\bm x}_1, {\bm x}_2, Y) \sim \ln^2 \left( |{\bm x}_1 -
    {\bm x}_2| \, Q_{s2} (Y) \right)
\end{align}
is a slowly varying function of rapidity $Y$. Using this in
\eq{dtrace_app} we conclude that at large separations between ${\bm
  x}_1, {\bm b}_1$ and ${\bm x}_2, {\bm b}_2$ the connected part of
the double-trace operator is a slowly-varying function of energy,
proportional to powers of the logarithm of energy. As we will see
shortly, the contribution where ${\bm x}_1, {\bm b}_1$ and ${\bm x}_2,
{\bm b}_2$ are close to each other grows as a power of energy, and is
thus dominant at high energies. Therefore, we can neglect the region
where ${\bm x}_1, {\bm b}_1$ and ${\bm x}_2, {\bm b}_2$ are far apart,
and concentrate on the region where all four transverse vectors in the
double-trace operator are within $1/Q_{s2} (Y)$ from each other.

Applying the Gaussian truncation, we see that to assess the energy
dependence of the cross section in \eq{eq:2glue_prod_main} we have to
expand the double-trace operators in it to the lowest non-trivial
order in $\Gamma_G$ (or, equivalently, in $D_1$, $D_2$, and $D_3$
defined in \eq{Ds}, but now for any rapidity $Y$), and then use
\eq{Gevol} to obtain the explicit dependence of the cross sections on
$Q_{s2} \left( Y \right)$. Namely we need to take
Eqs.~\eqref{Delta_exp} and \eqref{Ds} and expand $\Delta$ to the
lowest non-trivial order in $\Gamma_G$, and then substitute $\Gamma_G$
from \eq{Gevol} to obtain the energy dependence of the double-trace
correlator. Using the result in \eq{eq:2glue_prod_main} would then
give the energy dependence of the two-gluon production cross
section. However, the expansion of \eq{eq:2glue_prod_main} to the
lowest order in $D_i$'s was already constructed in
\cite{Kovchegov:2012nd} in order to reproduce the lowest-order
correlation function \eqref{eq:corr_LO} from
\cite{Dumitru:2008wn,Dusling:2009ni,Dumitru:2010iy}. We can,
therefore, use the results of the expansion from
\cite{Kovchegov:2012nd} simply replacing
\begin{align} 
\label{Gamma1}
  \Gamma_G \left( \bm x_1 , \bm x_2 , Y = 0 \right) = \frac{\left( \bm
      x_1 - \bm x_2 \right)^2 \, Q_{s2}^2}{4} \, \ln \frac{1}{|\bm x_1
    - \bm x_2| \, \Lambda} = Q_{s2}^2 \int \frac{d^2 l}{2 \pi}
  \frac{1}{|\bm l|^4} \left( 1 - e^{i \; \bm l \cdot \left( \bm x_1 -
        \bm x_2 \right)} \right)
\end{align}
by
\begin{align} 
\label{Gamma2}
\Gamma_G \left( \bm x_1 , \bm x_2 , Y \right) \propto \left( | \bm x_1
  - \bm x_2 | \, Q_{s2} \left( Y \right) \right)^{1+ 2 i \; \nu_0} =
c_0 \; Q_s^{1+ 2 i \; \nu_0} \left( Y \right) \int \frac{d^2 l}{2 \pi}
\frac{1}{|\bm l|^{3+ 2 i \; \nu_0}} \left( 1 - e^{i \; \bm l \cdot
    \left( \bm x_1 - \bm x_2 \right)} \right)
\end{align}
where 
\begin{align}
  \label{eq:c0}
  c_0 = - 2^{2 + 2 \, i \, \nu_0} \frac{\Gamma \left( \frac{3}{2} + i
      \, \nu_0 \right)}{\Gamma \left( - \frac{1}{2} - i \, \nu_0
    \right)} \approx 1.1255. 
\end{align}
We see that the primary differences between the two functions in
Eqs.~\eqref{Gamma1} and \eqref{Gamma2} is that the momentum in the
denominator along with the saturation scale are now taken to a
different power while the saturation scale is also rapidity-dependent.

With a good accuracy the rapidity dependence of the saturation scale
can be factored from the impact parameter dependence. Concentrating on
the powers of energy only, we write \cite{Gribov:1984tu,Iancu:2002tr}
\begin{align}
\label{QsY}
Q_{s2}^2 (Y) = e^{\lambda \, Y} \, Q_{s2}^2 (\bm b)
\end{align}
with $\lambda > 0$ a known quantity, the exact value of which is not
important to us here.

Proceeding to evaluate the energy-dependence of the correlated part of
\eq{eq:2glue_prod_main} we now employ the fact that at the lowest
quasi-classical order the interaction term in it (the sum of all the
$\Delta$'s in the square brackets of \eq{eq:square_cross_fact}) can be
written as
\begin{align} 
\label{eq:inter_square} 
Int_{square}( {\bm x}_1, {\bm y}_1 , {\bm b}_1 , {\bm x}_2 , {\bm
  y}_2, {\bm b}_2) & = \frac{Q_{s2}^4 ({\bm b})}{2 \, N_c^2 (2 \,
  \pi)^2} \int \frac{d^2 l}{l^4} \, \frac{d^2 l'}{l'^4} e^{i \; {\bm
    \Delta b} \cdot ({\bm l} - {\bm l'})} \notag \\ & \times \, \left[
  \left(1-e^{i \, {\bm l} \cdot {\bm x}_1} \right) \left(1-e^{- i \,
      {\bm l'} \cdot {\bm y}_1} \right) + \left(1-e^{-i \, {\bm l'}
      \cdot {\bm x}_1} \right) \left(1-e^{i \, {\bm l} \cdot {\bm
        y}_1} \right) \right] \notag \\ & \times \, \left[
  \left(1-e^{-i \, {\bm l} \cdot {\bm x}_2} \right) \left(1-e^{i \,
      {\bm l'} \cdot {\bm y}_2} \right) + \left(1-e^{i \, {\bm l'}
      \cdot {\bm x}_2} \right) \left(1-e^{-i \, {\bm l} \cdot {\bm
        y}_2} \right) \right]
\end{align}
with $l = |{\bm l}|$, $l' = |{\bm l'}|$. (This expression follows from
Eq.~(47) in \cite{Kovchegov:2012nd}.)  Generalizing this along the
steps justifying the transition from \eqref{Gamma1} to \eqref{Gamma2}
yields
\begin{align} 
\label{eq:inter_square_y} 
Int_{square}( {\bm x}_1, {\bm y}_1 , {\bm b}_1 , {\bm x}_2 , {\bm
  y}_2, {\bm b}_2) & \propto \frac{e^{\lambda \, (1+2 i \, \nu_0) \,
    Y} \, Q_{s2}^{2(1+2 i \, \nu_0)} ({\bm b})}{2 \, N_c^2 (2 \,
  \pi)^2} \int \frac{d^2 l}{l^{3+2 i \, \nu_0}} \, \frac{d^2
  l'}{{l'}^{3+2 i \, \nu_0}} e^{i \; {\bm \Delta b} \cdot ({\bm l} -
  {\bm l'})} \notag \\ & \times \, \left[ \left(1-e^{i \, {\bm l}
      \cdot {\bm x}_1} \right) \left(1-e^{- i \, {\bm l'} \cdot {\bm
        y}_1} \right) + \left(1-e^{-i \, {\bm l'} \cdot {\bm x}_1}
  \right) \left(1-e^{i \, {\bm l} \cdot {\bm y}_1} \right) \right]
\notag \\ & \times \, \left[ \left(1-e^{-i \, {\bm l} \cdot {\bm x}_2}
  \right) \left(1-e^{i \, {\bm l'} \cdot {\bm y}_2} \right) +
  \left(1-e^{i \, {\bm l'} \cdot {\bm x}_2} \right) \left(1-e^{-i \,
      {\bm l} \cdot {\bm y}_2} \right) \right].
\end{align}
Employing \eq{QsY} we conclude that the leading energy-dependence of
the connected part of the 'square' diagrams contribution to the
two-gluon production cross section is
\begin{align}
  \label{squareY}
  \frac{d \sigma_{square}}{d^2 k_1 dy_1 d^2 k_2 dy_2}
  \Bigg|_{connected} \propto e^{\lambda \, (1+2 i \, \nu_0) \, Y}.
\end{align}
This power-of-energy growth also justifies the approximation made
above in which we neglected large ($> 1/Q_{s2} (Y)$) transverse
separations.

We now study the 'crossed' diagrams contribution in
\eq{crossed_xsect}. Again the strategy is the same: use Gaussian
truncation to relate $S_G$ and $Q$ in \eq{crossed_xsect} to $\Gamma_G$
using Eqs.~\eqref{eq:SG_GM} , \eqref{eq:quad_MV} and \eqref{Ds}. Then
use $\Gamma_G$ from \eq{Gevol} to determine energy dependence of $S_G$
and $Q$. In the end, substituting this in \eq{crossed_xsect}, we would
obtain the energy dependence of this part of the two-gluon production
cross section.

From \eq{crossed_xsect_fac2} it follows that in the $k_1, k_2 \gg
Q_{s2} (Y)$ region the points ${\bm x}_1, {\bm b}_2$ are close to each
other and so are ${\bm x}_2, {\bm b}_1$. First assume that ${\bm x}_1,
{\bm b}_2$ are also close to ${\bm x}_2, {\bm b}_1$. In such a case we
can write the interaction term (the term in the second square brackets
of \eq{crossed_xsect}) as
\begin{align}
  \label{eq:intQ}
  Int_{crossed} ( {\bm x}_1, {\bm y}_1 , {\bm b}_1 , {\bm x}_2 , {\bm
    y}_2, {\bm b}_2) = Q_{s2}^4 \, \int \frac{d^2 l \, d^2 l'}{(2
    \pi)^2} \, \, \frac{1}{{l}^4} \, \frac{1}{{l}'^4} \, \bigg\{ e^{i
    \, ({\bm l} - {\bm l}') \cdot \Delta {\bm b}} \, \left( 1 - e^{i
      \, {\bm l}' \cdot {\bm {\tilde x}}_2} \right) \, \left( 1 - e^{-
      i \, {\bm l} \cdot {\bm {\tilde y}}_2} \right) \notag \\ \times
  \, \bigg[ \frac{1}{2} \, \left( 1 - e^{- i \, {\bm l}' \cdot {\bm
        {\tilde x}}_1} \right) \, \left( 1 - e^{i \, {\bm l} \cdot
      {\bm {\tilde y}}_1} \right) + \left( 1 - e^{i \, {\bm l} \cdot
      {\bm {\tilde x}}_1} \right) \, \left( 1 - e^{- i \, {\bm l}'
      \cdot {\bm {\tilde y}}_1} \right) \bigg] \notag \\ + \left( 1 -
    e^{i \, {\bm l} \cdot {\bm {\tilde x}}_1} \right) \, \left( 1 -
    e^{- i \, {\bm l}' \cdot {\bm {\tilde x}}_2} \right) \, \left( 1 -
    e^{- i \, {\bm l} \cdot {\bm {\tilde y}}_1} \right) \, \left( 1 -
    e^{i \, {\bm l}' \cdot {\bm {\tilde y}}_2} \right) \bigg\}
\end{align}
at the lowest non-trivial order in the quasi-classical approximation
(cf. Eq.~(56) in \cite{Kovchegov:2012nd}). Using the same substitution
as what led to \eqref{Gamma2} from \eqref{Gamma1} we obtain
\begin{align}
  \label{eq:intQ_y}
  Int_{crossed} ( {\bm x}_1, {\bm y}_1 , {\bm b}_1 , {\bm x}_2 , {\bm
    y}_2, {\bm b}_2) & \propto \frac{e^{\lambda \, (1+2 i \, \nu_0) \,
      Y} \, Q_{s2}^{2(1+2 i \, \nu_0)} (\bm b)}{(2 \, \pi)^2} \int
  \frac{d^2 l}{l^{3+2 i \, \nu_0}} \, \frac{d^2 l'}{{l'}^{3+2 i \,
      \nu_0}} \notag \\ & \times \bigg\{ e^{i \, ({\bm l} - {\bm l}')
    \cdot \Delta {\bm b}} \, \left( 1 - e^{i \, {\bm l}' \cdot {\bm
        {\tilde x}}_2} \right) \, \left( 1 - e^{- i \, {\bm l} \cdot
      {\bm {\tilde y}}_2} \right) \notag \\ & \times \, \bigg[
  \frac{1}{2} \, \left( 1 - e^{- i \, {\bm l}' \cdot {\bm {\tilde
          x}}_1} \right) \, \left( 1 - e^{i \, {\bm l} \cdot {\bm
        {\tilde y}}_1} \right) + \left( 1 - e^{i \, {\bm l} \cdot {\bm
        {\tilde x}}_1} \right) \, \left( 1 - e^{- i \, {\bm l}' \cdot
      {\bm {\tilde y}}_1} \right) \bigg] \notag \\ & + \left( 1 - e^{i
      \, {\bm l} \cdot {\bm {\tilde x}}_1} \right) \, \left( 1 - e^{-
      i \, {\bm l}' \cdot {\bm {\tilde x}}_2} \right) \, \left( 1 -
    e^{- i \, {\bm l} \cdot {\bm {\tilde y}}_1} \right) \, \left( 1 -
    e^{i \, {\bm l}' \cdot {\bm {\tilde y}}_2} \right) \bigg\},
\end{align}
such that
\begin{align}
  \label{crossedY}
  \frac{d \sigma_{crossed}}{d^2 k_1 dy_1 d^2 k_2 dy_2} \propto
  e^{\lambda \, (1+2 i \, \nu_0) \, Y}.
\end{align}

To properly justify \eq{crossedY} we also need to consider the case
when ${\bm x}_1, {\bm b}_2$ and ${\bm x}_2, {\bm b}_1$ from
\eq{crossed_xsect_fac2} are far apart. This is the same $D_1, D_2 \gg
1, D_3 =$~fixed regime considered earlier in the second half of
Sec.~\ref{sec:IRsafe}. Using \eq{eq:Qapprox} we see that the
Wilson-line correlator in \eq{crossed_xsect_fac2} becomes
\begin{align}
  \label{TrY}
  \left\langle \mbox{Tr} \left[ \left( \mathbb{1} - U_{{\bm x}_1}
        U_{{\bm b}_1}^\dagger \right) \left( \mathbb{1} - U_{{\bm
            x}_2} U_{{\bm b}_2}^\dagger \right) \right]
  \right\rangle_{A_2} (Y) \approx (N_c^2 - 1) \, \left[ 1 + S_G ({\bm
      x}_1, {\bm b}_2, Y) \, S_G ({\bm x}_2, {\bm b}_1, Y) \right]
  \notag \\ = (N_c^2 - 1) \, \left[ 2 + \ord{e^{\lambda \, (1+2 i \,
        \nu_0) \, Y}} \right],
\end{align}
where we have also utilized \eq{eq:SGevol}. We see that the
large-distance behavior in the 'crossed' diagrams case may not be
negligible, but grows with energy at most just as fast as the
short-distance contribution \eqref{crossedY}. (Note also that the term
in \eq{TrY}, when used in \eq{crossed_xsect_fac2}, also leads to $\sim
1/\Lambda^2_{\text{IR}}$ divergence after ${\bm \Delta
  b}$-integration, just like \eq{crossed_xsect_IR2}. If the saturation
effects in the projectile regulate this divergence, this would
generate a factor of $1/Q_{s1}^2$, which may also affect the
energy-dependence of such terms.)

Combining Eqs.~\eqref{squareY} and \eqref{crossedY} we conclude that
the net two-gluon production cross section in heavy-light ion
collisions scales with the center-of-mass energy as
\begin{align}
  \label{xsectY}
  \frac{d \sigma}{d^2 k_1 dy_1 d^2 k_2 dy_2} \propto
  e^{\lambda \, (1+2 i \, \nu_0) \, Y}.
\end{align}

To construct the correlator \eqref{corr_def2} we also need to
calculate the energy-dependence of the cross-sections for single-gluon
production. This cross-section can be written in terms of the gluon
dipole forward scattering amplitude as \cite{Kovchegov:2001sc}
\begin{align} 
\label{eq:single_gluoncross} 
\frac{d \sigma^{pA_2}}{d^2 k \, dy \, d^2 b} & = \frac{\as \, C_F}{4
  \, \pi^4} \int d^2 B \, T_1 ({\bm B} - {\bm b}) \, d^2 x \, d^2 y \,
e^{- i \; {\bm k} \cdot ({\bm x} - {\bm y})} \, \frac{ {\bm x} - {\bm
    b}}{ |{\bm x} - {\bm b} |^2 } \cdot \frac{ {\bm y} - {\bm b}}{
  |{\bm y} - {\bm b} |^2 } \notag \\ & \times \, \left[ N_G( {\bm x} ,
  {\bm b} , y) + N_G( {\bm b} , {\bm y} , y) - N_G( {\bm x} , {\bm y}
  , y) \right].
\end{align}
In the $k_T \gg Q_{s2} (Y)$ approximation we use Eqs.~\eqref{Gevol}
and \eqref{QsY} to write (see
e.g. \cite{KovchegovLevin,Jalilian-Marian:2005jf} for details of
similar integrations)
\begin{align} 
\label{eq:single_gluonint} 
\frac{d \sigma^{pA_2}}{d^2 k \, dy \, d^2 b} \propto c_0 \, e^{\lambda
  \, (1+2 i \, \nu_0) \, Y/2} \, \frac{\as \, C_F}{\pi^2} \frac{
  Q_{s2}^{1 + 2 i \, \nu_0} ({\bm b})}{2 \, k_T^{3 + 2 i \, \nu_0}}
\ln \left( \frac{k_T^2}{\Lambda^2} \right),
\end{align}
where we assume that $y \approx Y$, that is, the gluon is produced
near the projectile in rapidity, similar to our two-gluon production
case \eqref{eq_all}.  We see that the denominator of the first term of
the correlator \eqref{corr_def2} contains
\begin{align} 
\label{eq:denominator_propto} 
\frac{d \sigma}{d^2 k_1 d y_1} \, \frac{d \sigma}{d^2 k_2 d y_2}
\propto e^{\lambda \, (1+2 i \, \nu_0) \, Y}.
\end{align}

Substituting Eqs.~\eqref{xsectY} and \eqref{eq:denominator_propto}
into \eq{corr_def2} we conclude that, in the leading-power
approximation employed, the correlator is energy-independent, 
\begin{align}
  \label{CY}
  C ({\bm k}_1, y_1, {\bm k}_2, y_2) = \text{const} (Y).
\end{align}
Therefore, our two-gluon correlations are (almost) energy-independent.

%%%%%%%%%%%%%%%%%%%%%%%%%%%%%%%%%%%%%%%%%%%%%%%%%%%%%%%%%%%%%%%%%%%%%%%%%%%%%%%%%%

\section{Summary and Outlook}
\label{sec:conc}

In this paper we studied several properties of the two-gluon
production cross section in heavy-light ion collisions in the
saturation/CGC framework. We have constructed some qualitative
experimental predictions. The correlations were found to be almost
energy-independent. The CGC two-gluon long-range rapidity correlations
are stronger in tip-on-tip $U+U$ collisions, than in the side-on-side
ones. Detailed numerical predictions for the di-hadron correlation
function can be constricted by using the expressions
\eqref{eq:2glue_prod_main} and \eqref{crossed_xsect} further improved
by including running coupling corrections, but would require a
dedicated phenomenological effort.

On a more theoretical side we have constructed a new $k_T$-factorized
form of the expression given in \eq{eq:factorized_final}, involving
two new objects: the double-trace and quadrupole two-gluon Wigner
distributions \eqref{eq:doubletrace_dist} and
\eqref{eq:quad_dist}. This is by no means a proof of factorization for
two-gluon production: rather this simply is an observation that
two-gluon production cross section, calculated in the approximation
used above, can be written in this factorized form. It would be
interesting to see whether this (or any other) factorized form would
survive the inclusion of small-$x$ evolution correction in the
interval between the semi-dilute projectile and the produced
gluons. This is left for the future work.

We have also shown that the part of the two-gluon production cross
section given by \eq{crossed_xsect} contains a power-law IR
divergence, as shown in \eq{crossed_xsect_IR3}, despite all the target
wave function saturation effects included in the expression. Luckily
the azimuthal-angle dependence of the corresponding correlator is not
affected by this divergence. One could hope that saturation effects in
the projectile wave function would remove this divergence making the
final result for the two-gluon production IR-finite.

%%%%%%%%%%%%%%%%%%%%%%%%%%%%%%%%%%%%%%%%%%%%%%%%%%%%%%%%%%%%%%%%%%%%%%%%%%%%%%%%%%%%

\section*{Acknowledgments}

The authors are grateful to Miklos Gyulassy, Ulrich Heinz and Alfred
Mueller for discussions.

This research is sponsored in part by the U.S. Department of Energy
under Grant No. DE-SC0004286. \\

%%%%%%%%%%%%%%%%%%%%%%%%%%%%%%%%%%%%%%%%%%%%%%%%%%%%%%%%%%%%%%%%%%%%%%%%%%%%%%%%%%

\section*{Appendix}
\renewcommand{\theequation}{A\arabic{equation}}
  \setcounter{equation}{0}

\label{A}

Perhaps the most straightforward algebraic way to obtain evolution
equations for the correlators considered here is by applying JIMWLK
equation
\cite{Jalilian-Marian:1997dw,Jalilian-Marian:1997gr,Iancu:2001ad,Iancu:2000hn}
to the correlation functions. The application often involves tedious
algebra, but is conceptually straightforward. Below we list the
resulting leading-$\ln 1/x$ evolution equations for the expectation
values of the adjoint dipole, quadrupole, and the double-trace
operators (see e.g. \cite{KovchegovLevin} for a pedagogical
presentation of the technique we used to derive the evolution
equations).

The evolution equations below employ the following kernel, 
\begin{align}
 \label{Kernel}
 \mathcal{K}_{\bm x \bm z \bm y} \equiv \frac{\as \, N_c}{\pi^2} \,
 \frac{(\bm z - \bm x) \cdot (\bm z - \bm y)}{| \bm z - \bm x |^2 \,
   |\bm z - \bm y |^2}.
\end{align}

Defining the following expectation values of the Wilson line
correlators (normalized to one for the case of no interaction, $U = 1
= U^\dagger$)
\begin{subequations}
\label{Single_Green}
\begin{align}
  S_{{\bm x}_1 {\bm x}_2}^G & \equiv \frac{1}{N_c^2 -1} \,
  \left\langle \mbox{Tr}[ U_{{\bm x}_1} U_{{\bm x}_2}^\dagger ] \right\rangle
  \\
  S_{{\bm x}_1 {\bm x}_2 {\bm x}_3 {\bm x}_4}^{(4, dip)} & \equiv
  \frac{1}{N_c \, (N_c^2 -1)} \, \left\langle \mbox{Tr}[ T^a U_{{\bm x}_1}
    U_{{\bm x}_2}^\dagger T^a U_{{\bm x}_3} U_{{\bm x}_4}^\dagger ]
  \right\rangle
\end{align}
\end{subequations}
with $T^a$ the SU($N_c$) generators in the adjoint representation, we
write the evolution equation for the adjoint dipole correlator,
\begin{align}
\label{dip_JIMWLK}
\partial_Y S_{{\bm x}_1 {\bm x}_2}^G = \int d^2 z \bigg[ & \left(
  \mathcal{K}_{{\bm x}_1 {\bm z} {\bm x}_2} - \mathcal{K}_{{\bm x}_1
    {\bm z} {\bm x}_1} - \mathcal{K}_{{\bm x}_2 {\bm z} {\bm x}_2}
\right) \, S_{{\bm x}_1 {\bm x}_2}^G - \left( \mathcal{K}_{{\bm x}_1
    {\bm z} {\bm x}_2} - \mathcal{K}_{{\bm x}_1 {\bm z} {\bm x}_1}
\right) \, S_{{\bm z} {\bm x}_1 {\bm x}_1 {\bm x}_2}^{(4, dip)} \notag
\\ & - \left( \mathcal{K}_{{\bm x}_1 {\bm z} {\bm x}_2} -
  \mathcal{K}_{{\bm x}_2 {\bm z} {\bm x}_2} \right) \, S_{{\bm x}_2
  {\bm z} {\bm x}_1 {\bm x}_2}^{(4, dip)} + \mathcal{K}_{{\bm x}_1
  {\bm z} {\bm x}_2} \, S_{{\bm x}_2 {\bm x_1} {\bm x}_1 {\bm
    x}_2}^{(4, dip)} \bigg].
\end{align}

To write down the evolution equation for the quadrupole correlator we
will need the following definitions (also normalized to one for the
no-interaction case):
\begin{subequations}
\label{Quad_Green}
\begin{align}
  Q_{{\bm x}_1 {\bm x}_2 {\bm x}_3 {\bm x}_4} & \equiv \frac{1}{N_c^2
    -1} \, \left\langle \mbox{Tr}[ U_{{\bm x}_1} U_{{\bm x}_2}^\dagger
    U_{{\bm x}_3} U_{{\bm x}_4}^\dagger ] \right\rangle
  \\
  S_{{\bm x}_1 {\bm x}_2 {\bm x}_3 {\bm x}_4 {\bm x}_5 {\bm x}_6}^{(6,
    quad-1)} & \equiv \frac{1}{N_c \, (N_c^2 -1)} \, \left\langle [
    U_{{\bm x}_1} U_{{\bm x}_2}^\dagger ]^{ab} \, \mbox{Tr}[ T^a
    U_{{\bm x}_3} U_{{\bm x}_4}^\dagger T^b U_{{\bm x}_5} U_{{\bm
        x}_6}^\dagger ] \right\rangle
  \\
  S_{{\bm x}_1 {\bm x}_2 {\bm x}_3 {\bm x}_4 {\bm x}_5 {\bm x}_6}^{(6,
    quad-2)} & \equiv \frac{1}{N_c \, (N_c^2 -1)} \, \left\langle
    \mbox{Tr}[ T^a U_{{\bm x}_1} U_{{\bm x}_2}^\dagger T^a U_{{\bm x}_3}
    U_{{\bm x}_4}^\dagger U_{{\bm x}_5} U_{{\bm x}_6}^\dagger ]
  \right\rangle .
\end{align}
\end{subequations}
The adjoint quadrupole evolution equation reads
\begin{align}
\label{quad_JIMWLK}
\partial_Y Q_{{\bm x}_1 {\bm x}_2 {\bm x}_3 {\bm x}_4} & = \int d^2 z
\bigg[ \left( \mathcal{K}_{{\bm x}_1 {\bm z} {\bm x}_4} +
  \mathcal{K}_{{\bm x}_2 {\bm z} {\bm x}_3} - \mathcal{K}_{{\bm x}_1
    {\bm z} {\bm x}_1} - \mathcal{K}_{{\bm x}_2 {\bm z} {\bm x}_2} -
  \mathcal{K}_{{\bm x}_3 {\bm z} {\bm x}_3} - \mathcal{K}_{{\bm x}_4
    {\bm z} {\bm x}_4} \right) \, Q_{{\bm x}_1 {\bm x}_2 {\bm x}_3
  {\bm x}_4} \notag \\ & + \left( \mathcal{K}_{{\bm x}_1 {\bm z} {\bm
      x}_2} + \mathcal{K}_{{\bm x}_3 {\bm z} {\bm x}_4} -
  \mathcal{K}_{{\bm x}_1 {\bm z} {\bm x}_3} - \mathcal{K}_{{\bm x}_2
    {\bm z} {\bm x}_4} \right) \, S_{{\bm x}_1 {\bm x}_2 {\bm x}_3
  {\bm x}_4}^{(4, dip)} \notag \\ & + \left( \mathcal{K}_{{\bm x}_1
    {\bm z} {\bm x}_3} - \mathcal{K}_{{\bm x}_1 {\bm z} {\bm x}_2}
\right) \, S_{{\bm x}_1 {\bm z} {\bm x}_1 {\bm x}_2 {\bm x}_3 {\bm
    x}_4}^{(6, quad-1)} + \mathcal{K}_{{\bm x}_1 {\bm z} {\bm x}_2} \,
S_{{\bm x}_1 {\bm x}_2 {\bm x}_1 {\bm x}_2 {\bm x}_3 {\bm x}_4}^{(6,
  quad-1)} \notag \\ & + \left( \mathcal{K}_{{\bm x}_1 {\bm z} {\bm
      x}_3} - \mathcal{K}_{{\bm x}_3 {\bm z} {\bm x}_4} \right) \,
S_{{\bm z} {\bm x}_3 {\bm x}_1 {\bm x}_2 {\bm x}_3 {\bm x}_4}^{(6,
  quad-1)} - \mathcal{K}_{{\bm x}_1 {\bm z} {\bm x}_3} \, S_{{\bm x}_1
  {\bm x}_3 {\bm x}_1 {\bm x}_2 {\bm x}_3 {\bm x}_4}^{(6, quad-1)}
\notag \\ & + \left( \mathcal{K}_{{\bm x}_2 {\bm z} {\bm x}_4} -
  \mathcal{K}_{{\bm x}_1 {\bm z} {\bm x}_2} \right) \, S_{{\bm z} {\bm
    x}_2 {\bm x}_1 {\bm x}_2 {\bm x}_3 {\bm x}_4}^{(6, quad-1)} +
\mathcal{K}_{{\bm x}_3 {\bm z} {\bm x}_4} \, S_{{\bm x}_4 {\bm x}_3
  {\bm x}_1 {\bm x}_2 {\bm x}_3 {\bm x}_4}^{(6,
  quad-1)} \notag \\
& + \left( \mathcal{K}_{{\bm x}_2 {\bm z} {\bm x}_4} -
  \mathcal{K}_{{\bm x}_3 {\bm z} {\bm x}_4} \right) \, S_{{\bm x}_4
  {\bm z} {\bm x}_1 {\bm x}_2 {\bm x}_3 {\bm x}_4}^{(6, quad-1)} -
\mathcal{K}_{{\bm x}_2 {\bm z} {\bm x}_4} \, S_{{\bm x}_4 {\bm x}_2
  {\bm x}_1 {\bm x}_2 {\bm x}_3 {\bm x}_4}^{(6, quad-1)} \notag \\ & +
\left( \mathcal{K}_{{\bm x}_1 {\bm z} {\bm x}_1} - \mathcal{K}_{{\bm
      x}_1 {\bm z} {\bm x}_4} \right) \, S_{{\bm z} {\bm x}_1 {\bm
    x}_1 {\bm x}_2 {\bm x}_3 {\bm x}_4}^{(6, quad-2)} + \left(
  \mathcal{K}_{{\bm x}_4 {\bm z} {\bm x}_4} - \mathcal{K}_{{\bm x}_1
    {\bm z} {\bm x}_4} \right) \, S_{{\bm x}_4 {\bm z} {\bm x}_1 {\bm
    x}_2 {\bm x}_3 {\bm x}_4}^{(6, quad-2)} \notag \\ & +
\mathcal{K}_{{\bm x}_1 {\bm z} {\bm x}_4} \, S_{{\bm x}_4 {\bm x}_1
  {\bm x}_1 {\bm x}_2 {\bm x}_3 {\bm x}_4}^{(6, quad-2)} \notag \\ & +
\left( \mathcal{K}_{{\bm x}_2 {\bm z} {\bm x}_2} - \mathcal{K}_{{\bm
      x}_2 {\bm z} {\bm x}_3} \right) \, S_{{\bm x}_2 {\bm z} {\bm
    x}_3 {\bm x}_4 {\bm x}_1 {\bm x}_2}^{(6, quad-2)} + \left(
  \mathcal{K}_{{\bm x}_3 {\bm z} {\bm x}_3} - \mathcal{K}_{{\bm x}_2
    {\bm z} {\bm x}_3} \right) \, S_{{\bm z} {\bm x}_3 {\bm x}_3 {\bm
    x}_4 {\bm x}_1 {\bm x}_2}^{(6, quad-2)} \notag \\ & +
\mathcal{K}_{{\bm x}_2 {\bm z} {\bm x}_3} \, S_{{\bm x}_2 {\bm x}_3
  {\bm x}_3 {\bm x}_4 {\bm x}_1 {\bm x}_2}^{(6, quad-2)} \bigg].
\end{align}

Finally, for the evolution of the double-trace operator we need the
following definitions:
\begin{subequations}
\label{Dipole_Green}
\begin{align}
  D_{{\bm x}_1 {\bm x}_2 {\bm x}_3 {\bm x}_4} & \equiv \frac{1}{(N^2_c
    -1)^2} \, \left\langle \mbox{Tr}[ U_{{\bm x}_1} U_{{\bm
        x}_2}^\dagger ] \, \mbox{Tr}[ U_{{\bm x}_3} U_{{\bm
        x}_4}^\dagger ] \right\rangle
  \\
  S_{{\bm x}_1 {\bm x}_2 {\bm x}_3 {\bm x}_4}^{(4, double)} & \equiv
  \frac{1}{N_c \, (N^2_c -1)^2} \, \left\langle \mbox{Tr}[ T^a U_{{\bm
        x}_1} U_{{\bm x}_2}^\dagger ] \, \mbox{Tr}[ T^a U_{{\bm x}_3}
    U_{{\bm x}_4}^\dagger ] \right\rangle
  \\
  S_{{\bm x}_1 {\bm x}_2 {\bm x}_3 {\bm x}_4 {\bm x}_5 {\bm x}_6}^{(6,
    double - 1)} & \equiv \frac{1}{N_c \, (N^2_c -1)^2} \,
  \left\langle [ U_{{\bm x}_1} U_{{\bm x}_2}^\dagger ]^{ab} \,
    \mbox{Tr}[ T^a U_{{\bm x}_3} U_{{\bm x}_4}^\dagger ] \, \mbox{Tr}[
    T^b U_{{\bm x}_5} U_{{\bm x}_6}^\dagger ] \right\rangle
  \\
  S_{{\bm x}_1 {\bm x}_2 {\bm x}_3 {\bm x}_4 {\bm x}_5 {\bm x}_6}^{(6,
    double - 2)} & \equiv \frac{1}{N_c \, (N^2_c -1)^2} \,
  \left\langle \mbox{Tr}[ T^a U_{{\bm x}_1} U_{{\bm x}_2}^\dagger T^a
    U_{{\bm x}_3} U_{{\bm x}_4}^\dagger ] \mbox{Tr}[ U_{{\bm x}_5}
    U_{{\bm x}_6}^\dagger ] \right\rangle .
\end{align}
\end{subequations}
(Note that while $D_{{\bm x}_1 {\bm x}_2 {\bm x}_3 {\bm x}_4}$ and
$S_{{\bm x}_1 {\bm x}_2 {\bm x}_3 {\bm x}_4 {\bm x}_5 {\bm x}_6}^{(6,
  double - 2)}$ are normalized to one in the no-interactions case,
$S_{{\bm x}_1 {\bm x}_2 {\bm x}_3 {\bm x}_4}^{(4, double)}$ and
$S_{{\bm x}_1 {\bm x}_2 {\bm x}_3 {\bm x}_4 {\bm x}_5 {\bm x}_6}^{(6,
  double - 1)}$ actually vanish for $U = U^\dagger =1$, such that
their normalization is arbitrary, fixed here to match the
normalization of the other correlators.) Using the correlators in
\eqref{Dipole_Green} we obtain the evolution equation for the adjoint
double-trace operator
\begin{align}
\label{double_tr_JIMWLK}
\partial_Y D_{{\bm x}_1 {\bm x}_2 {\bm x}_3 {\bm x}_4} & = \int d^2 z
\, \bigg[ \left( \mathcal{K}_{{\bm x}_1 {\bm z} {\bm x}_2} +
  \mathcal{K}_{{\bm x}_3 {\bm z} {\bm x}_4} - \mathcal{K}_{{\bm x}_1
    {\bm z} {\bm x}_1} - \mathcal{K}_{{\bm x}_2 {\bm z} {\bm x}_2} -
  \mathcal{K}_{{\bm x}_3 {\bm z} {\bm x}_3} - \mathcal{K}_{{\bm x}_4
    {\bm z} {\bm x}_4} \right) \, D_{{\bm x}_1 {\bm x}_2 {\bm x}_3
  {\bm x}_4} \notag \\ & + \left( \mathcal{K}_{{\bm x}_1 {\bm z} {\bm
      x}_4} + \mathcal{K}_{{\bm x}_2 {\bm z} {\bm x}_3} -
  \mathcal{K}_{{\bm x}_1 {\bm z} {\bm x}_3} - \mathcal{K}_{{\bm x}_2
    {\bm z} {\bm x}_4} \right) \, S_{{\bm x}_1 {\bm x}_2 {\bm x}_3
  {\bm x}_4}^{(4, double)} \notag \\ & + \left( \mathcal{K}_{{\bm x}_1
    {\bm z} {\bm x}_3} - \mathcal{K}_{{\bm x}_1 {\bm z} {\bm x}_4}
\right) \, S_{{\bm x}_1 {\bm z} {\bm x}_1 {\bm x}_2 {\bm x}_3 {\bm
    x}_4}^{(6, double - 1)} + \mathcal{K}_{{\bm x}_1 {\bm z} {\bm
    x}_4} \, S_{{\bm x}_1 {\bm x}_4 {\bm x}_1 {\bm x}_2 {\bm x}_3 {\bm
    x}_4}^{(6, double - 1)} \notag \\ & + \left( \mathcal{K}_{{\bm
      x}_1 {\bm z} {\bm x}_3} - \mathcal{K}_{{\bm x}_2 {\bm z} {\bm
      x}_3} \right) \, S_{{\bm z} {\bm x}_3 {\bm x}_1 {\bm x}_2 {\bm
    x}_3 {\bm x}_4}^{(6, double - 1)} - \mathcal{K}_{{\bm x}_1 {\bm z}
  {\bm x}_3} \, S_{{\bm x}_1 {\bm x}_3 {\bm x}_1 {\bm x}_2 {\bm x}_3
  {\bm
    x}_4}^{(6, double - 1)} \notag \\
& + \left( \mathcal{K}_{{\bm x}_2 {\bm z} {\bm x}_4} -
  \mathcal{K}_{{\bm x}_2 {\bm z} {\bm x}_3} \right) \, S_{{\bm x}_2
  {\bm z} {\bm x}_1 {\bm x}_2 {\bm x}_3 {\bm x}_4}^{(6, double - 1)} +
\mathcal{K}_{{\bm x}_2 {\bm z} {\bm x}_3} \, S_{{\bm x}_2 {\bm x}_3
  {\bm x}_1 {\bm x}_2 {\bm x}_3 {\bm x}_4}^{(6, double - 1)} \notag \\
& + \left( \mathcal{K}_{{\bm x}_2 {\bm z} {\bm x}_4} -
  \mathcal{K}_{{\bm x}_1 {\bm z} {\bm x}_4} \right) \, S_{{\bm z} {\bm
    x}_4 {\bm x}_1 {\bm x}_2 {\bm x}_3 {\bm x}_4}^{(6, double - 1)} -
\mathcal{K}_{{\bm x}_2 {\bm z} {\bm x}_4} \, S_{{\bm x}_2 {\bm x}_4
  {\bm x}_1 {\bm x}_2 {\bm x}_3 {\bm x}_4}^{(6, double - 1)} \notag \\
& + \left( \mathcal{K}_{{\bm x}_1 {\bm z} {\bm x}_1} -
  \mathcal{K}_{{\bm x}_1 {\bm z} {\bm x}_2} \right) \, S_{{\bm z} {\bm
    x}_1 {\bm x}_1 {\bm x}_2 {\bm x}_3 {\bm x}_4}^{(6, double - 2)} +
\left( \mathcal{K}_{{\bm x}_2 {\bm z} {\bm x}_2} - \mathcal{K}_{{\bm
      x}_1 {\bm z} {\bm x}_2} \right) \, S_{{\bm x}_2 {\bm z} {\bm
    x}_1 {\bm x}_2 {\bm x}_3 {\bm x}_4}^{(6, double - 2)} \notag \\ &
+ \mathcal{K}_{{\bm x}_1 {\bm z} {\bm x}_2} \, S_{{\bm x}_2 {\bm x}_1
  {\bm x}_1 {\bm x}_2 {\bm x}_3 {\bm x}_4}^{(6, double - 2)} \notag \\
& + \left( \mathcal{K}_{{\bm x}_3 {\bm z} {\bm x}_3} -
  \mathcal{K}_{{\bm x}_3 {\bm z} {\bm x}_4} \right) \, S_{{\bm z} {\bm
    x}_3 {\bm x}_3 {\bm x}_4 {\bm x}_1 {\bm x}_2}^{(6, double - 2)} +
\left(\mathcal{K}_{{\bm x}_4 {\bm z} {\bm x}_4} - \mathcal{K}_{{\bm
      x}_3 {\bm z} {\bm x}_4} \right) \, S_{{\bm x}_4 {\bm z} {\bm
    x}_3 {\bm x}_4 {\bm x}_1 {\bm x}_2}^{(6, double - 2)} \notag \\ &
+ \mathcal{K}_{{\bm x}_3 {\bm z} {\bm x}_4} \, S_{{\bm x}_4 {\bm x}_3
  {\bm x}_3 {\bm x}_4 {\bm x}_1 {\bm x}_2}^{(6, double - 2)} \bigg].
\end{align}

None of the equations \eqref{dip_JIMWLK}, \eqref{quad_JIMWLK}, and
\eqref{double_tr_JIMWLK} are closed equations: they include
higher-order Wilson line correlators on their right-hand sides, as
expected from the equations in Balitsky hierarchy
\cite{Balitsky:1996ub,Balitsky:1998ya}.

%%%%%%%%%%%%%%%%%%%%%%%%%%%%%%%%%%%%%%%%%%%%%%%%%%%%%%%%%%%%%%%%%%%%%%%%%%%%%%

%\bibliographystyle{JHEP}
%\bibliography{references}

\providecommand{\href}[2]{#2}\begingroup\raggedright\endgroup

%%%%%%%%%%%%%%%%%%%%%%%%%%%%%%%%%%%%%%%%%%%%%%%%%%%%%%%%%%%%%%%%%%%%%%%%%%%%%%

\end{document}